\documentclass[3p,authoryear]{elsarticle}
\usepackage{epstopdf}
\usepackage{subfigure,graphicx,tabularx}
\usepackage{float}
\usepackage{amsmath, amsfonts, amssymb,mathrsfs}
\usepackage{amsthm}
\usepackage{epsf}
\usepackage{amsfonts}
\usepackage{stmaryrd}
\usepackage{amssymb}
\usepackage{leftidx}
\usepackage{color}
\usepackage{mathtools}
\usepackage{placeins}
\usepackage{booktabs}
\usepackage{enumitem}
\usepackage{caption}
\usepackage{multirow}
\usepackage{stackengine}
\usepackage[utf8]{inputenc}
\usepackage[english]{babel}
\usepackage{color}
\usepackage{pifont}
\usepackage{fontawesome}

\newcolumntype{?}{!{\vrule width 1pt}}

\theoremstyle{plain}
\newtheorem{theorem}{Theorem}[section]
\newtheorem{proposition}[theorem]{Proposition}
\newtheorem{lemma}[theorem]{Lemma}

\newtheorem{remark}[theorem]{Remark}
\newtheorem{definition}[theorem]{Definition}

\def\bfb{{\bf b}}

\def\bfs{{\bf s}}
\def\bfu{{\bf u}}
\def\bfv{{\bf v}}

\def\bfy{{\bf y}}

\def\bfA{{\bf A}}
\def\bfE{{\bf E}}

\def\bfC{{\bf C}}

\def\bfI{{\bf I}}

\def\bfS{{\bf S}}

\def\bfX{{\bf X}}
\def\bfF{{\bf F}}

\def\bfsig{{\boldsymbol \sigma}}

\def\bfe{{\bf e}}
\def\bfc{{\bf c}}

\def\bfs{{\bf s}}

\def\bftau{\mbox{\boldmath $\tau$}}

\def\obfF{\overline{\bfF}}

\def\obfS{\overline{\bfS}}

\def\obfS{\overline{\bfS}}
\def\obfF{\overline{\bfF}}

\newcommand\sts{s_{\texttt{ts}}}
\newcommand\scs{s_{\texttt{cs}}}
\newcommand\shs{s_{\texttt{hs}}}
\newcommand\sbs{s_{\texttt{bs}}}
\newcommand\sss{s_{\texttt{ss}}}

\newcommand{\as}{{\mathcal A}}
\newcommand{\hs}{{\mathcal H}}
\newcommand{\ks}{{\mathcal K}}

\newcommand{\leb}{{\mathcal L}}

\newcommand\oli{\overline}

\newcommand{\R}{{\mathbb R}}

\newcommand{\N}{{\mathbb N}}

\newcommand{\Msym}{{\rm M}^{N}_{\rm sym}}
\newcommand{\Mtwosym}{{\rm M}^{2}_{\rm sym}}

\newcommand\LDD{{\mathscr L\!\!\mathscr D}}

\newcommand{\Om}{\Omega}
\newcommand{\Omb}{\overline{\Omega}}

\newcommand{\weak}{\rightharpoonup}

\newcommand{\htwo}{\hs^{1}}

\newcommand{\hu}{\hs^1}

\newcommand{\bel}[1]{\begin{equation}\label{#1}}
\newcommand{\ee}{\end{equation}}

\newcommand{\pa}{\partial}

\definecolor{verde}{RGB}{50,150,80}

\begin{document}
\begin{frontmatter}

\title{Nucleation and propagation of brittle fracture as a constrained energy minimization problem}

\author[Illinois]{Oscar Lopez-Pamies}
\ead{pamies@illinois.edu}

\author[Illinois]{Farhad Kamarei}
\ead{kamarei2@illinois.edu}

\author[FlatIron]{Gilles A. Francfort}
\ead{gfrancfort@flatironinstitute.org}

\author[Brescia]{Alessandro Giacomini}
\ead{alessandro.giacomini@unibs.it}

\address[Illinois]{Department of Civil and Environmental Engineering, University of Illinois, Urbana--Champaign, IL 61801, USA \vspace{0.1cm}}

\address[FlatIron]{Flatiron Institute, 162 Fifth Avenue, New York, NY 10010, USA \vspace{0.1cm}}

\address[Brescia]{DICATAM, Sezione di Matematica, Università degli Studi di Brescia, Brescia, Italy  \vspace{0.1cm}}

\begin{abstract}

\vspace{-0.1cm}

This paper presents a macroscopic, or continuum, theory aimed at describing when, where, and why cracks nucleate and propagate in nominally elastic brittle materials under monotonic, quasi-static, but otherwise arbitrary mechanical loads. Motivated by recent insights, the proposed sharp theory posits that: \emph{cracks nucleate and propagate exclusively in regions where the strength surface of the material is exceeded, with their evolution dictated by the minimization of the sum of the potential --- the elastic energy minus the work done by the externally applied forces --- and surface energies.} While the theory applies, {\color{black} in principle,} to materials with any elasticity (linear or nonlinear) and any material symmetry (isotropic or anisotropic), attention is restricted here to the most basic case of isotropic elastic brittle materials. For demonstration purposes, the theory is confronted with a set of nine tests that span the entire range of well-settled experimental knowledge on fracture nucleation and propagation --- the so-called ``Nine Circles of Elastic Brittle Fracture'' --- on both a hard material (a silicate glass) and a soft material (a synthetic rubber).

\vspace{0cm}

\endkeyword

\end{abstract}

\end{frontmatter}

\section{Introduction}\label{Sec: Intro}

Over the past decade, we, together with various collaborators, have built a body of research \citep{KFLP18,KRLP18,FGLP19,KLP20,KBFLP20,KLP21,KRLP22,KLDLP24,LDLP24,KKLP24,SDLP24,KLP25,KZDLP26,SRLP26,SMMRLP26,LPK25} suggesting that a comprehensive continuum description of fracture nucleation and propagation in solids is within reach. Building on these efforts,  this paper takes a first step toward such an objective by initiating a series aimed at describing when, where, and why cracks nucleate and propagate in nominally elastic brittle materials subjected to mechanical loads. Recall that an elastic brittle material --- the most basic idealization of a solid capable of fracture --- is one that, in response to mechanical forces, either deforms elastically or creates new surfaces, the latter being the sole mechanism by which it dissipates energy. Specifically, this first paper addresses these questions for the case of monotonic quasi-static loading, deferring non-monotonic and dynamic loading to future work. While the proposed theory is applicable, {\color{black} in principle,} to any elastic constitutive behavior (linear or nonlinear) and material symmetry (isotropic or anisotropic), the present work focuses on the most common setting, that of isotropic elastic brittle materials.

Drawing on over two centuries of experimental observations across a diverse range of materials --- including rocks, concrete, ceramics, inorganic and organic glasses, and elastomers ---  the aforementioned studies have provided the following insights.

The first is a precise and complete definition of strength, a macroscopic property that has been historically a subject of mistreatment and misinterpretation; see Subsection \ref{Sec: History}. As in \citep{KLP20,KBFLP20} and as restated in Definition \ref{Def-Strength} below, the strength of an elastic brittle material is defined as the set of all critical stresses\footnote{Throughout this work, $\bfS$ denotes the first Piola-Kirchhoff stress tensor.} $\bfS$ at which fracture nucleates in a specimen\footnote{As for any macroscopic property to make sense, the specimen must be  large enough in all of its dimensions to be viewed as a homogeneous piece. What ``enough'' means for strength will be addressed later in Remark \ref{Remark_length}.} under a state of monotonically increasing, spatially uniform, but otherwise arbitrary stress. This collection of critical stress states defines a surface in stress space, $\mathcal{F}(\bfS)=0$, referred to as the \emph{strength surface} of the material.  

Then, that elastic brittle fracture necessitates three distinct macroscopic properties: ($i$) the elasticity of the material; ($ii$) its strength surface; and ($iii$) its fracture toughness or critical energy release rate. In particular, for the basic case of isotropic elastic brittle materials, any viable formulation must incorporate the isotropic elastic energy density
\begin{equation*}
W(\bfF)=\varphi(I_1,I_2,J)
\end{equation*}
of the material, its isotropic strength surface
\begin{equation*}
\mathcal{F}(\mathbf{S})=\widehat{\mathcal{F}}(s_1,s_2,s_3)=0,
\end{equation*}
and its scalar fracture toughness
\begin{equation*}
G_c.
\end{equation*}
In these expressions, $\bfF=\nabla\bfy$ denotes the gradient of the deformation field $\bfy$, $I_1={\rm tr}\,\bfC$, $I_2=\frac{1}{2}(I_1^2-{\rm tr}\,\bfC^2)$, and $J=\sqrt{\det\bfC}=\det\bfF$ stand for the classical principal invariants associated with the right Cauchy-Green deformation tensor $\bfC=\bfF^T\bfF$, and $s_1\geq s_2 \geq s_3$ are the principal nominal {\color{black}(Biot)} stresses.

Finally, that fracture itself is governed by a two-fold criterion:
\begin{enumerate}

\item{Cracks can nucleate or grow only in regions where the strength surface of the material, $\mathcal{F}(\bfS)=0$, has been exceeded.}
    
\item{Within the regions of strength violation, crack evolution is governed by a minimization of a functional, involving the competition between deformation and surface energies.}

\end{enumerate}
Note that the first part implies that the violation of the strength surface is a necessary, albeit \emph{not} sufficient, condition for the nucleation and propagation of fracture. The second part states that sufficiency is provided by an energetic competition within the strength violation regions. 

With this in mind, we present in Section \ref{Sec: The Theory} the central focus of this paper: a macroscopic, or continuum, theory aimed at describing when, where, and why sharp cracks nucleate and propagate in nominally elastic brittle materials under arbitrary monotonic quasi-static loading conditions.

At present, the theory is limited to time-discrete evolutions, and the corresponding time-continuous formulation remains an open challenge. This is so, in part, because the presence of strength in the theory impacts the amount of energy dissipated and possibly the minimizing principle in the time-continuous framework. The mathematical obstacles are numerous. A first set of mathematical results for well-posedness is presented in Section \ref{Sec: Math}.

As a preliminary, Section \ref{Sec: A representative case} analyzes three motivating experiments: a silicate glass specimen in simple tension under both displacement and force control, and the self-weight cantilever bending of a mortar beam. In Section \ref{Sec: Nine Circles}, we confront the theory with a set of nine tests spanning the complete spectrum of well-settled experimental knowledge on fracture nucleation and propagation --- the so-called ``Nine Circles of Elastic Brittle Fracture'' introduced in \citep{KZDLP26} --- for both a hard material (silicate glass) and a soft material (synthetic rubber). Finally, Section \ref{Sec: Final Comments} provides a few concluding comments.

\subsection{A brief overview of the history and current understanding of fracture from the continuum point of view}\label{Sec: History}

The following overview serves not only to situate the current work within the broader literature of solid mechanics, but also to ``$\acute{A}\pi\acute{o}\delta o\tau\varepsilon \ o\tilde{\upsilon}\nu \ \tau\grave{\alpha} \ K\alpha\acute{\iota}\sigma\alpha\rho o\varsigma \ K\alpha\acute{\iota}\sigma\alpha\rho\iota$ [Render unto Caesar the things which are Caesar's]''\footnote{Matthew 22:21.} and to underscore the tortuous path that research in this area has historically followed. The summary is divided into the three different types of mechanics that govern fracture, each restricted to the monotonic quasi-static loading conditions of interest in this paper: ($i$) the mechanics of deformation; ($ii$) the mechanics of strength; and ($iii$) the mechanics of toughness.

\subsubsection{The mechanics of elastic deformation} 

Fifteen years after \cite{Cauchy1823} had introduced the definition of stress, the equations that describe the deformation of a body made of a linear elastic material were established in their full generality by \cite{Green1838}. Prior to Green's work, \cite{Navier1827} and \cite{Cauchy1828}, among others, had written the equations, but with a degree of confusion as to  material symmetry; see, e.g., the historical account in the classical monograph  by 
\cite{Love1906}.  \cite{Lame1852}, for his part, did write the equations as they now stand. In an isotropic nonlinear setting, the correct equations were first written by \cite{Rivlin48I,Rivlin48IV} over a century later. Finally, for a body made of a fully anisotropic nonlinear elastic material, the equations were worked out by \cite{TN1965}.

\subsubsection{The mechanics of strength} 

Lam\'e was first to explicitly mention  that the equations of linear elastostatics cease to be valid at sufficiently large loads when the material fractures \citep{Lame1831}. He solved the boundary-value problems of a thick-walled circular cylinder and a thick-walled spherical shell, made of an isotropic linear elastic material, under inner and outer pressures and postulated that the specimens would fracture when the hoop stress at their inner boundary  reaches a critical value characteristic of the material, the ``tensile strength'' of the material. In other words, he posited that when the maximum tensile principal stress at a material point in a structure reaches that value, $s_1=\sts$ say, fracture would ensue at that material point. Fracture mechanics from the continuum point of view was born and it was all about nucleation. 

About two decades later, \cite{Rankine1858} generalized Lam\`e's criterion by postulating that fracture would ensue whenever either the maximum tensile or minimum compressive principal stress at a material point in a structure would reach critical values characteristic of the material, the ``tensile strength'' or ``compressive strength'', that is, whenever $s_1=\sts$ or $s_3=-\scs$. He applied this criterion, in particular, to beams subjected to transverse loading. A few years later, Saint-Venant  proposed the alternative view that fracture ensues at a material point whenever the maximum tensile principal strain --- as opposed to stress --- reaches a critical value \citep{Navier1864}. \cite{Mohr1900} noted that none of the aforementioned criteria could explain experimental observations in general. Instead, he proposed a criterion that depends on the maximum and minimum principal stresses, one  of the form $s_1-s_3=f(s_1+s_3)$, where the function $f(\cdot)$ is material specific. Following in his peers' footsteps, Mohr also posited that satisfaction of this criterion at a material point would result in fracture at that point. Mohr's approach was soon shown not to be experimentally viable either \citep{vonKarman1911}.

The notion of strength at material points continued to dominate the field until the early 1920s, when \cite{Griffith21,Griffith1924} steered the mechanics community in an entirely new direction, that of fracture propagation and the mechanics of toughness. The shift turned out to be so influential that theoretical studies of fracture nucleation and the mechanics of strength were \emph{de facto} abandoned for nearly a century. A resurgence of the topic took place in the early 2000s, prompted by investigations into fracture nucleation from notches in hard materials \citep[see also Remark \ref{Remark_FFM} below]{Leguillon2002,Cornetti2006,Mantic2014} and cavitation in elastomers \citep{LRLP15,BCLLP24}. The latter led to the following definition of strength introduced in \citep{KLP20,KBFLP20} and already  mentioned above:
\begin{definition}\label{Def-Strength} The strength of an elastic brittle material is the set of all critical stresses $\bfS$ at which the material fractures when it is subjected to a state of monotonically increasing, spatially uniform, but otherwise arbitrary stress. Such a set of critical stresses defines a surface $\mathcal{F}(\bfS)=0$ in stress space --- potentially, any star-shaped surface containing the origin $\bfS=\emph{\textbf{0}}$  --- which is referred to as the strength surface of the material.
\end{definition}
\noindent This definition redresses two early misconceptions. First, that violation of a stress (or strain) condition at a material point implies nucleation of fracture. Experiments show that, for fracture nucleation to occur, the violation of a stress condition has to take place over a sufficiently large subdomain whose size is material dependent. Then, that the stress condition has a universal form. Experimental evidence points to the contrary. 

As such, Definition \ref{Def-Strength} says nothing about when fracture nucleates in problems where the stress field is not spatially uniform and thus it says nothing about fracture nucleation at individual material points; as elaborated below, additional criteria are necessary to determine the onset of fracture when the stress state is spatially non-uniform. Definition \ref{Def-Strength} also says nothing about a particular form of the strength surface $\mathcal{F}(\bfS)=0$; it simply must be star-shaped and contain $\bfS=\textbf{0}$ in its interior.

\begin{remark}{\color{black}
Strength surfaces versus other surfaces in stress space. \emph{A recurring source of misunderstanding is the conflation of the strength surface $\mathcal{F}(\mathbf{S}) = 0$ in Definition \ref{Def-Strength} with other types of surfaces in stress space, most notably yield surfaces. While they share the same mathematical representation, their physical meanings are entirely distinct. Yield surfaces govern the onset of inelastic \emph{deformation}, a mechanism absent in elastic brittle materials, which deform solely elastically. The strength surface $\mathcal{F}(\mathbf{S}) = 0$, by contrast, has nothing to do with the mechanics of deformation. It is a macroscopic material property associated strictly with the mechanics of \emph{surface creation}.}}
\end{remark}

\subsubsection{The mechanics of toughness}\label{Toughness-History}

\medskip

\paragraph{The original idea of energy competition}  Aware of the inability of the prevailing strength conditions to explain experimental observations, \cite{Griffith21,Griffith1924} introduced an energy-based approach, positing that fracture is governed by a global energy balance. In his own words \cite[page 55]{Griffith1924}: ``\emph{If owing to the action of a stress a crack is formed, or a pre-existing crack is caused to extend, therefore, a quantity of energy proportional to the area of the new surface must be added, and the condition that this shall be possible is that such addition of energy shall take place without any increase in the potential energy of the system} [the sum of the potential energy of the applied forces, the strain energy, and the surface energy of the crack].'' Put differently, an elastic brittle body subjected to external loads can either deform elastically or undergo a combined process of elastic deformation and surface creation, both processes being consistent with energy conservation. The process with surface creation involves a redistribution of energy, from deformation energy to surface energy, with the latter being proportional to the area of the newly created surface. 

\cite{Griffith21,Griffith1924} applied his criterion to an elementary problem, which helped clarify three key aspects of his approach. The problem was that of an infinite plate, made of an isotropic linear elastic material, containing a pre-existing straight through-thickness crack, that is subjected to affine biaxial tensile tractions aligned with the crack. The goal was to determine when --- that is, at what value of the applied tractions --- the crack starts to grow, assuming a straight-ahead trajectory. In his own words \cite[page 57]{Griffith1924}: ``\emph{The condition that the crack may extend, that is, that rupture may occur, is then obtained by expressing the condition that the total energy} [the sum of the potential energy of the applied tractions, the strain energy and the surface energy of the crack] \emph{is unchanged by small variations in the length of the crack.}'' This reveals three significant aspects of his approach: first, the criterion applies strictly to the infinitesimal growth of pre-existing cracks rather than finite-size growth or fracture nucleation; second, it offers no predictive information regarding the resulting crack path; and third, its application amounts to computing variations of the elastic energy minus the work done by the applied forces with respect to infinitesimal add-cracks along a pre-supposed crack path, a quantity that is now classically denoted as $G$ and referred to as the energy release rate.

\paragraph{The physical meaning of $G_c$} \cite{Griffith21,Griffith1924} postulated that the constant of proportionality multiplying the area of the newly created surface in the global energy balance was twice the surface tension of the material (at interfaces with air). Recognizing that this choice often underestimated experimental measurements, \cite{Orowan48}, \cite{Irwin48}, and \cite{RT53} independently proposed a broader definition. They argued that the proportionality constant should represent the total energy, per unit undeformed area, expended  during the fracture process. This material property --- measurable through macroscopic testing --- is currently denoted as $G_c$ and referred to as the critical energy release rate or fracture toughness.

\paragraph{Equivalency of $G$ and stress intensity factors} In 1957, seeking to better understand the local implications of the global energy balance proposed by Griffith, within the setting of 2D isotropic linear elastostatics, \cite{Irwin57} established that the energy release rate $G$ --- again, associated with a pre-supposed crack path --- can be expressed in terms of the so-called stress intensity factors, the coefficients in the leading singular terms of the stress field at a crack front. The equivalency between the energy release rate (global perspective) and the crack front singularities (local perspective) was shown to hold true more generally in 3D and isotropic finite elastostatics in subsequent works \citep[e.g.,][]{Rice1966,KS1972,KS1973,BudyRice1973}. This insight provided by \cite{Irwin57} sparked a vast body of research dedicated to characterizing stress intensity factors across diverse geometries, materials, and loading conditions. Although predominantly focused on isotropic linear elastostatics, this pursuit remains active today and has resulted in reference handbooks; see, e.g., \citep{Tada73}. 

\paragraph{Equivalency of $G$ and the $J$-integral} In a related contribution published in 1968, within the setting of 2D isotropic small-strain but possibly nonlinear elastostatics, \cite{Rice1968} established that the energy release rate $G$ can additionally be expressed in terms of a path-independent integral, the $J$-integral, arising out of material invariances. Later works showed that such path-independent integrals can be worked out more generally in 3D and isotropic finite elastostatics \citep[e.g.,][]{KS1972,BudyRice1973}. Because it makes the determination of $G$ more palatable, the computation of the $J$-integral has become a standard feature in many commercial FE packages.

\paragraph{Crack path based on the stress field around the crack front} By the early 1960s, numerous comparisons with experiments on different materials had demonstrated an apparent validity of Griffith's idea of energy competition to describe and predict when a large\footnote{``Large'' refers to large relative to the characteristic size of the underlying heterogeneities in the material at hand. By the same token, ``small'' refers to sizes that are of the same order or just moderately larger than the sizes of the heterogeneities.} pre-existing crack starts to grow along a crack path that is known \emph{a priori}. The focus then turned to the prediction of the crack path. Starting with the work of \cite{Erdogan63}, who, motivated by their own experiments on PMMA, proposed the so-called Maximum Tangential Stress criterion, a slew of criteria based on the local stress fields around crack fronts were, and continue to be put forth; see, e.g., \citep{Sih74,Goldstein74,Nuismer75,Knauss78,Wu78,Theocaris82,Tian82} and the reviews \citep{Fatemi96,Hua23}. Most of them only apply to isotropic linear elastic brittle materials in 2D settings. Comparisons with the many experiments that have been carried out to probe nontrivial crack paths in a variety of materials seem to indicate that none of these have a universal character \citep{Panasyuk65-RUS,Knauss78,Williams72,Ueda83,Royer86,Richard85,Ravi89,Davenport93,Richard2014}. Also none are amenable to generalizations to nonlinear elastic and anisotropic materials in a 3D setting. 

\paragraph{The variational approach to fracture} At the end of the 1990s, \cite{Francfort98} introduced a new idea to describe crack path within the setting of linear elastic brittle materials subject to mixed Dirichlet and zero Neumann boundary conditions and in the absence of body forces. In a nutshell, the fracture state in a body is viewed as a global minimization problem for the sum of the elastic and surface energies with arbitrary add-cracks as test fields. This variational approach is in line with Griffith's original idea of energy competition and is amenable to generalizations to nonlinear elastic and anisotropic materials in a 3D setting \citep{Bourdin08}. In addition to describing the crack path, the formulation transcends the limitations of Griffith's approach in two additional ways. First, the growth of pre-existing cracks need not be infinitesimal, instead, finite-size crack growth is possible. Second, crack nucleation is also possible as a direct result of the global energy minimization. 

It became clear over time that the global energy minimization approach can yield results that are inconsistent with experimental evidence.  It also became clear that  replacing global energy minimization with some type of local minimization \citep[e.g.,][]{Chambolle2008,Chambolle2009,Chambolle2010,Larsen24}  was not a cure.  

A further limitation of the variational approach is its apparent inability to handle non-zero Neumann boundary conditions and/or body forces. An attempt at removing this limitation was recently proposed by \cite{Larsen21} but it yields a formulation that is inconsistent with experimental observations. Moreover, it does not reduce to the usual formulation in the case of zero Neumann boundary conditions.

\paragraph{The current view: Energy minimization constrained by strength} In 2025, guided by the two-fold criterion stated above, two of us \citep{LPK25} introduced the following  tenet: the growth of a large crack takes place within and only within regions where the strength surface of the material has been exceeded, this in a manner such that the sum of the potential and surface energies is minimized. 

The tenet for the growth of large cracks introduced there serves as a precursor to the theory of fracture nucleation and propagation proposed in this work.

\section{Illustrating the concept}\label{Sec: A representative case}

\subsection{The simple tension test of a silicate glass specimen: Fracture under displacement control}

One of the simplest possible illustrations of the proposed framework is that of a specimen that is subjected to simple tension until a crack nucleates and subsequently propagates severing the specimen into two pieces.

%
\begin{figure}[H]
\centering
\includegraphics[width=0.6\linewidth]{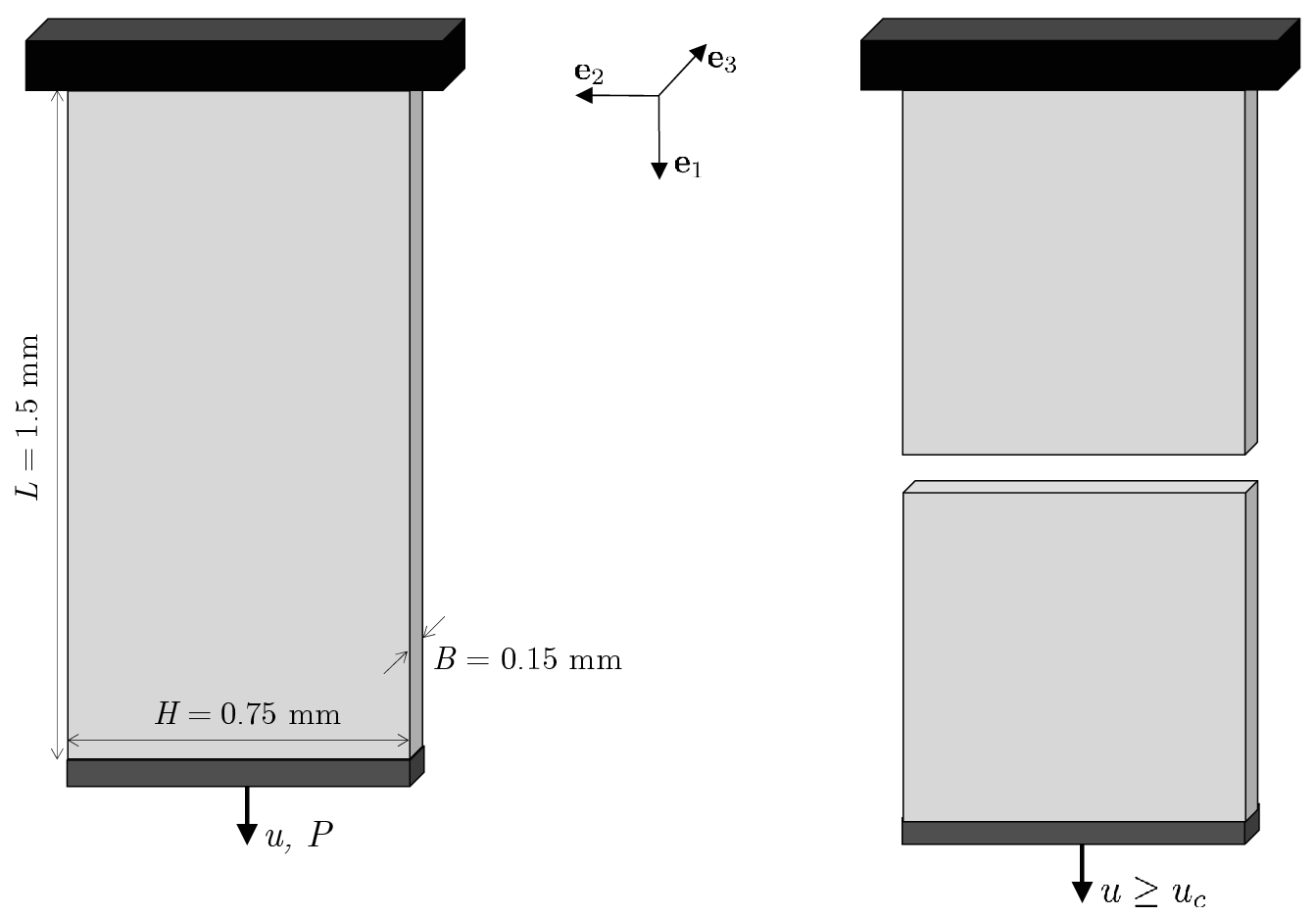}
\caption{{\small Schematic of the simple tension test on a silicate glass specimen examined in this and the next subsections: (left) initial undeformed configuration and applied loading; (right) fractured state following the nucleation and propagation of a crack orthogonal to the loading axis.}}\label{Fig1}
\end{figure}
%

For this type of test, experiments show a monotonic force-displacement ($P$ vs. $u$) response up to a critical state ($u_c$, $P_c$) at which point a crack  orthogonal to the loading direction appears and severs the specimen\footnote{Note that no direct experimental evidence has yet shown whether crack nucleation occurs simultaneously across the entire specimen or --- as is more likely --- begins in a subdomain and then propagates rapidly. This is due to the exact locus of crack nucleation being sensitive to spatial variations in strength (resulting from the inherent randomness of material defects), compounded by a fracture process that occurs too rapidly for conventional high-speed imaging to resolve.}  and this irrespective  of whether the specimen is pulled via an applied displacement $u$ or an applied force $P$; see Fig.~\ref{Fig1}. The post-critical behavior depends on how the specimen is pulled. Under displacement control, the force jumps from $P=P_c$ to $P=0$ at $u_c$. Under force control, it is the displacement that jumps  beyond $u=u_c$, leading to a loss of elastostatic equilibrium as the specimen fails. We examine displacement control in this subsection and force control in the next.

For definiteness, we consider the test schematically depicted in Fig.~\ref{Fig1}. It consists of a thin rectangular specimen, of length $L=1.5$ mm, width $H=0.75$ mm, and thickness $B=0.15$ mm, that is firmly gripped at the opposite ends of its length and pulled in tension. These small sizes are purposely selected to better illustrate key aspects of the problem. Denoting the laboratory Cartesian frame of reference by $\{\bfe_1,\bfe_2,\bfe_3\}$, the specimen occupies the open bounded domain $\Omega_0=\{\bfX:0<X_1<L,\,|X_2|<H/2,\,|X_3|<B/2\}$.  The specimen is assumed to be made of a typical silicate (soda-lime) glass with the classical elastic energy density 
\begin{equation}\label{W-Lin-elastic}
W(\bfF)=\mu\,{\rm tr}\,\bfE^2+\dfrac{\Lambda}{2}({\rm tr}\,\bfE)^2 
\end{equation}
of an isotropic linear elastic material, Drucker-Prager strength surface 
\begin{equation}
\mathcal{F}(\bfS)=\sqrt{\mathcal{J}_2}+\dfrac{s_{\texttt{cs}}-s_{\texttt{ts}}}
{\sqrt{3}\left(s_{\texttt{cs}}+s_{\texttt{ts}}\right)}\, \mathcal{I}_1-\dfrac{2s_{\texttt{cs}} s_{\texttt{ts}}}
{\sqrt{3}\left(s_{\texttt{cs}}+s_{\texttt{ts}}\right)}=0,\label{DP}
\end{equation}
and scalar fracture toughness $G_c$. The values of the material constants are in Table \ref{Table1}. In expressions (\ref{W-Lin-elastic}) and (\ref{DP}), $\bfE(\bfu)=1/2(\bfF+\bfF^T-2\bfI)=1/2(\nabla\bfu+\nabla\bfu^T)$ denotes the symmetrized gradient of the displacement field $\bfu=\bfy-\bfX$, or infinitesimal strain tensor, $\mu$ and $\Lambda$ are the Lam\'e constants, $\mathcal{I}_1=s_1+s_2+s_3$, $\mathcal{J}_2=1/3(s^2_1+s^2_2+s^2_3-s_1 s_2-s_1 s_3-s_2 s_3)$, while $s_{\texttt{ts}}$ and $s_{\texttt{cs}}$ stand for the uniaxial tensile and compressive strengths, that is, they denote the critical nominal stress values at which fracture nucleates under uniform states of monotonically increased uniaxial tension $\bfS={\rm diag}(s>0,0,0)$ and compression $\bfS={\rm diag}(-s<0,0,0)$, respectively.

\begin{table}[t!]\centering\small
\caption{Material constants for the silicate glass used throughout this work.}
\begin{tabular}{l?cccc}
\toprule
Elasticity constants & $\mu$ (GPa)&  $\Lambda$ (GPa) &  $E$ (GPa) &  $\nu$  \\
\midrule
                     & $28.7$   & $22.5$   & $70$ & 0.22     \\
\midrule
\midrule
Strength constants   & $s_{\texttt{ts}}$ (MPa) & $s_{\texttt{cs}}$ (MPa) \\
\midrule
                     & $40$                  & $1000$                 \\
\midrule
\midrule
Fracture toughness   & $G_c$ (N/m) & \\
\midrule
                               & $10$ & \\
\bottomrule
\end{tabular} \label{Table1}
\end{table}

Now, the two-fold criterion for fracture recalled in the Introduction suggests several plausible formulations of sharp fracture. The simplest among
these is the following: a sharp crack nucleates within and only within the region where the strength surface $\mathcal{F}(\bfS)=0$ of the material is exceeded in a way that the sum of the potential and surface energies 
\begin{equation}\label{E-DC} 
\mathcal{E}=\displaystyle\int_{\Omega_0\setminus\Gamma}W(\bfF)\,{\rm d}\bfX-\displaystyle\int_{\Omega_0\setminus\Gamma}\bfb\cdot\bfy\,{\rm d}\bfX-\displaystyle\int_{\partial\Omega^{\mathcal{N}}_0}\overline{\bfs}\cdot\bfy\,{\rm d}\mathcal{H}^{2}+G_c \mathcal{H}^{2}(\Gamma)
\end{equation}
is minimized among all possible test cracks $\Gamma$, this while elastostatic equilibrium is satisfied. Here, neglecting the minor effects of the body force due to gravity and the atmospheric pressure, $\bfb=\textbf{0}$ and $\overline{\bfs}=\textbf{0}$, while $\mathcal{H}^{2}(\Gamma)$ denotes the 2-dimensional Hausdorff measure (the surface measure) of the unknown crack.

\begin{figure}[t!]
\centering
\includegraphics[width=0.95\linewidth]{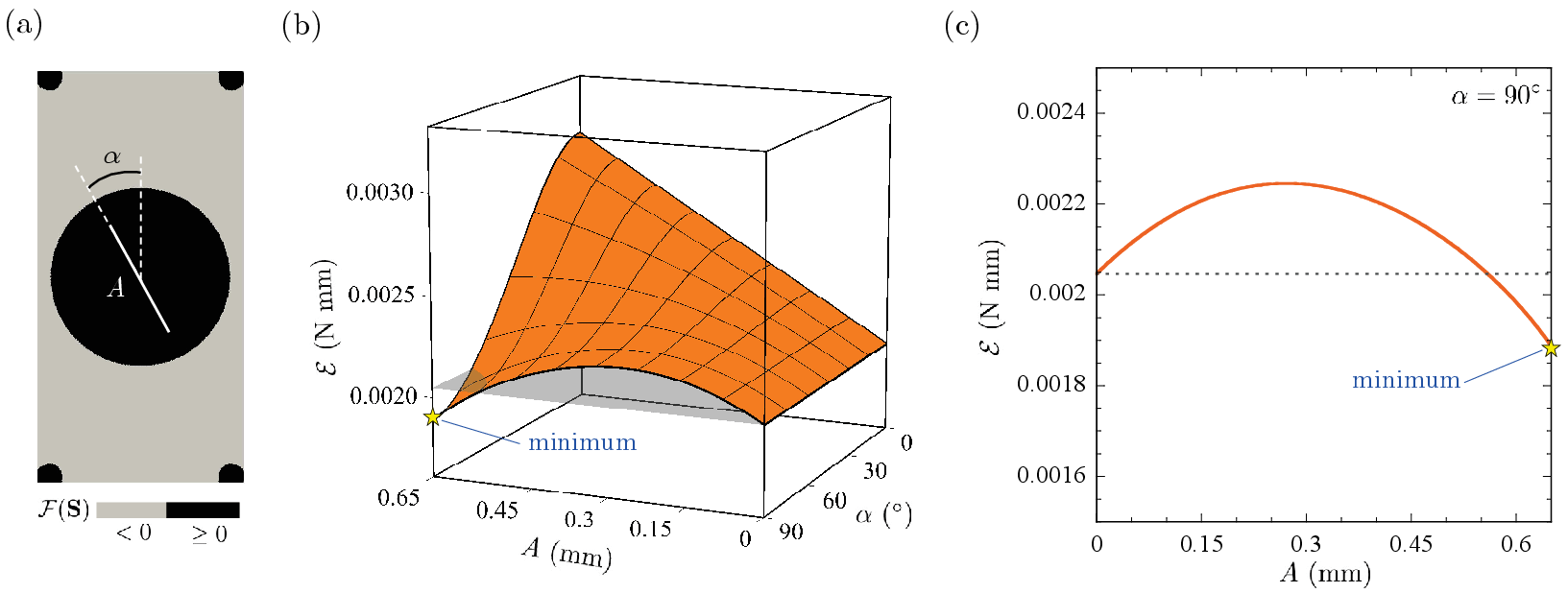}
\caption{{\small Elastic simulations of the tension test, at the applied global strain $u/L=5.7\times 10^{-4}$, on silicate glass containing test cracks of lengths $A\in[0,\,0.65\, {\rm mm}]$, thickness $B=0.15$ mm, oriented at angles $\alpha\in[0^\circ,\, 90^\circ]$ with respect to the direction of the applied tension. (a) Contour plot of the regions of the specimen where the stress field exceeds ($\mathcal{F}(\bfS)\geq0$) the strength surface of the material, depicting the test cracks considered within. (b) The energy $\mathcal{E}$ of the specimen plotted as a function of the test crack length $A$ and angle $\alpha$. (c) The cross section of $\mathcal{E}$ for fixed crack angle $\alpha=90^\circ$ as a function of $A$.}}\label{Fig2}
\end{figure}

We begin by numerically determining  (via finite elements\footnote{In this paper, all finite-element simulations for glass and mortar were carried out with quadratic simplicial elements. For rubber, they were carried out with conforming quadratic Crouzeix-Raviart elements \citep{CR1973}; see also the Appendix in \citep{LLP2017}.} and assuming plane-stress conditions) the critical global strain at which the strength condition $\mathcal{F}(\bfS)\geq0$ is first satisfied. For definiteness, we assume the presence of a weaker central circular region with a $0.65$ mm diameter where the uniaxial tensile strength is $10\%$ lower ($\sts = 40$ MPa) than the surrounding material ($\sts = 44$ MPa). The calculations show that the condition $\mathcal{F}(\mathbf{S})\geq0$ is first reached at the specimen's corners --- due to stress concentrations near the grips --- and then within the central circular region of reduced strength, the latter first occurring at $u/L=5.7 \times 10^{-4}$, as shown in Fig.~\ref{Fig2}(a). It is in this central region --- and not at the corners --- that a configuration with a crack can outcompete the initial uncracked configuration. At $u/L=5.7\times 10^{-4}$, we then numerically determine the energy (\ref{E-DC}) for specimens containing straight through-thickness test cracks, $\Gamma=\Gamma(A,\alpha)$, of crack length $A$, so that $\mathcal{H}^{2}(\Gamma)=A B$, and crack orientation  $\alpha$ relative to the applied tension; see Fig.~\ref{Fig2}(a). This crack topology is chosen for its consistency with experimental observations. Figure \ref{Fig2}(b) presents the numerical results obtained for the energy $\mathcal{E}$ as a function of $A$ and $\alpha$, while Fig.~\ref{Fig2}(c) presents a cross section of $\mathcal{E}$ as a function of $A$ at the fixed crack angle $\alpha=90^\circ$.

%
%

The results in Fig.~\ref{Fig2} offer the following insights. First, the crack of length $A=0.65$ mm --- the largest possible length that can be fitted within the strength violation region --- oriented at $\alpha=90^\circ$ minimizes the energy $\mathcal{E}$. This implies that such a crack nucleates at $u/L=5.7\times 10^{-4}$. Crucially, observe that the proposed constrained energy minimization delivers both the when ($u/L=5.7\times 10^{-4}$) and the where ($A=0.65$ mm and $\alpha=90^\circ$) of crack nucleation.

Second, if the strength violation occurs over a region with diameter smaller than $\ell_{\texttt{ts}}=0.54$ mm, the energy is minimized at $A=0$, meaning no crack nucleates. This threshold defines a length scale below which strength (in this case, the uniaxial tensile strength $\sts$) can no longer be treated as a macroscopic material property. This scale is comparable to, though distinct from, the classical material length scale $E G_c/\sts^2=0.44$ mm. As elaborated in the next section, such length scales arise from the dimensional interplay between elasticity ($force/length^2$), strength ($force/length^2$), and toughness ($force/length$) entering as material inputs in the constrained energy minimization.

Third, if the strength violation occurs over a region larger than 0.65 mm in diameter, the energy would be minimized again by the largest possible crack that fits within that region at $90^\circ$. In particular, if the region is the entire specimen, the energy would be minimized by a severing crack at $90^\circ$. 

Fourth, the minimizing energy ($\mathcal{E}=0.00189$ N mm) for the configuration with the nucleated crack is smaller than the elastic energy ($\mathcal{E}=0.00205$ N mm) for the configuration without the crack prior to nucleation; the latter is indicated by a grey plane in Fig.~\ref{Fig2}(b) and a dashed line in Fig.~\ref{Fig2}(c). An abrupt drop in the (quasi)-static energy functional occurs. It is not clear to us at present if the violation of the strength is accompanied by an additional dissipation which manifests itself whenever there is a jump in crack size (at the time of nucleation of a finite-size crack for example). In any case, in the setting of uniaxial traction, the nucleation of a severing crack will trigger dynamic effects that consume energy. If the material dissipates energy by deformation when loaded dynamically (e.g., via viscous processes) and/or the specimen is immersed in a dissipative surrounding medium like air, kinetic energy will eventually be dissipated. Without such dissipation mechanisms, the specimen will not
regain static equilibrium, rendering the quasi-static assumption invalid. 

%
\begin{figure}[t!]
\centering
\includegraphics[width=0.75\linewidth]{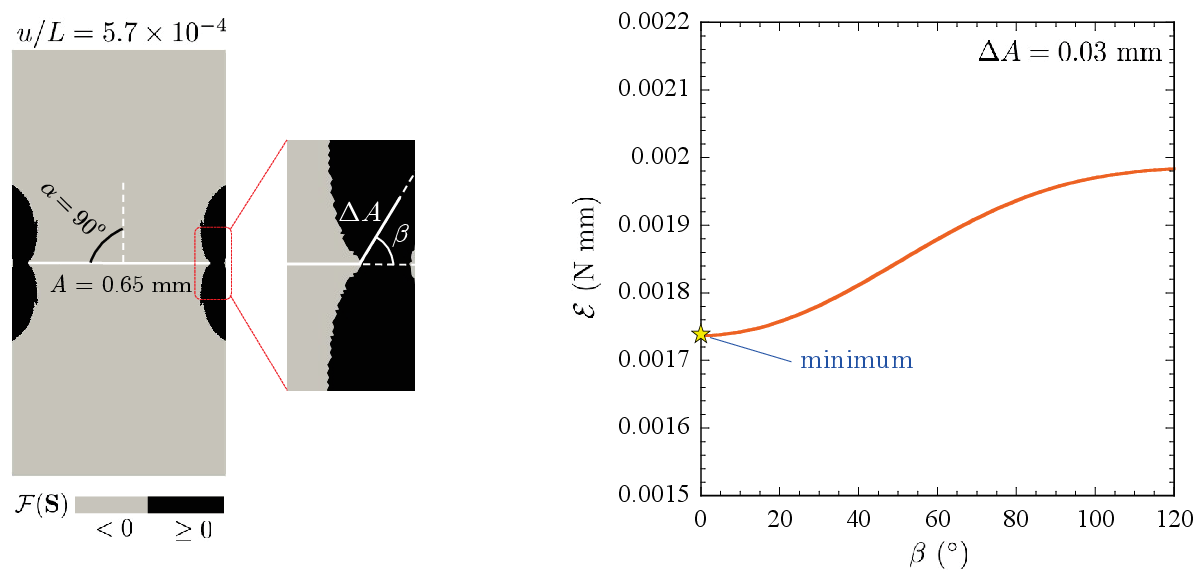}
\caption{{\small Elastic simulations of the tension test, at the applied global strain $u/L=5.7\times 10^{-4}$, on silicate glass containing the initial minimizing crack, $(A=0.65\,{\rm mm},\,\alpha=90^\circ)$, with test add-cracks of length $\Delta A$  oriented at angles $\beta$ with respect to the initial crack orientation. The contour plot illustrates the regions of the specimen where the stress field exceeds ($\mathcal{F}(\bfS)\geq0$) the strength surface of the material, depicting the test add-cracks considered within. The line plot shows the variation of the energy $\mathcal{E}$ with respect to the add-crack angle $\beta$ for the fixed add-crack length $\Delta A=0.03$ mm.}}\label{Fig3}
\end{figure}
%
%
\begin{figure}[t!]
\centering
\includegraphics[width=0.75\linewidth]{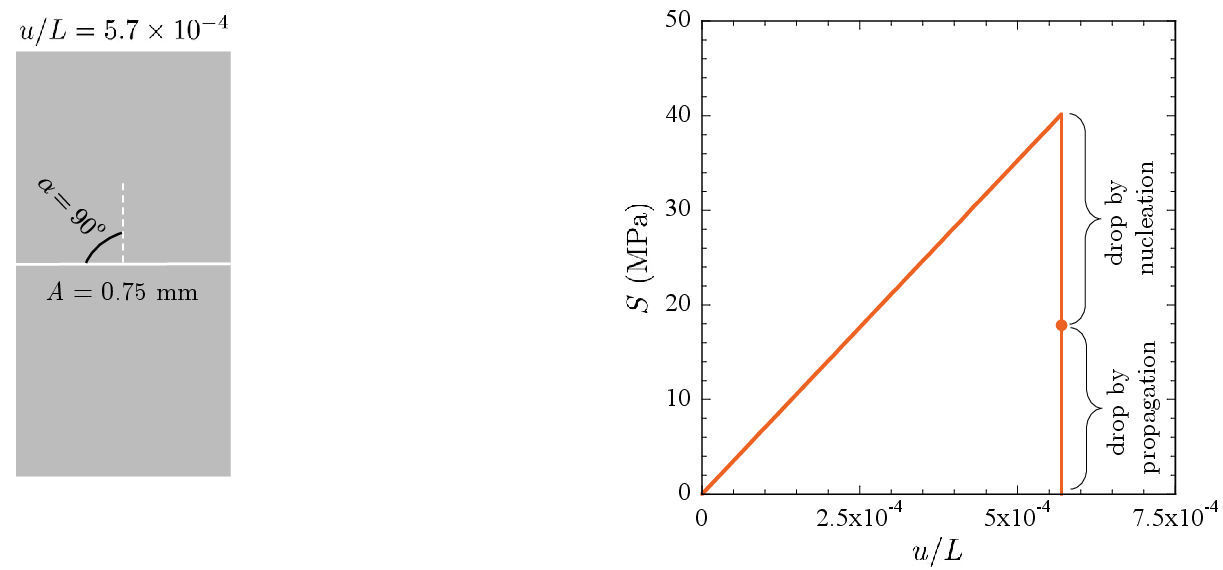}
\caption{{\small Theoretical prediction of the tension test on silicate glass under displacement control. The results show the global stress $S=P/(BH)$ as a function of the applied global strain $u/L$, as well as a depiction of the fractured state of the specimen following the nucleation and propagation of a crack orthogonal to the direction of applied tension.}}\label{Fig4}
\end{figure}
%

\begin{remark}
{\color{black}  Updating the strength violation set at fixed load versus increasing the load. \emph{Given the time-discrete nature of our approach, one can proceed with the calculation after crack nucleation in two ways: either by updating the region of strength surface violation without increasing the load, or by increasing the load first. Crucially, both procedures yield essentially the same result. We believe that updating the strength surface violation without increasing the load is the more physically sound choice, which is why we present it here. At any rate, this issue will be resolved by passing to the time-continuous limit, which for now we leave for future work.}}
\end{remark}

Assuming that the quasi-static assumption remains valid after fracture nucleation, we now examine the minimizing solution with crack length $A=0.65$ mm and crack angle $\alpha=90^\circ$. Following the previously established procedure, we numerically identify the regions where the strength criterion is exceeded ($\mathcal{F}(\bfS)\geq0$) at the applied global strain of $u/L=5.7\times 10^{-4}$. As illustrated by the contour plot in Fig.~\ref{Fig3}, the strength violation is now localized near the crack fronts. Within these updated violation regions, we consider straight through-thickness test add-cracks, $\Gamma=\Gamma(A+2\Delta A,\alpha,\beta)$, of add-crack length $\Delta A$, so that $\mathcal{H}^{2}(\Gamma)=(A+2\Delta A) B$, and add-crack orientation  $\beta$ relative to the initial crack orientation; see the schematic depictions in Fig.~\ref{Fig3}. Numerical results analogous to those presented in Fig.~\ref{Fig2}(b) show that the energy is minimized at $\Delta A=0.03$ mm and $\beta=0^\circ$. Fig.~\ref{Fig3} includes a cross section of $\mathcal{E}$ as a function of $\beta$ for the fixed add-crack length $\Delta A=0.03$ mm. That is, the constrained minimization dictates that the nucleated crack at $u/L=5.7\times 10^{-4}$ does not stay put, instead it propagates by a finite-size growth in the direction orthogonal to the applied tension.

Repeating the procedure once again leads to a crack, orthogonal to the direction of applied tension, that completely severs the specimen at $u/L=5.7\times 10^{-4}$. This is shown by Fig.~\ref{Fig4},  which also shows the associated global $S$ vs. $u/L$ response of the specimen. This result is consistent with experimental observations. Interestingly, it reveals that the manner in which a specimen under simple tension severs is by the nucleation of a finite-size crack in the region where the strength is weakest, followed by brutal crack propagation. 

In the scenario where the region of weakest strength spans the entire width of the specimen, the specimen would fracture via the nucleation of a severing crack.

\subsection{The simple tension test of a silicate glass specimen: Fracture under force control}

\begin{figure}[t!]
\centering
\includegraphics[width=0.95\linewidth]{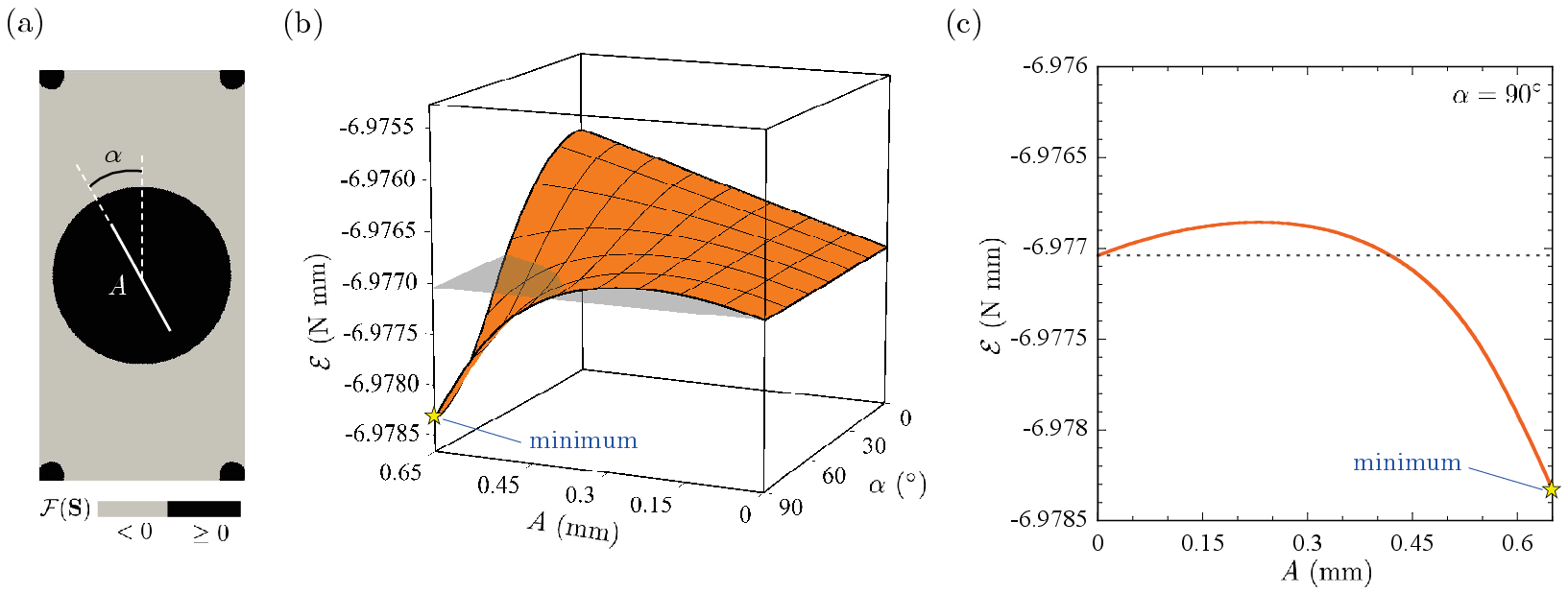}
\caption{{\small Elastic simulations of the tension test, at the applied global stress $S=40$ MPa, on silicate glass containing test cracks of lengths $A\in[0,\,0.65\, {\rm mm}]$, thickness $B=0.15$ mm, oriented at angles $\alpha\in[0^\circ,\, 90^\circ]$ with respect to the direction of the applied tension. (a) Contour plot of the regions of the specimen where the stress field exceeds ($\mathcal{F}(\bfS)\geq0$) the strength surface of the material, depicting the test cracks considered within. (b) The energy $\mathcal{E}$ of the specimen plotted as a function of the test crack length $A$ and angle $\alpha$. (c) The cross section of $\mathcal{E}$ for fixed crack angle $\alpha=90^\circ$ as a function of $A$.}}\label{Fig5}
\end{figure}

We now consider the test under force control, when the force $P$ is applied instead of the displacement $u$. In this case, the traction at the bottom grip $\{\bfX: X_1=L,\,|X_2|<H/2,\,|X_3|<B/2\}$  in the energy functional (\ref{E-DC}) is no longer zero but such that 
\begin{equation}\label{Traction-Work}
-\int_{\partial\Omega^{\mathcal{N}}_0}\overline{\bfs}\cdot\bfy\,{\rm d}\mathcal{H}^{2}=-P(u+L).
\end{equation}
Figure \ref{Fig5} shows results analogous to those presented in Fig.~\ref{Fig2} for applied displacement. The main difference is that the value of the energy functional (\ref{E-DC}) is negative due to the work done (\ref{Traction-Work}) by the applied boundary force. Otherwise, the energy functional is essentially the same in terms of the crack length $A$ and the crack angle $\alpha$. In particular, the crack of largest possible length $A=0.65$ mm oriented at $\alpha=90^\circ$ is again the minimizing configuration.

%
%

%
\begin{figure}[t!]
\centering
\includegraphics[width=0.75\linewidth]{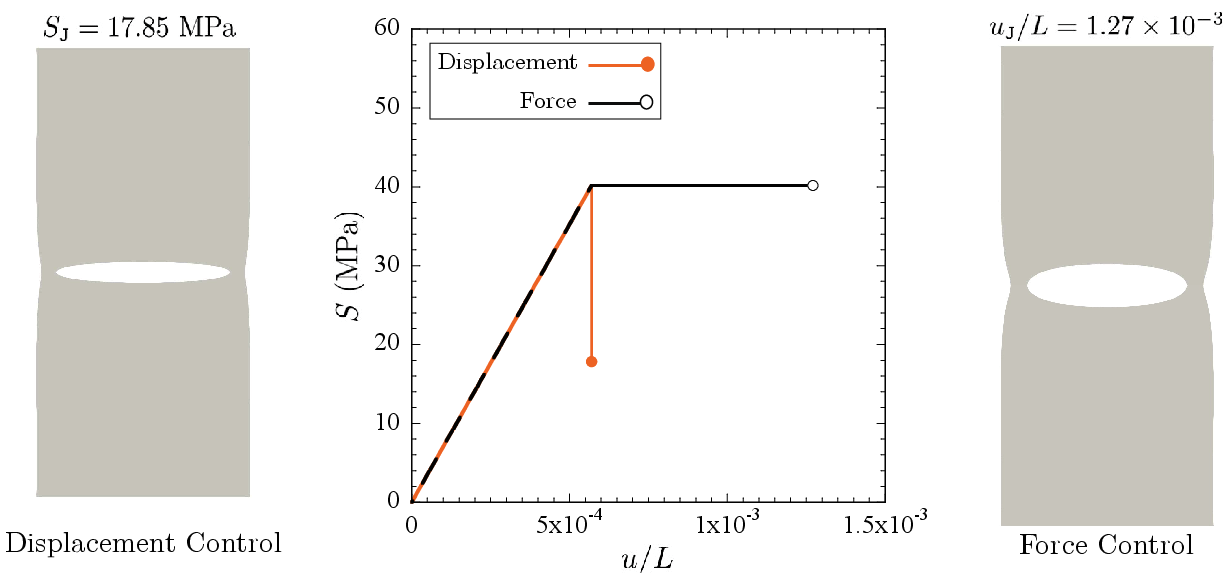}
\caption{{\small Comparison of the post-nucleation response in the tension test on silicate glass carried out under displacement and force control. The line plot shows that the global $S$ vs. $u/L$ results are identical up until crack nucleation $(A=0.65\, {\rm mm},\,\alpha=90^\circ$). Once the crack has nucleated, the global stress jumps from $S=40$ MPa to $S_{\texttt{J}}=17.85$ MPa at $u/L=5.7\times 10^{-4}$ under displacement control, while under force control it is the global strain that jumps from $u/L=5.7\times 10^{-4}$  to $u_{\texttt{J}}/L=1.27\times 10^{-3}$ at $S=40$ MPa. The left and right images show the deformed configurations (magnified $100\times$ for better visualization) of the specimens with the crack.}}\label{Fig6}
\end{figure}
%

Now, as previously noted, the response of the specimen is independent of the type of loading --- be it displacement or force --- up until crack nucleation. Post-nucleation, however, the response is fundamentally dictated by the specific boundary condition applied. Figure \ref{Fig6} compares the initial post-nucleation response of the tension test on silicate glass carried out under displacement and force control. Observe that the global stress jumps from $S=40$ MPa to $S_{\texttt{J}}=17.85$ MPa when the crack with length $A=0.65$ mm and angle $\alpha=90^\circ$ nucleates at $u/L=5.7\times 10^{-4}$ under displacement control. On the other hand, the global displacement jumps from $u/L=5.7\times 10^{-4}$  to $u_{\texttt{J}}/L=1.27\times 10^{-3}$ when the crack nucleates at $S=40$ MPa under force control. Figure \ref{Fig6} includes images --- magnified 100$\times$ for better visualization --- of the deformed configurations of both specimens, helping to visualize the disparity in the resulting jumps.

%
\begin{figure}[t!]
\centering
\includegraphics[width=0.75\linewidth]{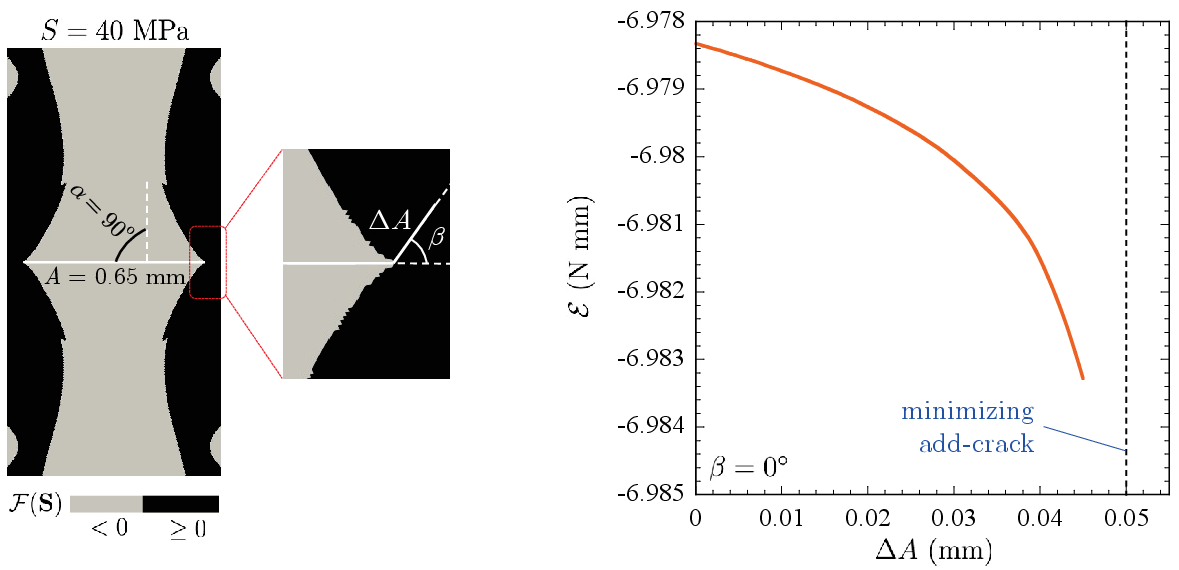}
\caption{{\small Elastic simulations of the tension test, at the applied global stress $S=40$ MPa, on silicate glass containing the initial minimizing crack, $(A=0.65\,{\rm mm},\alpha=90^\circ)$, with test add-cracks of length $\Delta A$  oriented at angles $\beta$ with respect to the initial crack orientation. The contour plot illustrates over the undeformed configuration the regions of the specimen where the stress field exceeds ($\mathcal{F}(\bfS)\geq0$) the strength surface of the material, depicting the test add-cracks considered within. The line plot shows the variation of the energy $\mathcal{E}$ with respect to the add-crack length $\Delta A$ for $\beta=0^\circ$.}}\label{Fig7}
\end{figure}
%

Figure \ref{Fig7} shows results analogous to those presented in Fig.~\ref{Fig3} for applied displacement. It is immediately apparent that the updated region of strength violation for the configuration with the minimizing crack $(A=0.65\, {\rm mm},\,\alpha=90^\circ$) is significantly larger than that found for the case of applied displacement. Notably, the region extends from the crack fronts all the way to the lateral boundary of the specimen. Within these updated violation regions, we consider straight through-thickness test add-cracks, $\Gamma=\Gamma(A+2\Delta A,\alpha,\beta)$, of add-crack length $\Delta A$, so that $\mathcal{H}^{2}(\Gamma)=(A+2\Delta A) B$, and add-crack orientation  $\beta$ relative to the initial crack orientation; see the schematic depictions in Fig.~\ref{Fig7}. The numerical results for the energy  $\mathcal{E}$ for $\beta=0^\circ$ are plotted in Fig.~\ref{Fig7} as a function of $\Delta A$ from $\Delta A=0$ to $\Delta A=0.045$ mm. While results for $\Delta A>0.045$ mm are difficult to compute reliably, they are such that $\mathcal{E}$ continues to decrease monotonically as $\Delta A$ increases, becoming unbounded as $\Delta A\nearrow 0.05$ mm --- that is, as the add-crack approaches the severing of the specimen. Consequently, the constrained minimization dictates that the nucleated crack at $S=40$ MPa does \emph{not} stay put, instead it propagates by a finite-size growth in the direction orthogonal to the applied tension, severing the specimen. This result is, again, consistent with experimental observations.  

As a final remark, following the complete severance of the specimen, the lower subdomain subjected to the applied force undergoes full elastic unloading and rigid-body translation, while the gripped upper subdomain unloads elastically. As elaborated below, in general, severed subdomains may or may not retain stored elastic energy, and could undergo further rigid-body translation and rotation. Regardless, each severed subdomain will subsequently deform and fracture independently. Consequently, they must be treated as separate, independent problems during any subsequent loading, should such loading be possible.

\subsection{The cantilever bending test of a mortar beam: Fracture under body forces (self-weight)}

Having considered examples where fracture is driven by an applied displacement and an applied traction, we now turn to an example where fracture is driven by a body force. 

\begin{figure}[t!]
\centering
\includegraphics[width=0.575\linewidth]{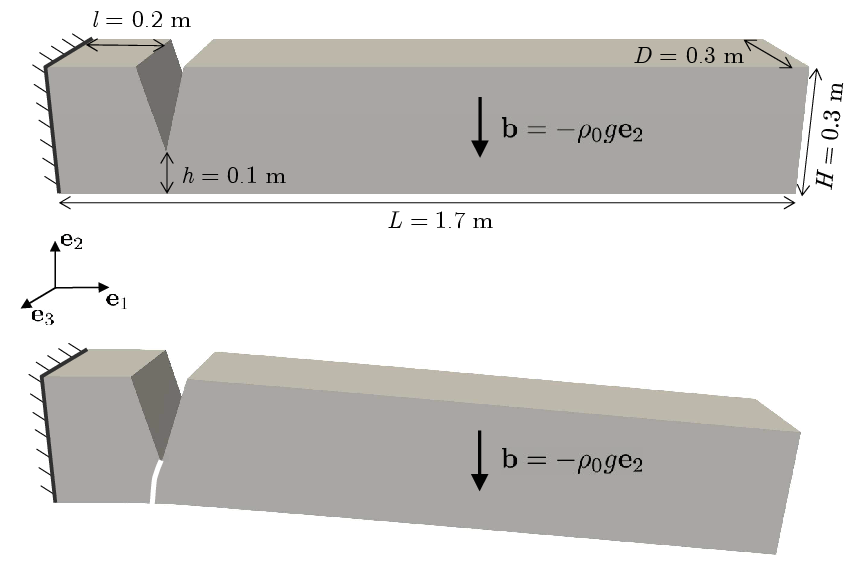}
\caption{{\small Schematic of the cantilever bending test of a mortar beam examined in this subsection: (top) initial undeformed configuration and applied gravitational body force; (bottom) fractured state following the nucleation of a crack from the notch front and subsequent propagation.}}\label{Fig8}
\end{figure}

Specifically, we consider the experiment recently introduced in \citep{KBRLP26}, schematically depicted in Fig.~\ref{Fig8}. The setup consists of a beam with length $L=1.7$ m, height $H=0.3$ m, and depth $D=0.3$ m, containing a $30^\circ$ V-notch located at $l=0.2$ m from the left boundary and $h=0.1$ m from the bottom edge. The beam is fully clamped at its left boundary to form a cantilever and is subjected to the body force of gravity, which in our frame of reference reads $\textbf{b}=-\rho_0 g \bfe_2$, where $\rho_0$ is the material mass density and $g$ is the gravitational acceleration. The setup is purposely designed so that a crack nucleates from the notch front and rapidly propagates toward the bottom edge, severing the beam; see Fig.~\ref{Fig8}. 

%
%

Here, the beam is assumed to be made of the 3D-printable mortar investigated in \citep{SMMRLP26}. It is characterized by the mass density $\rho_0$, elastic energy density (\ref{W-Lin-elastic}), strength surface (\ref{DP}), and fracture toughness $G_c$, with the values of the various material constants listed in Table \ref{Table2}. In this case, note that the contribution of the body force in the energy functional (\ref{E-DC}) specializes to 
\begin{equation*}
-\int_{\Omega_{0}}\bfb\cdot\bfy\,{\rm d}\bfX=\rho_0 g\int_{\Omega_{0}} y_2\,{\rm d}\bfX.
\end{equation*}
In the results that follow, for computational expediency, we apply the load by monotonically increasing the gravitational acceleration from $g=0$~m/s$^2$ and assume plane-strain conditions. In the actual experiment, given that the gravitational acceleration is fixed at $g=9.81$~m/s$^2$, the right free end of the beam is initially supported from the bottom. This support is gradually lowered and removed, allowing the beam to approach a state where it is subjected to its own self-weight. A 3D simulation of this exact loading process is reported in \citep{KBRLP26}, yielding results that are not fundamentally different from the 2D results for the simpler loading process presented here.
\begin{table}[t!]\centering\small
\caption{Material constants for the mortar used in this subsection.}
\begin{tabular}{l?cccc}
\toprule
Mass density &             $\rho_0$ (kg/m$^3$)&   &   &    \\
\midrule
                     & $2210$   &   &  &      \\
\midrule
\midrule
Elasticity constants & $\mu$ (GPa)&  $\Lambda$ (GPa) &  $E$ (GPa) &  $\nu$  \\
\midrule
                     & $10.7$   & $5.5$   & $25$ & 0.17     \\
\midrule
\midrule
Strength constants   & $s_{\texttt{ts}}$ (MPa) & $s_{\texttt{cs}}$ (MPa) \\
\midrule
                     & $5$                  & $48$                 \\
\midrule
\midrule
Fracture toughness   & $G_c$ (N/m) & \\
\midrule
                               & $20$ & \\
\bottomrule
\end{tabular} \label{Table2}
\end{table}
%

%
\begin{figure}[t!]
\centering
\includegraphics[width=0.95\linewidth]{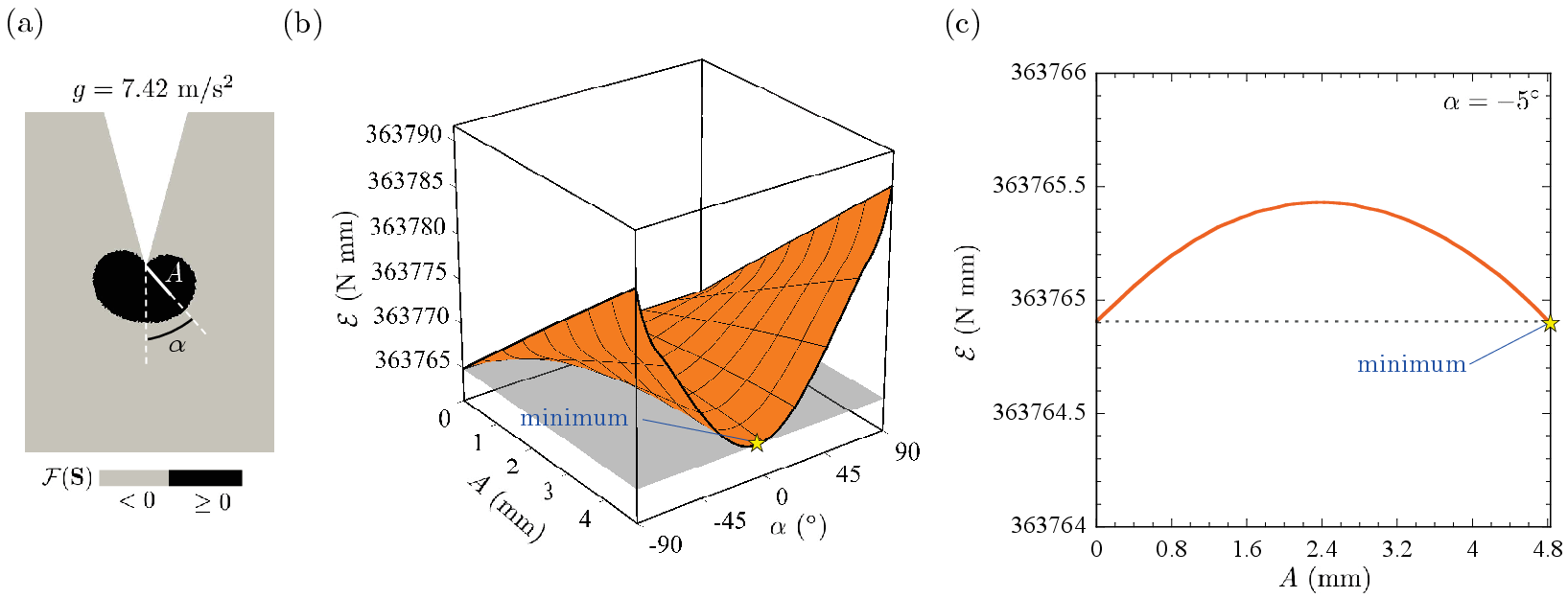}
\caption{{\small Elastic simulations of the cantilever bending test, at the applied gravitational acceleration $g=7.42$ m/s$^2$, on mortar containing test cracks of length $A$, thickness $D=0.3$ m, oriented at angles $\alpha$ with respect to the direction of the applied gravitational body force. (a) Contour plot of a magnified region of the specimen where the stress field exceeds ($\mathcal{F}(\bfS)\geq0$) the strength surface of the material, depicting the test cracks considered within. (b) The energy $\mathcal{E}$ of the beam plotted as a function of the test crack length $A$ and angle $\alpha$. (c) The cross section of $\mathcal{E}$ for fixed crack angle $\alpha=-5^\circ$ as a function of $A$.}}\label{Fig9}
\end{figure}
%

As in the previous two examples, we incrementally increase the load, in this case the gravitational acceleration $g$, to numerically identify the regions where the strength surface is violated, $\mathcal{F}(\mathbf{S}) \geq 0$. Such regions are, as expected, localized around the front of the notch as well as near the clamped edges of the beam. At each increment, we determine the energy (\ref{E-DC}) for configurations containing straight through-thickness test cracks, $\Gamma=\Gamma(A,\alpha)$, of crack length $A$, and hence surface area $\mathcal{H}^2(\Gamma) = A D$, emanating from the notch front and oriented at an angle $\alpha$ relative to the direction of gravity; see Fig.~\ref{Fig9}. Again, this crack topology is chosen for its consistency with experimental observations. Figure~\ref{Fig9}(b) presents the numerical results obtained for the energy $\mathcal{E}$ as a function of $A$ and $\alpha$, while Fig.~\ref{Fig9}(c) presents a cross section of $\mathcal{E}$ as a function of $A$ at the fixed crack angle $\alpha = -5^\circ$. These results pertain to $g = 7.42\,\text{m/s}^2$, which is the critical value at which a configuration containing a crack first minimizes (\ref{E-DC}). The minimizing crack is one of length $A=4.83$ mm oriented at $\alpha = -5^\circ$. Again, this implies that such a crack nucleates at $g = 7.42\,\text{m/s}^2$.

%
\begin{figure}[t!]
\centering
\includegraphics[width=0.85\linewidth]{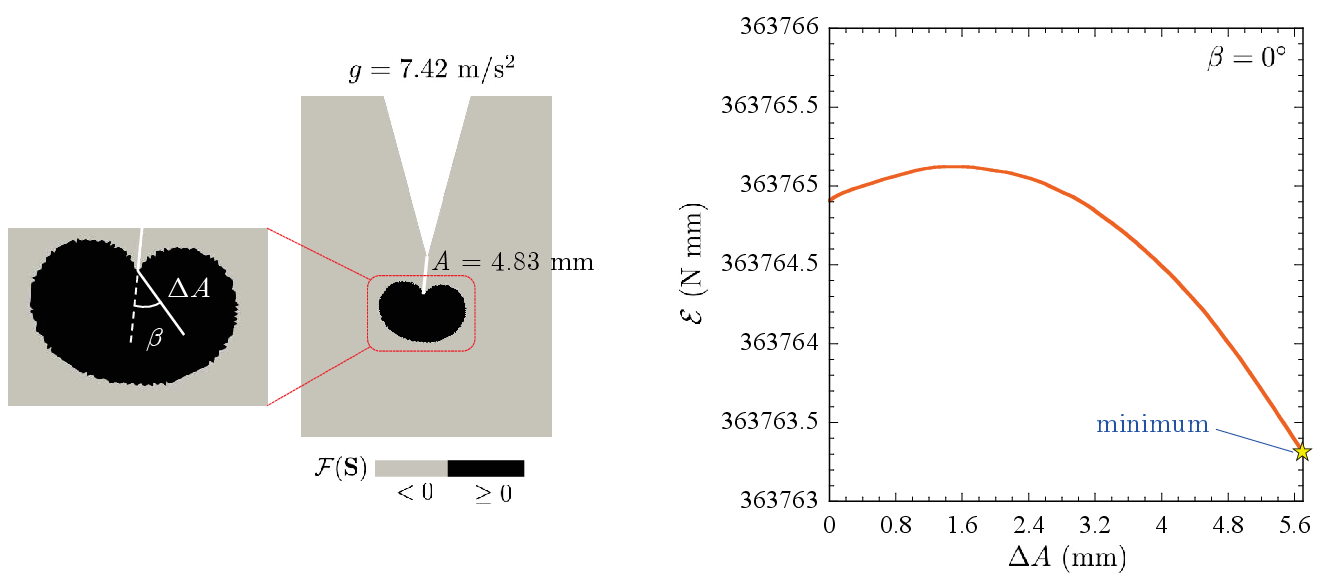}
\caption{{\small Elastic simulations of the cantilever bending test, at the applied gravitational acceleration $g=7.42$ m/s$^2$, on mortar containing the initial minimizing crack, $(A=4.83\,{\rm mm},\alpha=-5^\circ)$, with test add-cracks of lengths $\Delta A$ oriented at angles $\beta$ with respect to the initial crack orientation. The contour plot illustrates over the undeformed configuration a magnified region of the specimen where the stress field exceeds ($\mathcal{F}(\bfS)\geq0$) the strength surface of the material, depicting the test add-cracks considered within. The line plot shows the variation of the energy $\mathcal{E}$ with respect to the add-crack length $\Delta A$ for $\beta=0^\circ$.}}\label{Fig10}
\end{figure}
%

We next analyze the minimizing solution with crack length $A=4.83$ mm oriented at $\alpha=-5^\circ$. At the critical gravitational acceleration $g = 7.42\,\text{m/s}^2$, we numerically evaluate the regions where the stress field violates the strength surface, $\mathcal{F}(\mathbf{S}) \geq 0$. As shown in Fig.~\ref{Fig10}, this strength violation localizes around the crack front. Within this updated domain of strength violation, we evaluate straight through-thickness test add-cracks, $\Gamma=\Gamma(A+\Delta A,\alpha,\beta)$, of add-crack length $\Delta A$ oriented at an angle $\beta$ relative to the original crack orientation, resulting in a total crack surface area of $\mathcal{H}^2(\Gamma) = (A + \Delta A) D$. Our energy calculations indicate that $\mathcal{E}$ is minimized for $\Delta A = 5.7\,\text{mm}$ and $\beta = 0^\circ$; see Fig.~\ref{Fig10}. This result indicates that upon nucleation at $g = 7.42\,\text{m/s}^2$, the crack does not stay put, but instead experiences finite-size growth along its initial trajectory. Subsequent calculations demonstrate that the crack continues its downward propagation until fully severing the beam, in agreement with experimental observations \citep{KBRLP26}.

\section{The theory: A constrained energy minimization problem}\label{Sec: The Theory}

As anticipated in the preceding section, we propose the following principle:

\medskip

\noindent \emph{Cracks nucleate and propagate exclusively in regions where the strength surface of the material is exceeded, with their evolution dictated by the minimization of the sum of the potential and surface energies.}

\medskip

\noindent At present, we cannot offer a time-continuous formulation, most notably because the propriety of introducing a strength-related dissipation that would be activated at discontinuity times of the crack evolution is yet undecided. Thus the formulation will only be made explicit at a finite set of times $t_k\in\{0=t_0 <t_1,...<t_m<t_{m+1},...<t_M=T\}$.  

In the case of an isotropic elastic brittle material occupying an open bounded domain $\Omega_0\subset \mathbb{R}^3$ in its undeformed and stress-free configuration, the theory, in its time-discretized version, can be cast as follows. We denote by $\bfb(\bfX,t)$  the time-dependent body force, $\overline{\bfs}(\bfX,t)$ the time-dependent  nominal traction on the Neumann part $\partial\Omega_0^\mathcal{N}$ of the boundary $\partial\Omega_0$, both of which we assume to be dead loads for simplicity sake, and $\overline{\bfy}(\bfX,t)$ the time-dependent deformation prescribed on the Dirichlet part $\partial\Omega_0^\mathcal{D}$ of the boundary $\partial\Omega_0$. We write 
\begin{align}\label{Potential-Energy}
\mathcal{P}(\bfy,\Gamma):=\int_{\Omega_0\setminus\Gamma}W(\nabla\bfy)\,{\rm d}\bfX-\displaystyle\int_{\Omega_0\setminus\Gamma}\bfb_k\cdot\bfy\,{\rm d}\bfX-\displaystyle\int_{\partial\Omega^{\mathcal{N}}_0}\overline{\bfs}_k\cdot\bfy\,{\rm d}\mathcal{H}^{2}
\end{align}
for the potential energy, at times $t_k$. Then, the pair  deformation field-crack set $(\bfy_k=\bfy_k(\bfX),$ $\Gamma_k)$ minimizes
\begin{align}
\mathcal{E}(\bfy,\Gamma):=\mathcal{P}(\bfy,\Gamma)+G_c\,\mathcal{H}^{2}\left(\Gamma\right)\label{Fracture-Energy-Min}
\end{align}
among all $(\bfy,\Gamma)$ with
\begin{equation}\label{Gamma-Admi}
\left\{\begin{array}{l}
\bfy(\bfX)=\overline{\bfy}(\bfX,t_k),\quad \bfX\in \partial\Omega_0^\mathcal{D} \\[2mm] 
\bfy(\bfX) \textrm{ satisfies material impenetrability} \\[2mm] 
\Gamma\setminus\Gamma_{k-1}\subset \mathcal{V}_{\mathcal{F}}(t_{k-1})\\[2mm] 
\Gamma\supset\Gamma_{k-1}\\[2mm] 
\Gamma \cap\{\mbox{a specified neighborhood of } \partial\Omega_0^\mathcal{D}\cup {\;\rm supp\;}\overline{\bfs}_{k}\}=\emptyset \end{array}\right. .
\end{equation}
In these expressions, $\mathcal{H}^{2}(\Gamma)$ denotes, again, the $2$-dimensional Hausdorff measure of the unknown crack $\Gamma$ in the undeformed configuration and
\begin{equation}\label{Viol-Set}
\mathcal{V}_{\mathcal{F}}(t)=\{\bfX\in\Om_0\setminus\Gamma: \mathcal{F}(\bfS(\bfX,t))\ge0\}
\end{equation}
is the set of all material points at which the strength surface $\mathcal{F}(\bfS)=0$ of the material is exceeded at time $t$, referred to as the strength violation set.

Of course, at such a level of generality and without further specifying the form and regularity of the various ingredients that enter the minimization process, it is impossible to assert the existence of such a minimizing pair. In particular, if $\bfb_k$ and/or $\overline{\bfs}_k\not\equiv 0$ and if
\begin{equation*}
\mbox{ the infimum of \eqref{Fracture-Energy-Min} is }-\infty,
\end{equation*}
then one should be extremely cautious.  This will be so for example if it is possible to sever $\Omega_0$ into two pieces $\Omega_1, \Omega_2$ with a crack $\Gamma$ with $\Omega_1\cap\pa\Omega_0^{\mathcal D}=\emptyset$ and if the torsor
$$
\left(\int_{\Omega_1} \bfb_k \,{\rm d}\bfX+\int_{\pa\Omega_1} \oli\bfs_k {\rm d}\mathcal{H}^{2}, \int_{\Omega_1} \bfX\wedge\bfb_k \,{\rm d}\bfX+\int_{\pa\Omega_1} \bfX\wedge\oli\bfs_k \,{\rm d}\mathcal{H}^{2}\right)\not\equiv 0
$$  
because then, severing the piece $\Omega_1$ and moving it by a rigid body deformation will decrease the potential energy at will, at least provided this does not contradict material impenetrability. As of now, we do not have an all encompassing view of that case, except in the 2D case without tractions or body forces. This will be the object of the next section.

\begin{remark}\label{rem:ub} The unbreakable boundary. \emph{The unbreakable neighborhood of $\partial\Omega_0^\mathcal{D}\cup{\;\rm supp\;}\overline{\bfs}_{k-1}$ is not a material characteristic, but rather the recognition that a loaded boundary, be it by imposing a deformation or applying a force, necessitates an ad hoc reinforcement which makes it more difficult to break than the rest of the sample. Modeling the corresponding attachments is not part and parcel of a cogent theory of brittle fracture and would require the import of additional ingredients (like adhesion or gripping) that are external contributions that we do not wish to address herein.
}
\end{remark}

\begin{remark} Fracture irreversibility and frictionless crack contact. \emph{The formulation (\ref{Potential-Energy})-(\ref{Gamma-Admi}) assumes that fracture is an irreversible process --- precluding crack healing --- and that contact between crack surfaces is frictionless. The former assumption is particularly well-suited  to  monotonic loading conditions.  The latter may be called into question when dealing with problems with pre-existing cracks under large compressive and shear loads.}
\end{remark}

\begin{remark}\label{Remark_length} A critical material length scale $\ell_{c}$ and an explicit upper bound. \emph{Consider the special case when a body free of pre-existing cracks is under a state of spatially uniform stress, $\bfS(\bfX,t)=\obfS(t)$, as the result of an affine deformation $\overline{\bfy}(\bfX,t)=\obfF(t)\bfX$ applied over the entirety of its boundary ($\partial\Omega_0^\mathcal{D}=\partial\Omega_0$). According to the admissibility constraint (\ref{Gamma-Admi})$_3$, no crack can nucleate until the time at which $\obfS(t)$ reaches the strength surface of the material. Denote by $t=t_{k-1}$ that time. Then, according to the Definition \ref{Def-Strength} of strength, a crack should nucleate at $t=t_{k}$. This implies that the minimizing energy for the configuration with the nucleated crack is less than or equal to the energy for the configuration without the crack at $t=t_{k}$, that is,
\begin{equation}\label{Constraint-Omega-Per}
\mathcal{E}(\bfy_k,\Gamma_k)=\int_{\Omega_0\setminus\Gamma}W(\nabla\bfy_k)\,{\rm d}\bfX+G_c\,\mathcal{H}^{2}\left(\Gamma_k\right)\leq \int_{\Omega_0}W(\obfF(t_k))\,{\rm d}\bfX=|\Omega_0|W(\obfF(t_k)).
\end{equation}
This inequality unveils an important constraint: the domain $\Omega_0$ occupied by the body cannot be arbitrarily small. Specifically, there exists a critical material length scale, $\ell_c$, below which this inequality is violated. Physically, this means that at scales smaller than $\ell_c$, the strength surface $\mathcal{F}(\bfS)=0$ ceases to be a valid macroscopic material property. Although the computation of $\ell_c$ is generally intractable analytically, establishing an explicit upper bound, $\ell_{bound}$, is straightforward. To this end, we first note that
\begin{equation*}
\mathcal{E}(\bfy_k,\Gamma_k)\leq G_c \, {\rm Per}(\Omega_0),
\end{equation*}
where ${\rm Per}(\Omega_0)$ denotes the perimeter of the domain $\Omega_0$. It follows that domains that satisfy the condition 
\begin{equation*}
\ell_c\leq\displaystyle\max_{\mathcal{G}(\bfF)=0}\dfrac{G_c}{W(\bfF)}=:\ell_{bound} \leq \dfrac{|\Omega_0|}{{\rm Per}(\Omega_0)},\quad {\rm where} \quad \mathcal{G}(\bfF)\equiv\mathcal{F}\left(\dfrac{\partial W}{\partial\bfF}(\bfF)\right),
\end{equation*}
satisfy the condition (\ref{Constraint-Omega-Per}). As an example, we note that for the silicate glass considered throughout this work, with the material properties listed in Table \ref{Table1}, the smallest elastic energy density at strength is attained at $\bfS={\rm diag}(28\,{\rm MPa}, 26.5\,{\rm MPa}, 10.2\, {\rm MPa})$, yielding the bound $\ell_{bound}=1.37$ mm. 
}
\end{remark}

\begin{remark} Connection with the variational formulation of Francfort and Marigo and the classical Griffith energy competition. \emph{The proposed formulation (\ref{Potential-Energy})-(\ref{Gamma-Admi}) builds directly upon the variational framework of \cite{Francfort98} by introducing two critical constraints. First and foremost, the fundamental constraint (\ref{Gamma-Admi})$_3$ ensures that competing add-cracks lie entirely within the strength violation set (\ref{Viol-Set}). Second, the technical constraint (\ref{Gamma-Admi})$_4$ dictates that fracture at boundary regions --- when subjected to prescribed deformations and non-zero loads --- requires explicit modeling of the relevant interfacial mechanics.} 

\emph{Additionally, in the limiting case where a large pre-existing crack undergoes stable infinitesimal growth in opening mode, the formulation (\ref{Potential-Energy})-(\ref{Gamma-Admi}) reduces identically to the classical Griffith criterion $G=G_c$, subject to the interpretation of $G_c$ detailed in Subsection \ref{Toughness-History}.}
\end{remark}


\begin{remark}\label{Remark_FFM}
{\color{black} Difference with ``finite fracture mechanics''. \emph{Although it should be amply clear from the foregoing development, persistent confusion in the literature warrants stating explicitly that the proposed theory (\ref{Potential-Energy})-(\ref{Gamma-Admi}) is \textit{not} the approach commonly referred to as ``finite fracture mechanics'' or the ``coupled stress and energy criterion'' \citep{Leguillon2002,Cornetti2006,Mantic2014,Doitrand2024}. Beyond the superficial similarity of invoking both stress and energy, the two frameworks have little in common and differ in two fundamental respects.}

\emph{First, finite fracture mechanics relies on a vague, \textit{ad hoc} stress condition --- typically requiring the normal or shear traction along a putative crack surface to exceed a scalar critical value, either pointwise \citep{Leguillon2002} or on average \citep{Cornetti2006} --- that is not defined from experimental observations and bears no relation to the macroscopic strength surface $\mathcal{F}(\mathbf{S})=0$ of Definition \ref{Def-Strength}. Whereas $\mathcal{F}(\mathbf{S})=0$ is an intrinsic macroscopic material property, the stress condition in finite fracture mechanics is tied to a pre-selected plane and lacks experimental basis under general loading.}

\emph{Second, finite fracture mechanics says nothing about the crack path: a trajectory must be prescribed \textit{a priori}, with the stress and energy conditions merely evaluated along it. By contrast, in the proposed theory (\ref{Potential-Energy})--(\ref{Gamma-Admi}), the crack path, orientation, and size are not assumed beforehand, but emerge directly as predictions from the minimization of the energy functional (\ref{Fracture-Energy-Min}).}}
\end{remark}

\begin{remark} Non-convexity and computational tractability. \emph{The energy functional (\ref{Fracture-Energy-Min}) is generically not convex, even if the elastic energy density $W$ is taken to be convex. This is because the surface energy contribution is strictly monotonically increasing in $\Gamma$, while the potential energy is monotonically decreasing in $\Gamma$. While this non-convexity makes the minimization challenging, the admissibility constraint (\ref{Gamma-Admi})$_3$ --- which restricts competing add-crack to lie entirely within the strength violation set (\ref{Viol-Set}) --- is expected to be restrictive enough to render the problem more computationally  tractable.}
\end{remark}

\begin{remark}\label{rmk:inv}Frame indifference, material symmetry, and strength. \emph{As an intrinsic macroscopic material property, strength has frame indifference and material symmetry properties. So, we will assume henceforth that the set $\mathcal{F}(\bfS(\bfX,t))\le 0$ is actually a closed star-shaped subset $\ks$ of $\R^3$ containing $\textbf{0}$, its center, in its interior,  the boundary of which must be attained by the eigenvalues of the monotonically increasing in time and spatially constant stress tensor when a crack nucleates.}
\end{remark}

\section{First mathematical results: The case of isotropic linear elastic brittle materials in 2D}\label{Sec: Math} 

In this section, we specialize the theory to the case of a 2D homogeneous isotropic linear elastic brittle material under pure displacement boundary conditions. We do so because we can be more mathematically precise in such a setting, both in terms of the proposed formulation and of the currently achievable results.  

Notationwise, we will be using the standard notation of isotropic linear elasticity. In particular the Lam\'e coefficients, $\Lambda$ and $\mu$, and the Young's modulus and Poisson's ratio, $E$ and $\nu$, will be used. We recall that, classically in a $N$-dimensional setting,  $\mu>0, N\Lambda+2\mu>0$, $E>0$ and $\nu\in ]-1,1/2[$ if $N=2$, $\nu\in ]-1/(N-2),1/2[$ if $N\ge 3$. Further $\Lambda=\nu(1+\nu)/(E(1-(N-1)\nu)),\; \mu=(1+\nu)/(2E)$, and
\bel{eq.pose}\Lambda ({\rm tr\;}\bfE)^2+2\mu \,\bfE\cdot \bfE\ge \alpha |\bfE|^2 \mbox{ for some }\alpha>0.
\ee

For this section, we consider a bounded, open, connected, Lipschitz domain $\Om_0:=\Om$ with boundary $\pa\Om:=\partial\Omega^\mathcal{D}\cup\pa\Om^{\mathcal N}\cup\Delta$ where $\partial\Omega^\mathcal{D}$ and $\partial\Omega^\mathcal{N}$ are disjoint relatively open sets and the relative boundary $\Delta$ has $0$ $\htwo$-measure. We also abandon requirement \eqref{Gamma-Admi}$_5$ for simplicity and also because it is  not present in the by now classical description of variational fracture evolution.

\subsection{Length scale}

We first revisit and expand Remark \ref{Remark_length} in the particular context detailed above. Here, as in Section \ref{Sec: A representative case},  $\overline{\bfy}(\bfX,t)=\bfX+ \bfu(\bfX,t)$ where $\bfu$ is the displacement field. The elastic energy density $W$ is a function of $\bfE(\bfu):=1/2(\nabla \bfu+\nabla \bfu^T)$ and the first Piola-Kirchhoff stress tensor $\bfS$ identifies with the Cauchy stress tensor $\bfsig$. 

Because $\bfsig$ is constant throughout $\Om$, the  elasticity problem for any time $t$  is
\bel{eq:sys-el}
\begin{cases}
\bfu= t\bfe\bfX \mbox{ on }\pa\Om\\[2mm]\displaystyle \bfe=\frac1E((1+\nu)t\bfsig-\nu {\rm tr\;\!}(t\bfsig) \mathfrak i) \mbox{ in }\Om\\[2mm]
{\rm Div\;}(t\bfsig)=\textbf{0} \mbox{ in }\Om 
\end{cases},
\ee
where  the fourth equation above is automatically satisfied because $\bfsig$ is constant and $\mathfrak i$ stands for the identity mapping on $\R^3$.\footnote{Because of frame indifference and material symmetry, we do not  need to add to the displacement field a linearized rigid body motion, that is a term of the form $\bfA\bfX+\bfc$ where $\bfA$ is an arbitrary skew-symmetric matrix and $\bfc$ an arbitrary vector.}
The potential energy associated with \eqref{eq:sys-el} is
$$
\dfrac{1}{2} t^2 \leb^2(\Om) \left(\frac{1+\nu}{E} \bfsig\cdot\bfsig-\frac\nu E ({\rm tr\;\!}\bfsig)^2\right).
$$
Then, whenever $t$ is such that $t\bfsig\in \pa K$, say at $t:=t_\bfsig$, it must be so that
\begin{equation}\label{eq:nuc-nc}
\dfrac{1}{2} t_\bfsig^2 \leb^2(\Om) \left(\frac{1+\nu}{E} \bfsig\cdot\bfsig-\frac\nu E ({\rm tr\;\!}\bfsig)^2\right)\ge\frac{1}{2}\int_{\Om\setminus\Gamma}
\left(\frac{1+\nu}E \bftau\cdot\bftau-\frac\nu E ({\rm tr\;\!}\bftau)^2\right){\rm d}\bfX+ G_c \htwo(\Gamma)
\end{equation}
for some closed $\oli\Omega\supset\Gamma\ne \emptyset$, where $\bftau(\bfX)$ is the  stress field solution to
$$\begin{cases}
\bfu= t_\bfsig \bfe\bfX\mbox{ on }\partial\Omega^\mathcal{D}\setminus\Gamma\\[2mm] \bfE(\bfu)(\bfX)=\dfrac1E((1+\nu)\bftau-\nu {\rm tr\;\!}(\bftau) \mathfrak i) \mbox{ in }\Om\setminus\Gamma\\[2mm]
{\rm Div\;}(\bftau)=\textbf{0} \mbox{ in }\Om\setminus\Gamma
\end{cases} .$$

Now, in particular, it is sufficient to take $\Gamma=\pa\Om$, provided that $\Om$ has finite perimeter in which case we obtain as sufficient condition
\bel{eq:ineq-nuc}
\dfrac{1}{2} t_\bfsig^2 \leb^2(\Om) \left(\frac{1+\nu}{E} \bfsig\cdot\bfsig-\frac\nu E ({\rm tr\;\!}\bfsig)^2\right)\ge  G_c {\rm Per\;\!}(\Omega).
\ee
Note that the isoperimetric inequality \citep[e.g.,][Theorem 3.46]{AFP}, namely that ${\rm Per\;\!}(\Omega)\ge  \sqrt {4\pi}
(\leb^2(\Om))^{\frac12}$ does not add any useful information to \eqref{eq:ineq-nuc}. In any case, since we can span all of $\pa K$ by varying $\bfsig$ we obtain, upon setting ${\boldsymbol\tau}:=t_\bfsig\bfsig$ in \eqref{eq:ineq-nuc}, that, for a domain $\Om$ with finite perimeter made up of that material to be admissible for at least one nucleation event, it suffices that
\bel{eq:sc-nuc}
\dfrac{1}{2} \max_{{\boldsymbol\tau} \in \pa K} \left\{\frac{1+\nu}{E} {\boldsymbol\tau}\cdot{\boldsymbol\tau}-\frac\nu E ({\rm tr\;\!}{\boldsymbol\tau})^2\right\}\ge G_c \frac{{\rm Per\;\!}(\Omega)}{\leb^2(\Om)}.
\ee
Recalling Remark \ref{rmk:inv}, we thus obtain from \eqref{eq:sc-nuc} that the following proposition holds true.
\begin{proposition}\label{prop:dom-suf}
If the bounded open set $\Om$ has finite perimeter, then a crack can nucleate in $\oli\Om$ provided that
\bel{eq:dom-suf}
\frac{\leb^2(\Om)}{{\rm Per\;\!}(\Omega)}\ge 2G_cE\min_{(\lambda_1,\lambda_2)\in \ks} \frac1{(\lambda_1^2+\lambda_2^2)-2\nu\lambda_1\lambda_2}.
\ee
\end{proposition}
\noindent This proposition introduces a minimal length-scale for {${\leb^2(\Om)}/{{\rm Per\;\!}(\Omega)}$ which, if satisfied, will ensure that some kind of nucleation event can take place. In particular, it points to a notion of fatness of the domain. Indeed a very thin and long domain would generically violate \eqref{eq:dom-suf}, at least provided that the right hand-side is not $0$. 

\begin{remark}
\emph{Note that, by appealing to a result of Federer \citep[e.g.,][Theorem 3.46, Section 5.11]{evans.gariepy}, any bounded set with Lipschitz boundary has finite perimeter. The same would hold true of a Lipschitz domain $\Om$ with a (smooth) pre-crack $\Gamma$ such that $\htwo(\Gamma\cap\pa\Om)<\infty$.}
\end{remark}

\begin{remark}Difference with cohesive fracture.\emph{
The interested reader will remark that, in the setting of cohesive fracture, nucleation must occur when, in essence, the stress state reaches the slope at $0$ of the surface energy \citep[e.g.,][Section 4.2]{Bourdin08} and this at a single material point. There, nucleation  is not connected to any kind of minimality, but only to a first order necessary condition for any kind of minimality, whereas here, global minimality plays a decisive role in obtaining \eqref{eq:nuc-nc}.}

\emph{Further, there is no intrinsic length scale in the cohesive notion of strength. In other words, the meaningful domains from the standpoint of fracture are not constrained by the strength. Our approach, to the contrary, imparts a length scale which operates as a selection criterion for domains for which fracture can take place. This we view as both a positive and a distinguishing feature.}
\end{remark}

\subsection{The discrete evolution}

Let us now rewrite the constrained energy minimization problem put forth in Section \ref{Sec: The Theory} in the specific setting at hand and when no tractions or body forces are present.

As mentioned before, we are unable at this time to put forth a time-continuous evolution for an elastic brittle domain subjected to a quasi-static loading process. This is so because the nature of the dissipation carried by the strength eludes us. As was seen in Section  \ref{Sec: A representative case}, there might be a drop of energy when the crack experiences a sudden jump in length. That drop is not seen in the classical formulation \citep{francfort.larsen, DMFT} where the only non-zero boundary conditions are on a Dirichlet part of the boundary.
  
We will take, for $t_0=0\le t_1\le...\le t_N=t$ the loading process to be of the form
$$
\oli\bfu_k   \mbox{ a $\bfX$-dependent imposed displacement on }\pa\Om^{\mathcal{D}}
$$
with adequate regularity (say $C^1$ for example).



The formulation is detailed below. It starts with the definition of the admissible test cracks and displacement fields.
\begin{definition}[\bf The admissible configurations]\label{def:min-sch} We set all quantities with index $-1$ to be $0$. Knowing $\Gamma_{k-1}, \bfu_{k-1}, \bfsig_{k-1}:=\Lambda {\rm Div\;\!}\bfu_{k-1}\mathfrak i+2\mu \bfE(\bfu_{k-1})$ the stress field associated to $\bfu_{k-1}$, we define
\bel{def:sigi-1}
\Sigma_{k-1}:=\overline{\{\bfX\in \Om\setminus\Gamma_{k-1}: \bfsig_{k-1}(\bfX)\notin\mathring\ks\}}
\ee
where 
$\ks$ was defined in Remark \ref{rmk:inv}.

We say that the pair $(\bfu,\Gamma)$ is an admissible configuration at time $t_k$ for the boundary displacement $\oli\bfu_k\in H^1(\Om;\R^2)$ if
$$
\Gamma\in \ks_m^k(\Omb)
\mbox{ 
and }
\bfu\in \LDD(\Om\setminus \Gamma)\;\text{with}\; \bfu=\oli\bfu_k\text{ on }\partial\Omega^\mathcal{D} \setminus \Gamma,
$$
where
$m\ge 1$ is a fixed number, where,
for any open set $A$,
$$
\LDD(A):=\{\bfv \in L^2_{loc}(A;\mathbb\R^2): \bfE(\bfu)\in L^2(A;\mathbb R^{2\times2})\}
$$
and where
\begin{multline*}
\ks_m^k(\Omb):=\{\Gamma \mbox{ closed subset of $\bar\Om$ with at most $m$ connected components},\\
\Gamma_{k-1}\subset\Gamma\subset \Gamma_{k-1}\cup \Sigma_{k-1}
\text{\;and $\hu(\Gamma)<+\infty$}\}.
\end{multline*}

We will write $(\bfu,\Gamma)\in \as(\oli\bfu_k)$.
The pair $(\bfu,\bfE(\bfu))$ can be thought of as an element of $L^2_{loc}(\Om;\R^2)\times L^2(\Om;\Mtwosym)$ since $\Gamma$ has null Lebesgue measure.
\end{definition}

\noindent The family of closed sets in $\R^N$ can be endowed with the Hausdorff metric $d_H$ defined by
\begin{equation*}
\label{eq:def-dh}
d_H(K_1,K_2) := \max \left\{ \sup_{x \in K_1} {\rm dist}(x,K_2), \sup_{y \in
K_2} {\rm dist}(y,K_1)\right\}
\end{equation*}
with the conventions ${\rm dist}(x, \emptyset)=+\infty$ and $\sup \emptyset=0$, so that $d_H(\emptyset, K)=0$ if $K=\emptyset$ and $d_H(\emptyset,K)=+\infty$ if $K \not=\emptyset$. The Hausdorff metric has good compactness properties \citep[e.g.,][Theorem 4.4.15]{ambrosio.tilli}.

\begin{proposition}[\bf Compactness]
\label{prop:blashke}
Let $(\Gamma_n)_{n\in\N}$ be a sequence of compact sets contained in a fixed compact set $C$  of $\R^N$. Then there exists a compact set $\Gamma\subseteq C$ such that up to a subsequence
$$
\Gamma_{n}\to \Gamma
\qquad\text{in the Hausdorff metric}.
$$
\end{proposition}

\noindent Then the formulation hinges on the following minimization scheme. 

\noindent 
\begin{definition}[\bf The minimization scheme]\label{def.minsch}
Compute
\begin{equation*}
\mathcal I:= \inf_{(\bfu,\Gamma)}\int_{\Om\setminus\Gamma}
\dfrac{1}{2}\left(\Lambda ({\rm Div\:\!}\bfu(\bfX))^2+2\mu \bfE(\bfu(\bfX))\cdot \bfE(\bfu(\bfX))\right){\rm d}\bfX
+ G_c \htwo(\Gamma)
\end{equation*}
over all pairs $(\bfu,\Gamma)\in \as(\oli\bfu_k)$. Denote by $(\bfu_{k+1},\Gamma_{k+1})$ a minimizer.

 
 
 

\end{definition}




\noindent If we restrict the setting to  cracks that are closed sets with a uniformly bounded number of connected components, say $m$, then the incremental problem is well-posed. Indeed,  consider a minimizing sequence $(\bfu_n,\Gamma_n)$. Since $\Gamma_{k-1}\cup\Sigma_{k-1}$ with $\Sigma_{k-1}$ defined in \eqref{def:sigi-1} is a compact set we can apply Proposition \ref{prop:blashke} to $\Gamma_n$ and we obtain that, up to a subsequence (not relabeled), $\Gamma_n$ Hausdorff converges to $\Gamma\in \ks_m^k(\Omb)$. Further, in view of \eqref{eq.pose}, we can also assume that $\bfE(\bfu_n)$ converges weakly in $L^2(\Om;\Msym)$ to some $\Phi$.

With this the following compactness result \citep[Lemma 5.2]{FGLP19} applies to the setting at hand.

\begin{lemma}
\label{lem:config-closed-lin}
Let $\oli\bfu_n,\oli \bfu\in H^1(\Om;\R^2)$ be such that
$$
\oli\bfu_n\to \oli\bfu\qquad\text{strongly in }H^1(\Om;\R^2).
$$
Assume that $(\bfu_n,\Gamma_n)\in \as(\oli\bfu_n)$ with
$$
\bfE(\bfu_n) \weak \Phi\qquad\text{weakly in }L^2(\Om;\Msym),
$$
and
$$
\Gamma_n\to \Gamma\qquad\text{in the Hausdorff metric}.
$$
Then there exists $(\bfu,\Gamma)\in \as(\oli\bfu)$ such that $\Phi=\bfE(\bfu)$ on $\Om\setminus \Gamma$.
\end{lemma}

\noindent It now suffices to apply {\fontencoding{T1}\selectfont Go\l\k{a}b}'s theorem \citep[e.g.,][Theorem 4.4.17]{ambrosio.tilli}.

\begin{theorem}[\bf {\fontencoding{T1}\selectfont Go\l\k{a}b}'s]
\label{thm:golab}
Let $(K_n)_{n\in\N}$ be a sequence of compact connected sets in $\R^N$ such that
$$
K_n\to K\qquad\text{in the Hausdorff metric.}
$$
Then $K$ is connected and for every open set $A\subseteq \R^N$
$$
\hu(K\cap A)\le \liminf_{n\to\infty} \hu(K_n\cap A).
$$
\end{theorem}

\noindent Note that the lower semicontinuity in {\fontencoding{T1}\selectfont Go\l\k{a}b}'s Theorem still holds when $K_n$ has a uniformly bounded number of connected components. The conclusion is reached thanks to Lemma \ref{lem:config-closed-lin}, to {\fontencoding{T1}\selectfont Go\l\k{a}b}'s Theorem and to obvious arguments of weak lower-semicontinuity.

\begin{remark}\label{rem:force-loads}\emph{In the case where either non-zero traction forces and/or non-zero body forces are present, the situation is more complex and several issues may arise: the lack of bounds on the crack length of the minimizing sequences, the lack of $L^1$ regularity of the test displacement fields, etc. This may occur whether the infimum of (\ref{Fracture-Energy-Min}) is finite or not. In particular, the crack may sever $\Omega$ into pieces that --- non-interpenetrability permitting --- can be moved to infinity as rigid bodies, generically producing an infimum equal to $-\infty$ as mentioned earlier in Section \ref{Sec: The Theory}. This will be the object of our next contribution.}
\end{remark}

\section{Compliance with \emph{``The Nine Circles of Elastic Brittle Fracture''} on glass and rubber}\label{Sec: Nine Circles}

In a recent contribution, \cite{KZDLP26} introduced ``The Nine Circles of Elastic Brittle Fracture,'' a comprehensive obstacle course of nine challenge problems designed to assess the viability of any proposed fracture theory. As defining features, these problems can be carried out experimentally with standard testing equipment and, more critically, span the full spectrum of well-established experimental knowledge for both fracture nucleation and propagation. Specifically, they examine nucleation dominated by material strength under uniform stress, by the Griffith energy competition from large pre-existing cracks, and by the more complex  interplay between strength and Griffith energy competition under non-uniform stress states, be they singular or not. For propagation, the problems examine crack growth under opening (Mode I) and tearing (Mode III) conditions. Failure to accurately resolve even a single ``circle'' disqualifies a theory as a viable framework for describing, and hence predicting, fracture in general.

\begin{table}[b!]\centering\small
\caption{Material constants for the synthetic rubber used throughout this work.}
\begin{tabular}{l?cc}
\toprule
Elasticity constants & $\mu$ (MPa)& $\Lambda$ (MPa) \\
\midrule
                     & $0.52$   & $85.77$        \\
\midrule
\midrule
Strength constants   & $s_{\texttt{ts}}$ (MPa) & $s_{\texttt{hs}}$ (MPa) \\
\midrule
                     & $0.3$                  & $1$          \\
\midrule
\midrule
Fracture toughness   & $G_c$ (N/m) &  \\
\midrule
                               & $41$ &   \\
\bottomrule
\end{tabular} \label{Table3}
\end{table}
In this section we subject the proposed theory (\ref{Potential-Energy})-(\ref{Gamma-Admi}) to ``The Nine Circles of Elastic Brittle Fracture.'' Two circles have already been successfully tested. Section~\ref{Sec: A representative case} addresses the $1^{\text{st}}$ circle (nucleation under uniaxial tension), and in \citep{LPK25}, it was tested against the $4^{\text{th}}$ circle (nucleation from a large pre-existing crack under opening mode). In that study, the theory was also directly compared with classical experiments on nucleation from large pre-existing cracks under complex loading conditions to validate its predictions of non-trivial crack paths. Consequently, we focus here on the seven remaining challenge problems. Much like in \citep{KZDLP26}, the results pertain to the same hard silicate glass investigated in Section~\ref{Sec: A representative case}, as well as to a soft synthetic rubber with Neo-Hookean elastic energy density
\begin{equation*}
W(\bfF)=\dfrac{\mu}{2}\left(I_1-3\right)-\mu\ln J+\dfrac{\Lambda}{2}(J-1)^2,
\end{equation*}
Drucker-Prager strength surface (\ref{DP}), scalar fracture toughness $G_c$, and the material constants listed in Table~\ref{Table3}; note that in terms of the hydrostatic strength, $\shs$, the uniaxial compressive strength is given by $\scs=3\shs\sts/(3\shs-2\sts)$. Specifically, the poker-chip ($7^{\text{th}}$ circle) and trousers ($9^{\text{th}}$ circle) tests are carried out on the synthetic rubber, whereas the remaining five are conducted on the silicate glass.

\subsection{Biaxial tension test}

We begin by considering the $2^{\text{nd}}$ circle, a circular plate of radius $R = 5$~mm and thickness $B = 0.25$~mm that is subjected to equi-biaxial tension via a uniform radial displacement $u$ applied at its lateral boundary $\bfX\in\partial\Omega_l=\{\bfX:X_1^2+X_2^2=R^2,\,-B/2<X_3<B/2\}$, while its top and bottom boundaries remain traction-free. Here, $\{\bfe_1,\bfe_2,\bfe_3\}$ stands for the laboratory frame indicated in the test schematic shown in Fig.~\ref{Fig11}(a). This boundary condition generates a spatially uniform biaxial strain, $u/R$, and a corresponding uniform biaxial stress, $S$, within the plate. Precisely,
\begin{equation*}
\bfE=\dfrac{u}{R}(\bfe_1\otimes\bfe_1+\bfe_2\otimes\bfe_2)+e_t\bfe_3\otimes\bfe_3\quad {\rm and}\quad \bfS=S(\bfe_1\otimes\bfe_1+\bfe_2\otimes\bfe_2),
\end{equation*}
where $e_t$ denotes the transverse strain induced by the Poisson effect. This uniform state of stress and strain persists until the stress $S$ reaches the material's biaxial tensile strength $\sbs$; for the silicate glass of interest in this work, $\sbs=2\sts\scs/(3\scs-2\sts)=27$~MPa. At this critical threshold, a through-thickness crack abruptly nucleates and rapidly propagates, severing the plate into multiple fragments. Due to the inherent spatial variations in strength, the exact location of crack nucleation is inherently arbitrary.

%
\begin{figure}[t!]
\centering
\includegraphics[width=0.925\linewidth]{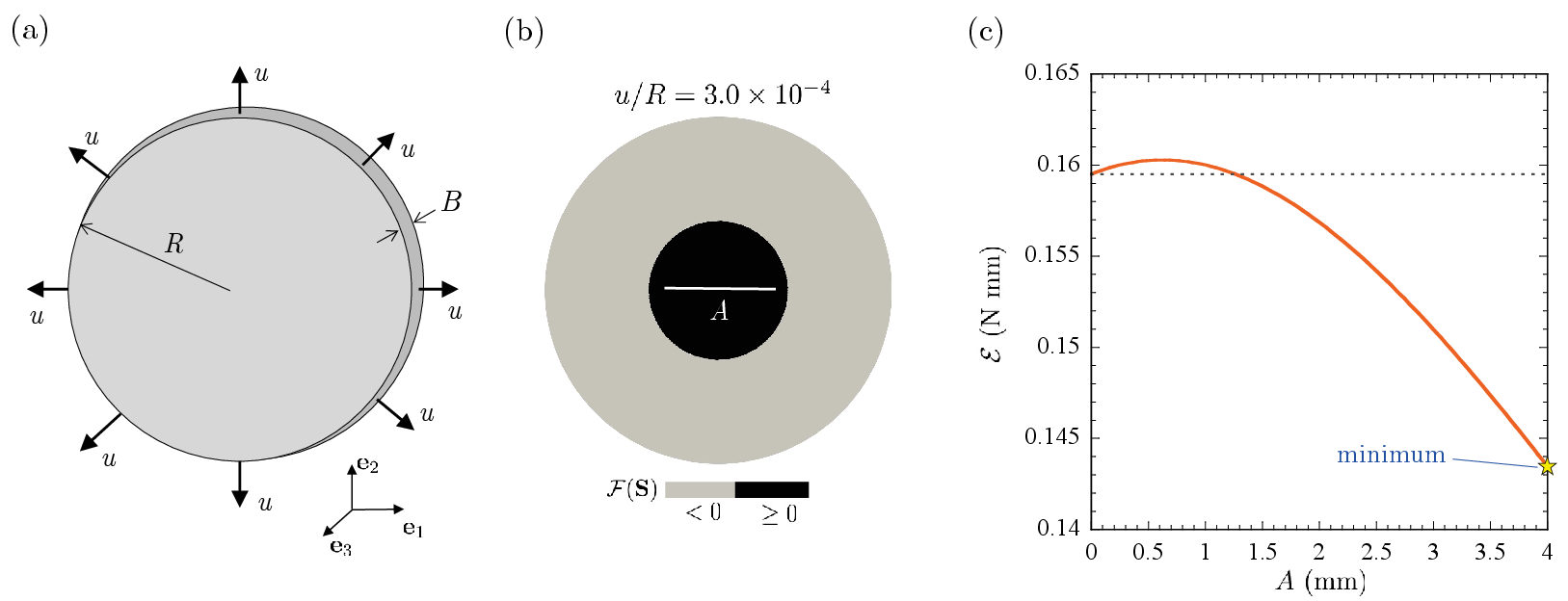}
\caption{{\small (a)  Schematic of the biaxial tension test on silicate glass. (b) Contour plot of the region of the specimen where the stress field first exceeds ($\mathcal{F}(\bfS)\geq0$) the strength surface of the material, depicting the test cracks considered within, at $u/R=3.0\times 10^{-4}$. (c) The corresponding energy $\mathcal{E}$ of the specimen plotted as a function of the test crack length $A$.}}\label{Fig11}
\end{figure}
%

%
\begin{figure}[t!]
\centering
\includegraphics[width=0.925\linewidth]{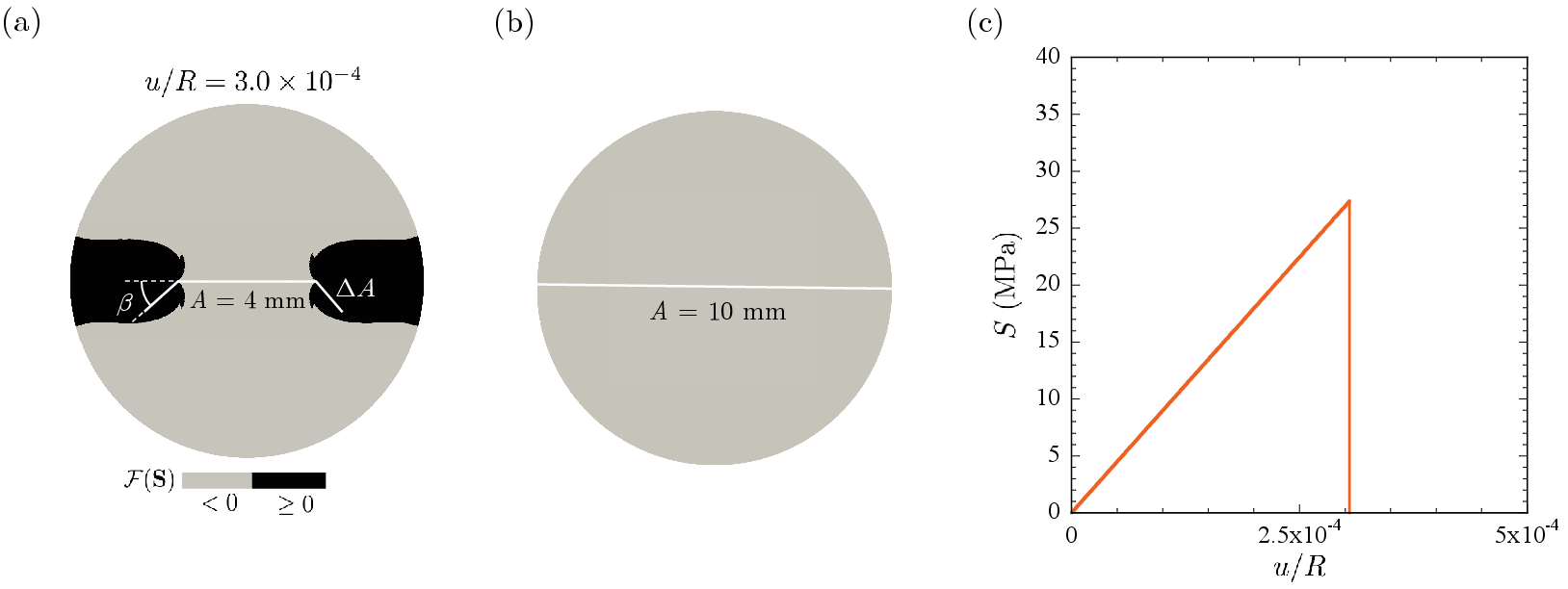}
\caption{{\small (a) The specimen at the applied strain $u/L=3.0\times 10^{-4}$ containing the initial minimizing crack ($A=4\,\mathrm{mm}$), depicting the regions where the stress field exceeds ($\mathcal{F}(\bfS)\geq0$) the strength surface of the material along with the test add-cracks of length $\Delta A$ oriented at angles $\beta$ considered within. (b) Schematic of the severed state of the specimen following the brutal propagation of the crack at $\beta=0^\circ$ dictated by the theory. (c) The global stress $S$ as a function of the applied strain $u/R$ predicted by the theory.}}\label{Fig12}
\end{figure}
%

As in Section \ref{Sec: A representative case},  we deploy the proposed theory (\ref{Potential-Energy})-(\ref{Gamma-Admi}) by first numerically determining the strain $u/R$ at which the strength condition $\mathcal{F}(\bfS)\geq 0$ is initially satisfied within some region of the plate. For definiteness, we consider the presence of a weaker central circular region with a 4 mm diameter where the uniaxial tensile strength is 10\% lower ($\sts=40$ MPa) than the surrounding material ($\sts=44$ MPa). The strength condition $\mathcal{F}(\bfS)\geq 0$ is first satisfied in such a weaker region at $u/R=3.0\times 10^{-4}$. At this strain, we then numerically determine the energy (\ref{Fracture-Energy-Min}) for plates containing straight through-thickness test cracks, $\Gamma=\Gamma(A)$, of crack length $A$, so that $\mathcal{H}^{2}(\Gamma)=A B$; see Fig.~\ref{Fig11}(b). Due to loading symmetry and material isotropy, different crack orientations need not be considered. Figure \ref{Fig11}(c) presents the resulting energy $\mathcal{E}$ as a function of $A$ assuming plane-stress conditions. The main observation from this figure is that the crack of length $A=4$ mm --- the largest possible length that can be fitted within the strength violation region --- minimizes the energy $\mathcal{E}$. According to the theory (\ref{Potential-Energy})-(\ref{Gamma-Admi}), this implies that such a crack nucleates at $u/R=3.0\times 10^{-4}$, consistent with the reference result. It is also worth noting that no crack nucleates --- because the energy is minimized by $A=0$ --- if the strength violation is confined to an area with a diameter under $\ell_{\texttt{bs}}=1.25$ mm. This defines the length scale below which the biaxial tensile strength $\sbs$ can no longer be taken as a macroscopic material property; see Remark \ref{Remark_length}.

We now examine the minimizing solution with crack length $A=4$ mm. At the critical applied strain of $u/R=3.0\times 10^{-4}$, we numerically identify the regions where the stress field violates the strength surface, $\mathcal{F}(\bfS)\geq0$. As the contour plot in Fig.~\ref{Fig12}(a) illustrates, this strength violation is localized near the crack fronts. Within these specific regions, we introduce straight through-thickness test add-cracks. These are characterized by an added length $\Delta A$ --- yielding a total crack surface of $\mathcal{H}^{2}(\Gamma)=(A+2\Delta A) B$ --- and an orientation $\beta$ relative to the initial crack, as depicted schematically in Fig.~\ref{Fig12}(a). Our results indicate that the energy $\mathcal{E}$ is minimized at $\Delta A=3\,\mathrm{mm}$ and $\beta=0^\circ$. That is, the proposed constrained minimization dictates that the nucleated crack at $u/R=3.0\times 10^{-4}$ does \emph{not} stay put, instead it propagates by a finite-size growth in the same initial crack direction severing the specimen. Figures \ref{Fig12}(b) and (c) display the resulting severed configuration and the complete global stress-strain response predicted by the theory, both of which align with the reference result.

\subsection{Torsion test}

We now consider the $3^{\text{rd}}$ circle, a thin-walled circular tube with an initial length $L=5$ mm, an inner radius $R_i=2.85$ mm, and an outer radius $R_o=3$ mm. The tube is subjected to torsion via an angle of twist $\phi$ applied at one end, while the opposite end remains rigidly fixed; see Fig.~\ref{Fig13}(a). Under these conditions, the tube experiences a state of shear strain and shear stress of the form
\begin{equation}
\bfE=\dfrac{\phi X_1}{2L}(\bfe_2\otimes\bfe_3+\bfe_3\otimes\bfe_2)-\dfrac{\phi X_2}{2L}(\bfe_1\otimes\bfe_3+\bfe_3\otimes\bfe_1)=\gamma(\bfe_{\Theta}\otimes\bfe_Z+\bfe_Z\otimes\bfe_{\Theta})\label{E-gamma}
\end{equation}
and
\begin{align}
\bfS=\dfrac{2T X_1}{\pi(R_o^4-R_i^4)}(\bfe_2\otimes\bfe_3+\bfe_3\otimes\bfe_2)-\dfrac{2TX_2}{\pi(R_o^4-R_i^4)}(\bfe_1\otimes\bfe_3+\bfe_3\otimes\bfe_1)
=\tau(\bfe_{\Theta}\otimes\bfe_Z+\bfe_Z\otimes\bfe_{\Theta})\label{S-tau}
\end{align}
for $\bfX\in\Omega=\{\bfX:R_i^2<R^2<R_o^2,\,0<X_3<L\}$ with $R=\sqrt{X_1^2+X_2^2}$. Here, $\{\bfe_1,\bfe_2,\bfe_3\}$ is the laboratory frame depicted in the figure, whereas $\{\bfe_R,\bfe_{\Theta},\bfe_Z\}$ denotes the associated cylindrical frame of reference, $\gamma=\phi R/(2L)$ is the shear strain, $T$ denotes the resultant torque at the tube ends, and $\tau=2T R/(\pi(R_o^4-R_i^4))$ stands for the corresponding shear stress. 

Because the tube is thin ($t=R_o-R_i=0.15$ mm relative to $R_o=3$ mm), the stress (\ref{S-tau}) and strain (\ref{E-gamma}) fields are nearly uniform, with average values of $S=T (R_i+R_o)/(\pi(R_o^4-R_i^4))=T(R_i-t)/(2\pi R_i^3 t)+O(t)$ and $\phi (R_i+R_o)/(4L)=\phi(2R_i+t)/(4L)$.  This quasi-uniform state persists until the macroscopic shear stress $S$ reaches the material's shear strength $\sss$; for the silicate glass of interest in this work, $\sss=2\sts\scs/(\sqrt{3}(\scs+\sts))=44.4$~MPa. At this threshold, abrupt cracking severs the tube. While material stochasticity causes these cracks to form at arbitrary locations, experimentally they consistently exhibit a $45^\circ$ orientation relative to the symmetry axis $\bfe_3$ --- an orientation classically attributed to maximum principal stress alignment; see, e.g., Chapter 3 in the popular textbook \cite{BeerJohnston2019} and Remark \ref{rmk: Torsion}.

%
\begin{figure}[t!]
\centering
\includegraphics[width=0.925\linewidth]{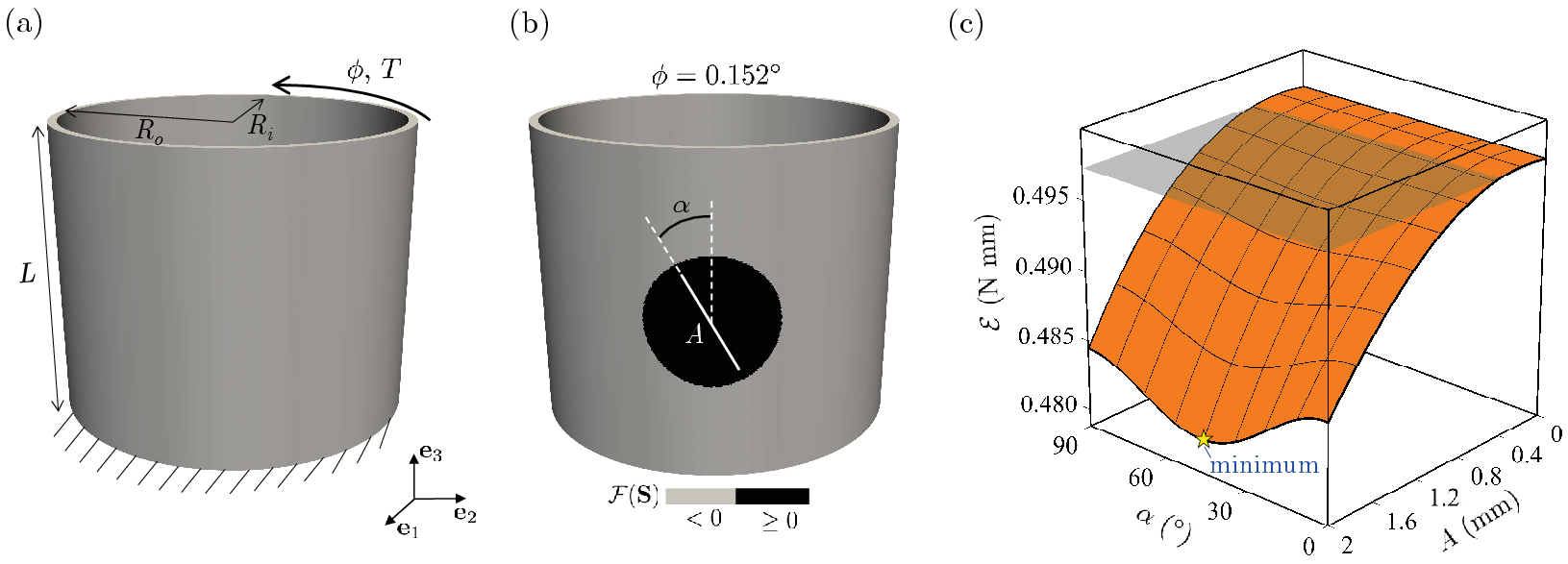}
\caption{{\small (a)  Schematic of the torsion test on silicate glass. (b) Contour plot of the region of the specimen where the stress field first exceeds ($\mathcal{F}(\bfS)\geq0$) the strength surface of the material, depicting the test cracks considered within, at $\phi=0.152^\circ$. (c) The corresponding energy $\mathcal{E}$ of the specimen plotted as a function of the test crack outer arc length $A$ and angle $\alpha$.}}\label{Fig13}
\end{figure}
%
%
\begin{figure}[t!]
\centering
\includegraphics[width=0.85\linewidth]{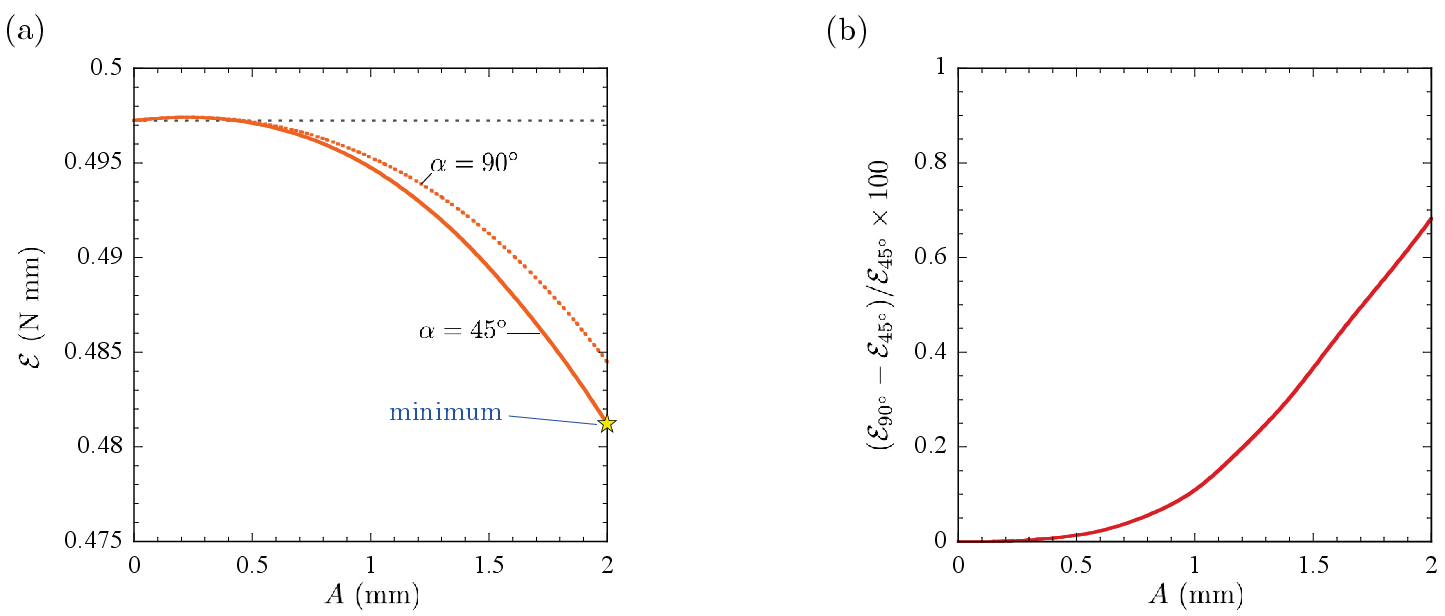}
\caption{{\small (a) The variation of the energy $\mathcal{E}$ as a function of crack length $A$ for fixed crack angles $\alpha=45^\circ$ and $90^\circ$. (b) The corresponding percentage difference in $\mathcal{E}$ between the two cases.}}\label{Fig14}
\end{figure}
%

As with the previous cases, we start by determining the angle of twist $\phi$ at which the strength condition $\mathcal{F}(\bfS)\geq 0$ is first satisfied within some region of the tube. For definiteness, we consider the presence of a weaker through-thickness cylindrical region with a 2 mm diameter where the uniaxial tensile strength is 10\% lower ($\sts=40$ MPa) than the surrounding material ($\sts=44$ MPa). The strength criterion $\mathcal{F}(\bfS)\geq 0$ is initially met in this weaker region at $\phi=0.152^\circ$.  At this critical angle of twist, we numerically determine the energy (\ref{Fracture-Energy-Min}) for tubes containing straight through-thickness test cracks of  outer arc length $A$, inner arc length $A R_i/R_o$, and thickness $t=R_o-R_i$ --- yielding a crack surface area of $\mathcal{H}^{2}(\Gamma)=At (2R_i+t)/(2(R_i+t))=A t+O(t^2)$ --- and an orientation $\alpha$ relative to the symmetry axis of the tube; see Fig.~\ref{Fig13}(b). Figure~\ref{Fig13}(c) presents the resulting energy $\mathcal{E}$ as a function of $A$ and $\alpha$. To visualize the quantitative aspects of the results more clearly, Fig.~\ref{Fig14}(a) plots $\mathcal{E}$ as a function of $A$ for fixed orientations $\alpha = 45^\circ$ and $90^\circ$, while Fig.~\ref{Fig14}(b) plots the percentage difference in $\mathcal{E}$ between these two cases. 

%
\begin{figure}[t!]
\centering
\includegraphics[width=0.925\linewidth]{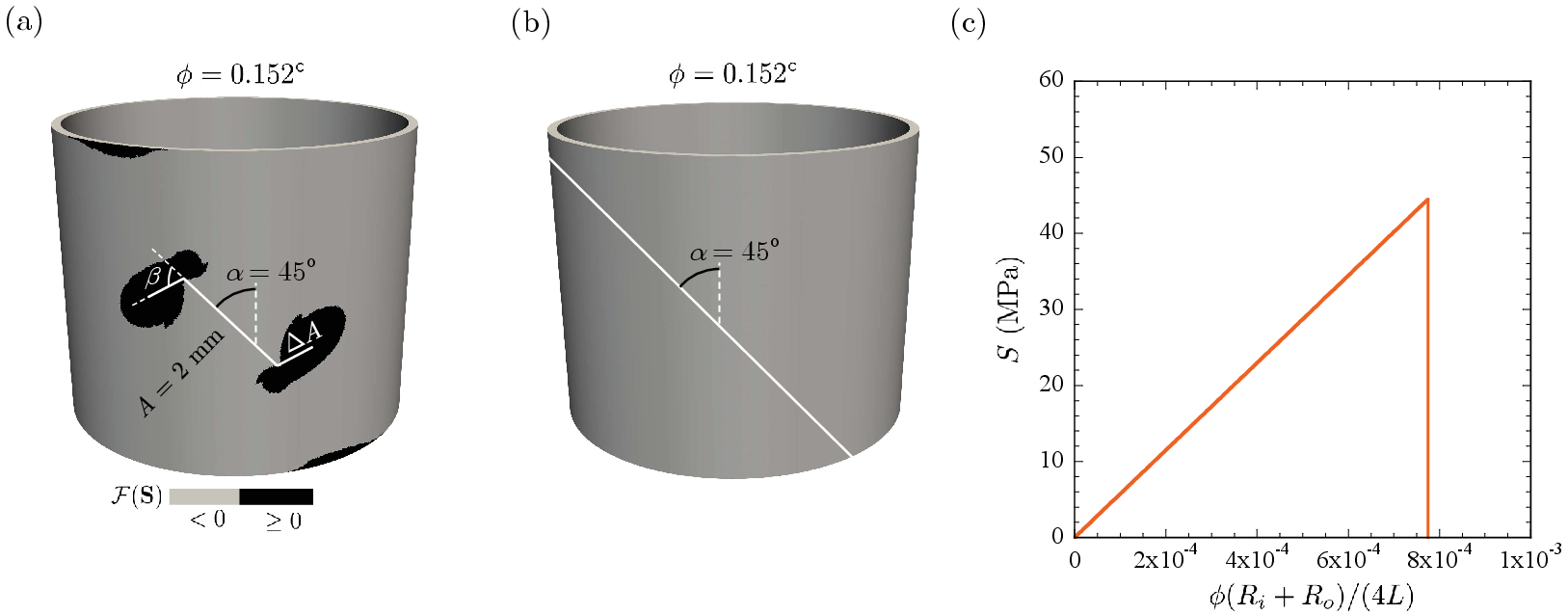}
\caption{{\small (a) The specimen at the critical applied angle of twist $\phi=0.152^\circ$ containing the initial minimizing crack $(A=2\,\mathrm{mm},\alpha=45^\circ)$, depicting the regions where the stress field exceeds ($\mathcal{F}(\bfS)\geq0$) the strength surface of the material along with the test add-cracks of length $\Delta A$ oriented at angles $\beta$ considered within. (b) Schematic of the severed state of the specimen following the brutal propagation of the crack at $\beta=0^\circ$ dictated by the theory. (c) The global stress $S$ as a function of the global strain $\phi(R_i+R_o)/(4L)$ predicted by the theory.}}\label{Fig15}
\end{figure}
%

Figures~\ref{Fig13}(c) and \ref{Fig14} suggest the following. First and foremost, the energy $\mathcal{E}$ is minimized by a crack with an outer arc length $A=2$ mm --- the largest crack that can be accommodated within the strength-violation region --- oriented at $\alpha=45^\circ$. According to the theory (\ref{Potential-Energy})-(\ref{Gamma-Admi}), this indicates that a crack nucleates at an angle of twist of $\phi=0.152^\circ$ along this orientation, in agreement with the reference result. Second, if the strength violation is restricted to a region with a diameter smaller than $\ell_{\texttt{ss}}=0.44$ mm, the energy minimum shifts to $A=0$, meaning no crack nucleates. As detailed in Remark~\ref{Remark_length}, this establishes the critical length scale below which the shear strength $\sss$ ceases to be treatable as a macroscopic material property.  

We next examine the minimizing solution with crack length $A=2$ mm oriented at $\alpha=45^\circ$. At the critical applied angle of twist $\phi=0.152^\circ$, we numerically identify the regions where the stress field violates the strength surface, $\mathcal{F}(\bfS)\geq 0$. As illustrated by the contour plot in Fig.~\ref{Fig15}(a), this strength violation is localized near the crack fronts, as well as around small regions on the top and bottom boundaries. Within the former, we consider straight through-thickness test add-cracks defined by an additional outer arc length $\Delta A$ and an orientation angle $\beta$ relative to the initial crack direction, yielding a total crack surface area of $\mathcal{H}^{2}(\Gamma)=(A+2\Delta A)t(2R_i+t)/(2(R_i+t))=(A+2\Delta A)t+O(t^2)$. Our computations show that the energy $\mathcal{E}$ is minimized at $\Delta A=0.31$ mm and $\beta=0^\circ$. This implies that the crack nucleated at $\phi=0.152^\circ$ does not remain stationary, but instead undergoes finite-size growth along its original trajectory. Subsequent computations reveal that the crack continues to propagate along this path until the tube is severed. Figures \ref{Fig15}(b) and (c) display the resulting fractured configuration and the predicted global stress-strain response, both of which are in agreement with the reference results. 

\begin{remark}\label{rmk: Torsion} \emph{The proposed theory demonstrates that in elastic brittle tubes subjected to torsion, cracks nucleate and propagate at a $45^\circ$ angle because this trajectory minimizes the energy (\ref{Fracture-Energy-Min}). In this particular setting, the energy-minimizing direction coincidentally aligns with the plane orthogonal to the maximum principal stress.  Because of this, it has become a mantra that orthogonality to the maximum principal stress direction is the physical explanation for the observed fracture angle \citep[e.g.,][]{BeerJohnston2019}. However the maximum principal stress criterion does not agree with many other experimental observations.}
\end{remark}

\subsection{Single edge notch test}

In contrast to the previous circles which considered fracture nucleation under spatially (quasi-)uniform stress states, this subsection and the subsequent two subsections shift focus to the $5^{\text{th}}$ through $7^{\text{th}}$ circles, which examine fracture nucleation under spatially non-uniform stress fields. 

%
\begin{figure}[t!]
\centering
\includegraphics[width=0.9\linewidth]{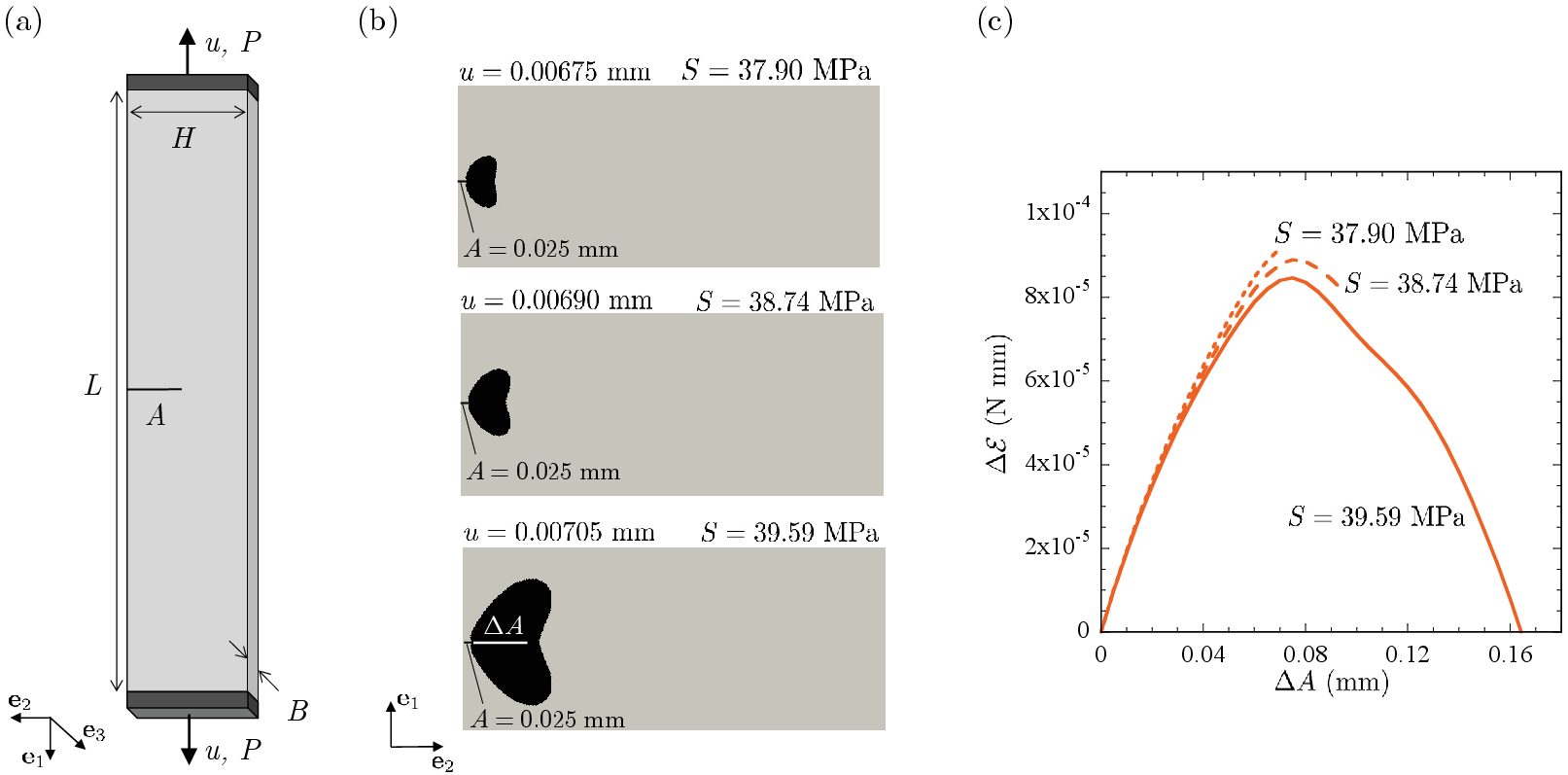}
\caption{{\small (a) Schematic of the single edge notch test on silicate glass. (b) Contour plots of magnified regions where the strength surface is violated ($\mathcal{F}(\mathbf{S})\geq0$) for a specimen with a pre-existing crack of length $A=0.025$~mm, illustrating the considered test add-cracks, at three applied displacements  $u$ and global stresses $S=P/(BH)$. (c) The corresponding energy difference $\Delta\mathcal{E}=\mathcal{E}-\mathcal{E}\vert{}_{\Delta A=0}$ plotted as a function of the add-crack length $\Delta A$.}}\label{Fig16}
\end{figure}
%
%
\begin{figure}[t!]
\centering
\includegraphics[width=0.85\linewidth]{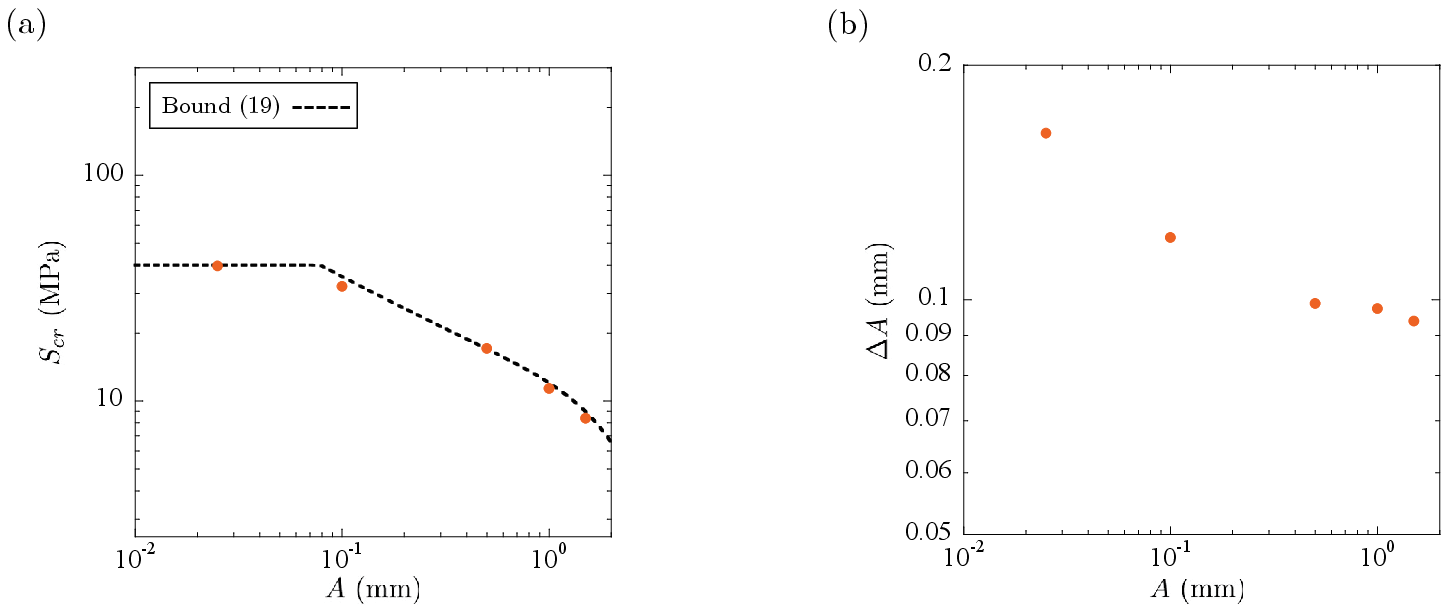}
\caption{{\small(a) The critical global stress $S_{cr}$ and (b) the associated add-crack length $\Delta A$ as a function of the pre-existing crack length $A$, as predicted by the theory.}}\label{Fig17}
\end{figure}
%

We proceed with the $5^{\text{th}}$ circle, the single edge notch test depicted in Fig.~\ref{Fig16}(a). The specimens consist of thin strips with a length $L = 25\text{ mm}$, width $H = 5\text{ mm}$, and thickness $B = 0.25\text{ mm}$. Each specimen features a pre-existing, through-thickness edge crack of various lengths $A=0.025, 0.1, 0.5, 1,$ and $1.5\text{ mm}$. The specimens are rigidly clamped along their top and bottom boundaries and pulled apart by a prescribed global displacement $u$, generating a resultant reaction force $P$. We denote the corresponding nominal global stress by $S = P/(BH)$. As the applied displacement $u$ increases, the specimens respond elastically until reaching a critical threshold $u_{cr}$, at which point the pre-existing crack grows straight ahead. For silicate glass, the critical global stress at the onset of this fracture nucleation event, denoted by $S_{cr}$, is bounded from above as follows, according to \citep{KZDLP26},
\begin{equation}
S_{cr}\leq\min\left\{\sts,S_{G}\right\}\quad {\rm with}\quad S_{G}=\dfrac{\cos\left(\dfrac{\pi A}{2H}\right)\sqrt{\dfrac{G_c E}{\pi A}}}{\left(0.752+1.0431\dfrac{A}{H}+0.6076\left(1-\sin\left(\dfrac{\pi A}{2H}\right)\right)^3\right)\sqrt{\dfrac{2H}{\pi A}\tan\left(\dfrac{\pi A}{2H}\right)}}.\label{Scr-SENT-glass}
\end{equation}
In particular, the equality $S_{cr}=S_{G}$ is approached for sufficiently large cracks, while $S_{cr}=\sts$ is approached for sufficiently small cracks.

Once again, we incrementally increase the displacement $u$ to numerically identify the regions where the strength surface is violated, $\mathcal{F}(\mathbf{S}) \geq 0$. Such regions are invariably localized around the front of the pre-existing crack, as well as around the corners at the grips. At each increment, we determine the length $\Delta A$ of a straight through-thickness add-crack --- with add-crack surface area of $\mathcal{H}^2(\Gamma)=\Delta A B$ --- emanating from the pre-existing crack front that minimizes the energy  (\ref{Fracture-Energy-Min}). We consider only test add-cracks aligned with the pre-existing cracks and with straight fronts through the thickness, having previously checked that neither misaligned trajectories nor cracks nucleating from the corners yield energy minimizers, and that curved crack fronts have a minor effect on $\mathcal{E}$ due to the small specimen thickness ($B = 0.25\text{ mm}$).

Figures~\ref{Fig16}(b) and (c) present three representative results for the case of the smallest pre-existing crack, $A = 0.025$ mm. Specifically, Fig.~\ref{Fig16}(b) shows three magnified contour plots of the strength violation set at the applied displacements $u = 0.00675$ mm, $0.00690$ mm, and $0.00705$ mm --- which correspond to global stresses $S = 37.90$ MPa, $38.74$ MPa, and $39.59$ MPa --- depicting the add-cracks considered. Fig.~\ref{Fig16}(c) plots the corresponding energy $\mathcal{E}$ as a function of $\Delta A$. More specifically, for better visualization, the figure plots the difference $\Delta\mathcal{E} = \mathcal{E} - \mathcal{E}\vert{}_{\Delta A=0}$ between the energy with the add-crack and the corresponding energy without it ($\Delta A = 0$). The results show that the pre-existing crack starts to grow when $u = u_{cr}= 0.00705$ mm at a global stress of $S_{cr} = 39.59$ MPa, consistent with the result (\ref{Scr-SENT-glass}). The crack growth is of finite length $\Delta A = 0.16351$ mm.

Figure \ref{Fig17} compiles the theoretical predictions for the critical global stress $S_{cr}$ and the associated add-crack length $\Delta A$ for all five pre-existing crack lengths $A$. The main observation is that the theoretical predictions are once again in agreement with the reference result. It is also worth remarking that the length $\Delta A$ of the add-crack decreases monotonically with increasing $A$.

\subsection{Indentation test of a block with a flat-ended cylindrical punch}

Figure \ref{Fig18}(a) shows a schematic of the $6^{\text{th}}$ circle, a cylindrical block of height $L=25$ mm and radius $R_b=25$ mm, whose bottom boundary is firmly bonded to a rigid substrate. The block is frictionlessly indented along its axis of symmetry by a rigid, flat-ended cylindrical punch of radius $R_i=1$ mm subjected to a prescribed displacement $u$. For silicate glass, the block initially deforms elastically. Upon reaching a critical displacement $u_{cr}$, a shallow ring crack nucleates from the surface of the specimen just outside the perimeter of the indenter ($R > R_i$). Further loading causes the crack to propagate at a roughly constant angle with respect to the surface, evolving into a cone (frustum) crack.

%
\begin{figure}[b!]
\centering
\includegraphics[width=0.95\linewidth]{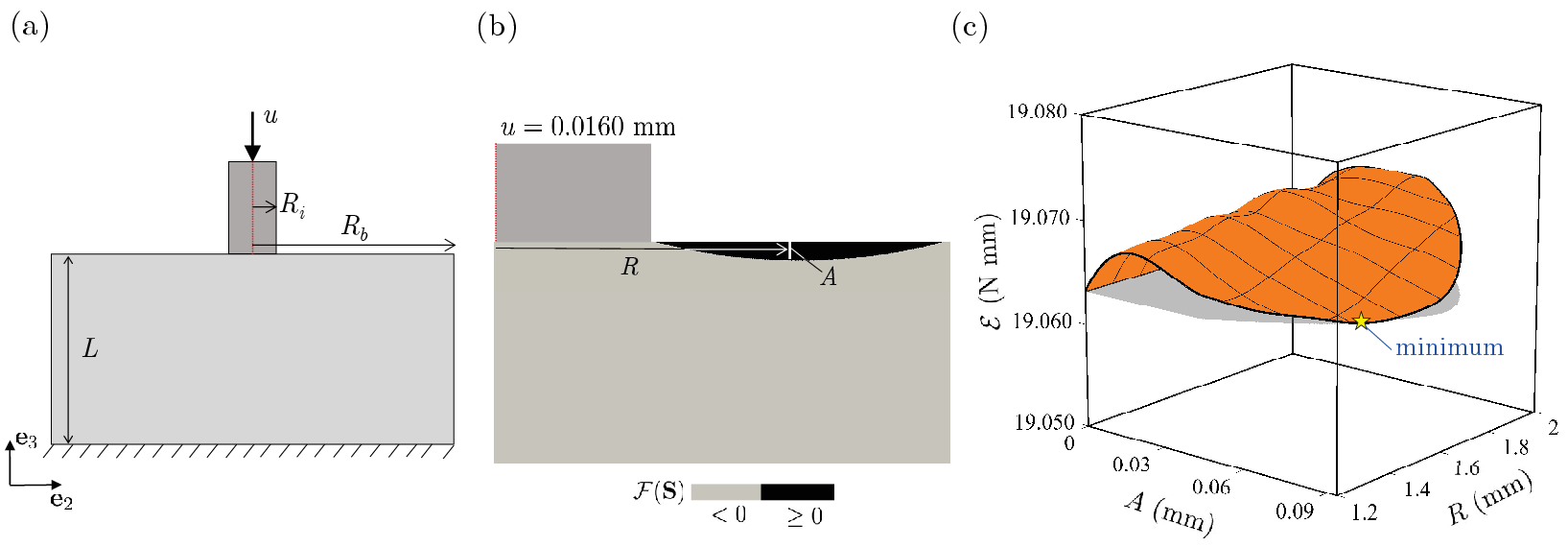}
\caption{{\small (a)  Schematic of the indentation test on silicate glass. (b) Contour plot of a magnified region of the specimen where the stress field exceeds the strength surface of the material ($\mathcal{F}(\bfS)\geq0$). The plot depicts the test cracks considered within this region and corresponds to $u=u_{cr}=0.0160$ mm, the critical applied displacement at which a cracked configuration first minimizes the energy $\mathcal{E}$. (c) The corresponding energy $\mathcal{E}$ of the specimen plotted as a function of the test crack distance $R$ from the symmetry axis and depth $A$.}}\label{Fig18}
\end{figure}
%
%
\begin{figure}[t!]
\centering
\includegraphics[width=0.65\linewidth]{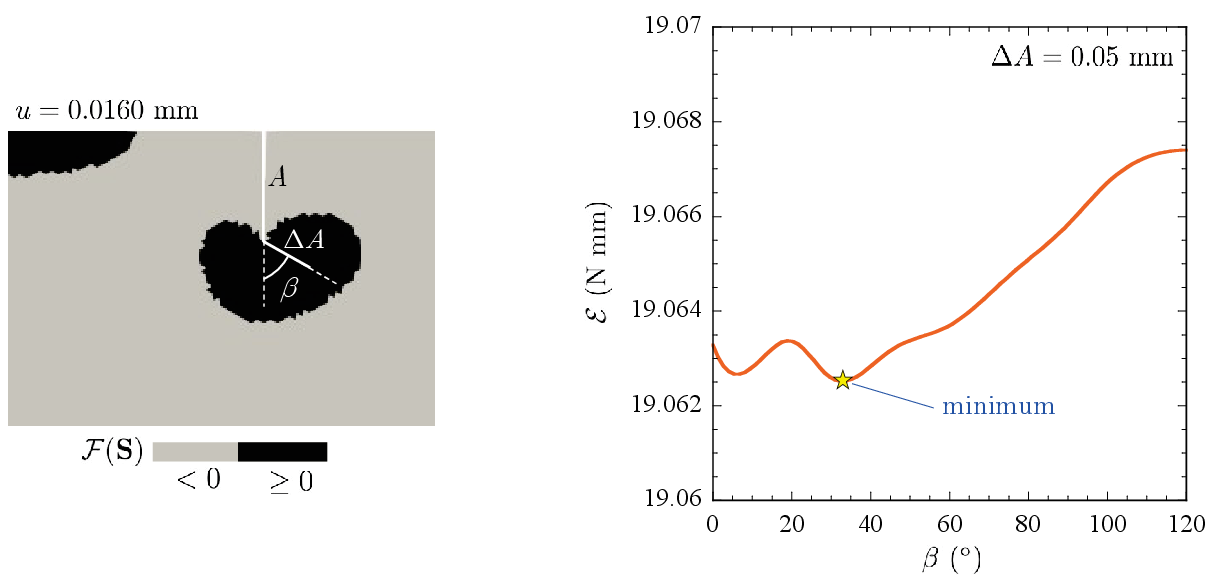}
\caption{{\small The specimen at the critical applied displacement $u=u_{cr}=0.0160$ mm containing the initial minimizing crack $(A=0.0810\,{\rm mm},R=1.4311\,{\rm mm})$, depicting a magnified region where the stress field exceeds ($\mathcal{F}(\bfS)\geq0$) the strength surface of the material along with the test add-cracks of length $\Delta A$ oriented at angles $\beta$ considered within. The line plot shows the variation of the energy $\mathcal{E}$ with respect to the add-crack angle $\beta$ for $\Delta A=0.05$ mm.}}\label{Fig19}
\end{figure}
%

Once again, we incrementally increase the displacement $u$ to numerically identify regions where the strength surface is violated, $\mathcal{F}(\mathbf{S}) \geq 0$. These regions emanate from the top surface of the specimen, adjacent to the indenter. Within these regions, for each displacement increment, we compute the energy (\ref{Fracture-Energy-Min}) for specimens containing axisymmetric ring test cracks. These cracks have a depth $A$ at a radial distance $R$ from the symmetry axis, yielding a crack surface area of $\mathcal{H}^2(\Gamma)=2\pi R A$; see Fig.~\ref{Fig18}(b). Figure \ref{Fig18}(c) plots the resulting energy $\mathcal{E}$ as a function of $A$ and $R$ at $u=u_{cr}=0.0160$ mm, the critical displacement at which a cracked configuration first minimizes (\ref{Fracture-Energy-Min}). The results show that a ring crack of length $A=0.0810$ mm at the distance $R=1.4311$ mm from the symmetry axis minimizes the energy $\mathcal{E}$ at $u=u_{cr}=0.0160$ mm. This is consistent with the reference result. 

We now examine the minimizing solution with crack length $A=0.0810$ mm at distance $R=1.4311$ mm from the symmetry axis. At the critical applied displacement $u=u_{cr}=0.0160$ mm, we numerically identify the region where the stress field violates the strength surface, $\mathcal{F}(\mathbf{S})\geq0$. Within this region, we introduce axisymmetric conical add-cracks of length $\Delta A$, oriented at an angle $\beta$ relative to the initial crack direction. This geometry yields a total crack surface area of $\mathcal{H}^{2}(\Gamma)=2\pi R A+\pi\Delta A(2R+\Delta A\sin\beta)$. Our computations reveal that the energy $\mathcal{E}$ is minimized at $\Delta A=0.05$ mm and $\beta=32^\circ$, indicating that the nucleated crack at $u=u_{cr}$ does not stay put but grows into a cone crack. Figure \ref{Fig19} illustrates the violation of the strength surface, depicts the add-cracks considered within, and plots the corresponding energy $\mathcal{E}$. Further computations show that subsequent crack propagation is stable, requiring an increase in the displacement $u$ for the crack to continue growing. All of these findings are in agreement with the reference result.

\subsection{Poker-chip test}

%
\begin{figure}[b!]
\centering
\includegraphics[width=0.925\linewidth]{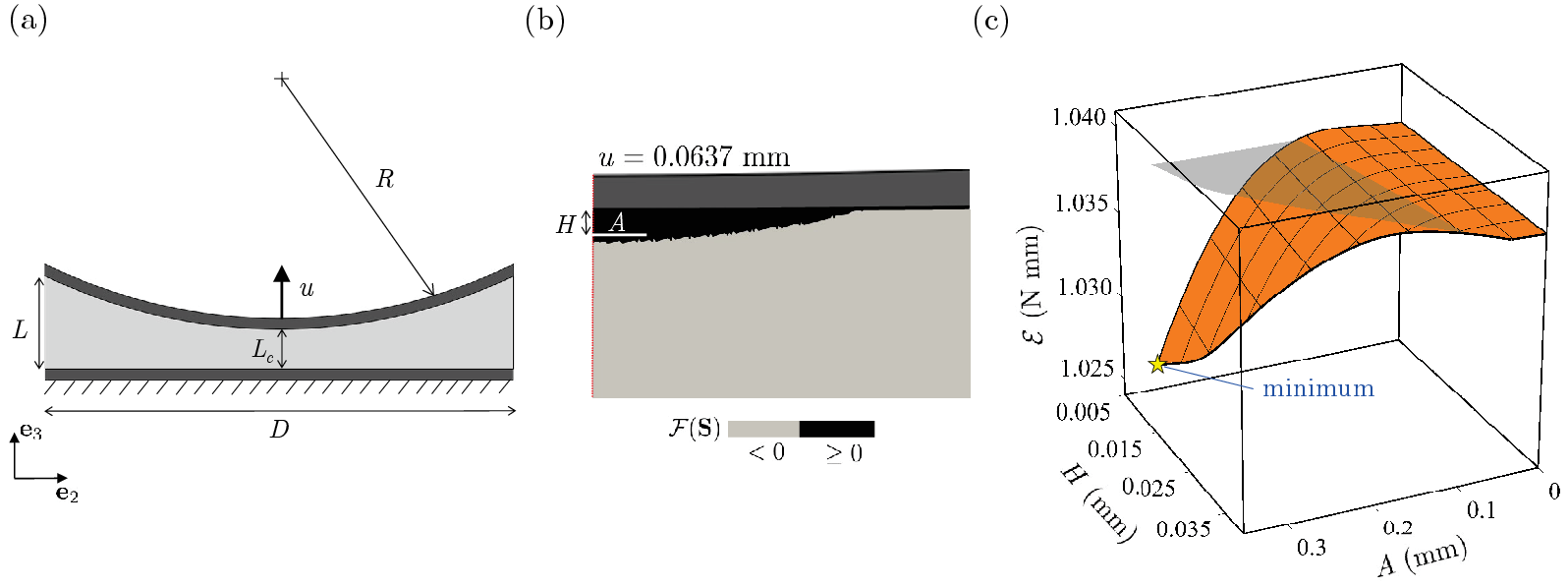}
\caption{{\small (a)  Schematic of the poker-chip test on synthetic rubber. (b) Contour plot of a magnified region of the specimen where the stress field exceeds the strength surface of the material ($\mathcal{F}(\bfS)\geq0$). The plot depicts the test cracks considered within this region and corresponds to $u=u_{cr}=0.0637$ mm, the critical applied displacement at which a cracked configuration first minimizes the energy $\mathcal{E}$. (c) The corresponding energy $\mathcal{E}$ of the specimen as a function of the test crack radius $A$ and distance $H$ from the top boundary.}}\label{Fig20}
\end{figure}
%

Next, we turn to the $7^{\text{th}}$ circle, the poker-chip test schematically illustrated in Fig.~\ref{Fig20}(a). The specimen consists of a circular disk of diameter $D = 10$ mm, whose bottom boundary is fixed to a flat substrate and top boundary is firmly bonded to a spherical fixture of radius $R = 18.2$ mm. The thickness of the disk is therefore non-uniform, increasing radially from a centerline thickness of $L_c = 1$ mm to $L = 1.7$ mm at the outer edge. The spherical fixture is subjected to a prescribed vertical displacement $u$. For a nearly incompressible material --- such as the synthetic rubber considered in this work --- the applied displacement produces a non-homogeneous stress field within the disk characterized by a large hydrostatic component around its centerline. Under this stress state, a penny-shaped crack perpendicular to the loading direction nucleates near the centerline at a critical displacement $u_{cr}$.

Again, we begin by incrementally increasing the applied displacement $u$ to numerically locate regions where the strength surface is violated, $\mathcal{F}(\mathbf{S}) \geq 0$. Such regions develop from the top boundary around the centerline of the specimen. For each displacement increment within these violation regions, the energy (\ref{Fracture-Energy-Min}) is evaluated for specimens containing axisymmetric penny-shaped test cracks of radius $A$ and at a distance $H$ from the top boundary, corresponding to a crack surface area of $\mathcal{H}^2(\Gamma)=\pi A^2$; see Fig.~\ref{Fig20}(b). Figure \ref{Fig20}(c) presents the resulting energy $\mathcal{E}$ as a function of $A$ and $H$ evaluated at the critical displacement $u=u_{cr}=0.0637$ mm, the point at which a cracked configuration first minimizes the energy (\ref{Fracture-Energy-Min}). Specifically, the energy at $u=u_{cr}=0.0637$ mm is minimized by a crack of radius $A=0.31$ mm located at a distance $H=0.005$ mm from the top boundary, a finding that is in agreement with the reference result.

\subsection{The double cantilever beam test}

Having addressed all the circles in \citep{KZDLP26} pertaining to fracture nucleation, we now turn to the $8^{\text{th}}$ and $9^{\text{th}}$ circles, which focus on fracture propagation under opening and tearing modes.

%
\begin{figure}[b!]
\centering
\includegraphics[width=0.99\linewidth]{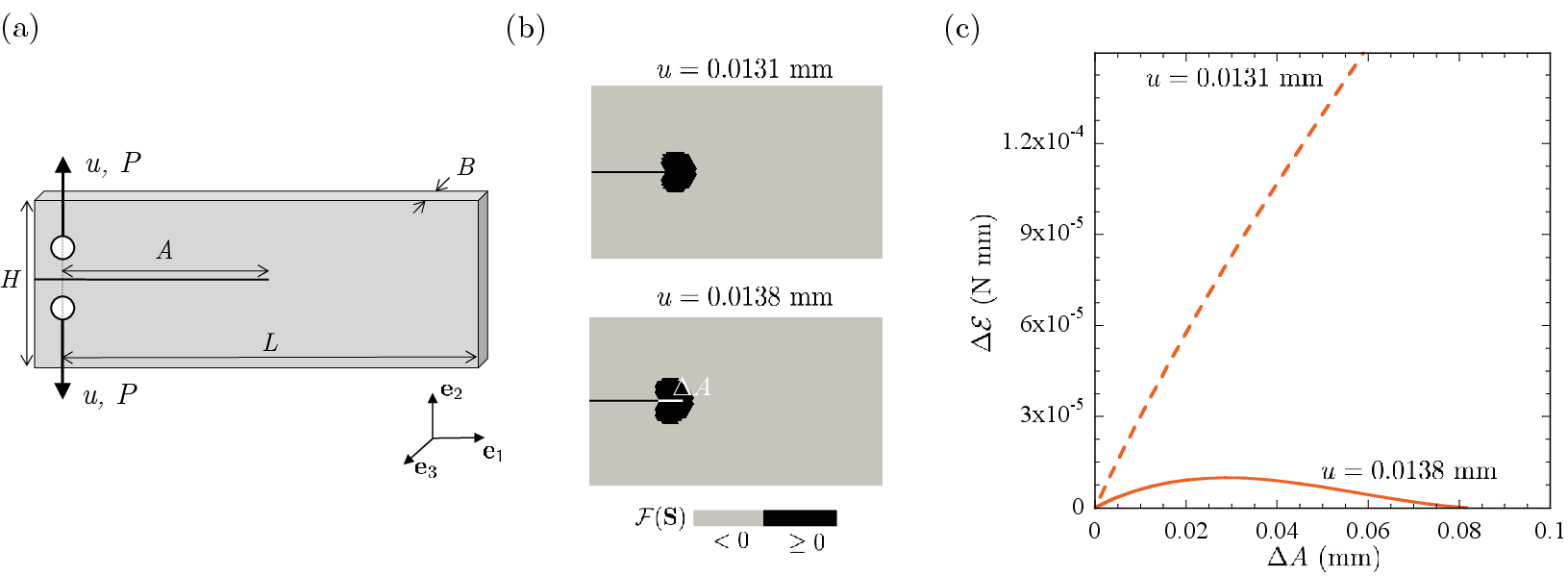}
\caption{{\small (a) Schematic of the double cantilever beam test on silicate glass. (b) Contour plots of magnified regions where the strength surface is violated ($\mathcal{F}(\mathbf{S})\geq0$), illustrating the considered test add-cracks, at two applied displacements  $u$. (c) The corresponding energy difference $\Delta\mathcal{E}=\mathcal{E}-\mathcal{E}\vert{}_{\Delta A=0}$ plotted as a function of the add-crack length $\Delta A$.}}\label{Fig21}
\end{figure}
%

%
\begin{figure}[t!]
\centering
\includegraphics[width=0.9\linewidth]{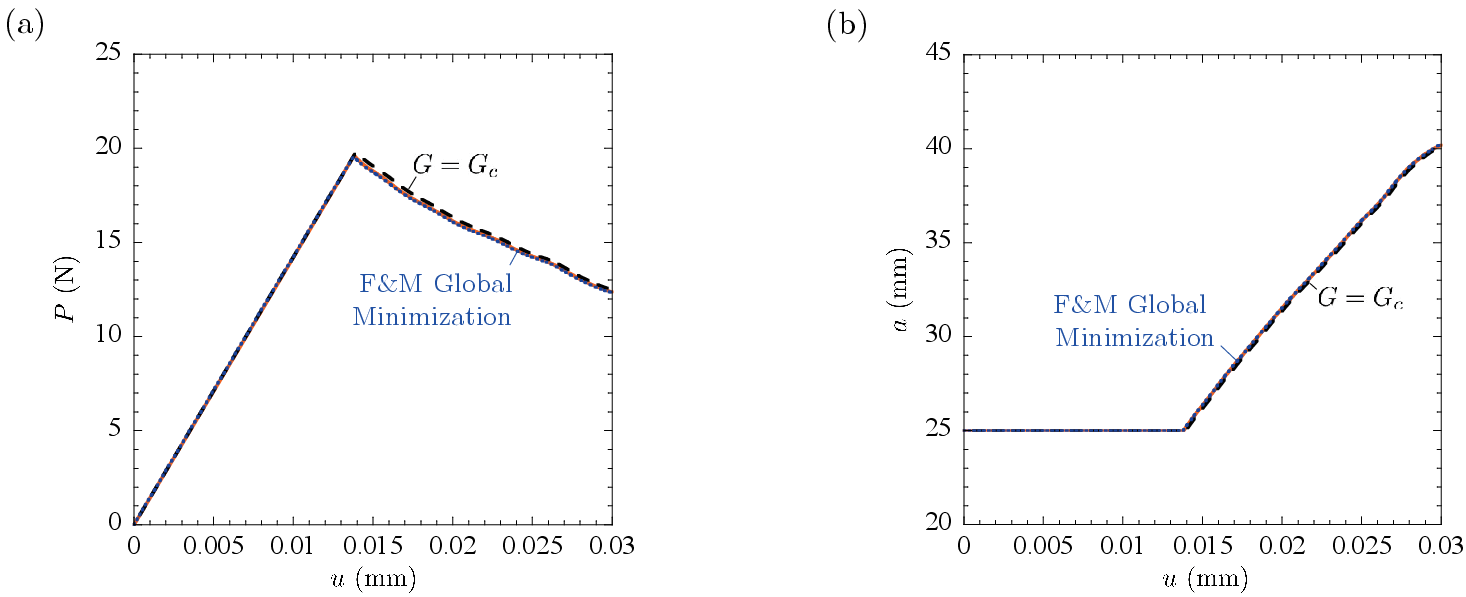}
\caption{{\small Theoretical predictions (solid lines) for the double cantilever beam test. (a) The force-displacement response. (b) The associated evolution of crack length $a$. For direct comparison, both plots include the predictions (dashed and dotted lines) associated with the classical Griffith criterion, $G=G_c$, and the Francfort-Marigo global energy minimization.}}\label{Fig22}
\end{figure}
%

Figure \ref{Fig21}(a) schematically illustrates the $8^{\text{th}}$ circle, a double cantilever beam test. The specimen is a prismatic bar with dimensions $L=55$ mm, $H=20$ mm, and $B=2.5$ mm, featuring a pre-existing edge crack of length $A=25$ mm. To drive the growth of the crack, a displacement $u$ is applied in the $\bfe_2$ direction through two pinholes. The corresponding resultant force is denoted by $P$. The pinholes have a radius of $B/8=0.3125$ mm, are positioned $1.5$ mm inward from the left boundary, and are set at a vertical spacing of $H/3=6.6667$ mm between their centers. Increasing the applied displacement $u$ causes the specimen arms to deform like cantilever beams. Once $u$ reaches a critical value $u_{cr}$, self-similar crack growth ensues in the $\bfe_1$ direction. Further increase in the applied displacement $u$ results in the stable propagation of the crack along the $\bfe_1$ direction. 

%
%
%
%
%
%

To determine whether the proposed theory (\ref{Potential-Energy})-(\ref{Gamma-Admi}) yields predictions in agreement with the reference results, we begin by incrementally applying the displacement $u$ and numerically identifying the regions where the strength surface is violated, $\mathcal{F}(\mathbf{S}) \geq 0$. As expected, these regions consistently localize around the crack front. At each displacement step, we determine the length $\Delta A$ of a straight, through-thickness add-crack (with surface area $\mathcal{H}^2(\Gamma)=\Delta A B$) that minimizes the energy (\ref{Fracture-Energy-Min}). The search is restricted to self-similar test add-cracks with straight fronts. We checked that kinked trajectories do not yield energy minimizers and that, due to the small specimen thickness ($B=2.5$ mm), curved crack fronts through the thickness have a minor effect on $\mathcal{E}$. Figure \ref{Fig21} shows contour plots of magnified strength-violation regions for two displacements, $u=0.0131$ mm and $0.0138$ mm, and depicts the test add-cracks considered within these regions. As shown in Fig.~\ref{Fig21}(c), the value $u=0.0138$ mm corresponds to the critical displacement $u_{cr}$ at which an add-crack first minimizes (\ref{Fracture-Energy-Min}) --- that is, the displacement at which the pre-existing crack begins to grow. Importantly, the minimizing crack is small but of a finite length $\Delta A=0.08$ mm.

Repeating the computation on the updated geometry with a crack length of $a=A+\Delta A=25.08$ mm reveals that the crack propagates by another jump of $\Delta A=0.08$ mm as the displacement increases to $u=0.0139$ mm. This stable propagation via small jumps --- and \emph{not}  via a time-continuous smooth growth as commonly assumed --- continues as $u$ increases further. The results, plotted in Fig.~\ref{Fig22}, confirm that these predictions are in agreement with the reference results. 

For direct comparison, Fig.~\ref{Fig22} includes the predictions of the classical Griffith criterion --- which dictates that crack propagation occurs when the energy release rate reaches the fracture toughness, $G=G_c$ --- alongside those generated by the global energy minimization approach of \cite{Francfort98}. We emphasize that the classical Griffith criterion restricts add-cracks to be infinitesimally small, assuming that crack growth is a time-continuous smooth process. By contrast, global energy minimization places no restrictions on the size of add-cracks. Accordingly, it allows for growth by jumps --- much like the theory proposed herein --- which is precisely what it predicts for this test (specifically, jumps of $0.10\text{ mm}$). 

A key observation from Fig.~\ref{Fig22} is that all three sets of predictions are qualitatively and quantitatively similar for this specific problem, as the jumps predicted by both the proposed theory and global energy minimization are small. Consequently, isolated use of this type of double cantilever beam test cannot provide meaningful experimental validation.

\subsection{The trousers test}

Finally, Fig.~\ref{Fig23}(a) shows a schematic of the $9^{\text{th}}$ circle, a trousers test. The specimen, made of the synthetic rubber investigated in this work, consists of a rectangular sheet of length $L=100$ mm, width $H=40$ mm, and thickness $B=1$ mm containing a pre-existing edge crack of initial length $A=50$ mm. Loading is initiated by bending the two legs in opposite directions to align them in a coplanar configuration. The lower ends of the legs are firmly gripped and subjected to an imposed separation $l$ along the $\bfe_2$ direction, generating a resultant reaction force $P$. As the separation $l$ increases, the legs of the trousers undergo elastic stretching, bending, and twisting. Once a critical separation between the grips, $l_{cr}$, is reached, the crack begins to propagate in a self-similar manner along the $\bfe_3$ direction. Further increases in $l$ result in stable crack growth along $\bfe_3$.
%
\begin{figure}[t!]
\centering
\includegraphics[width=0.99\linewidth]{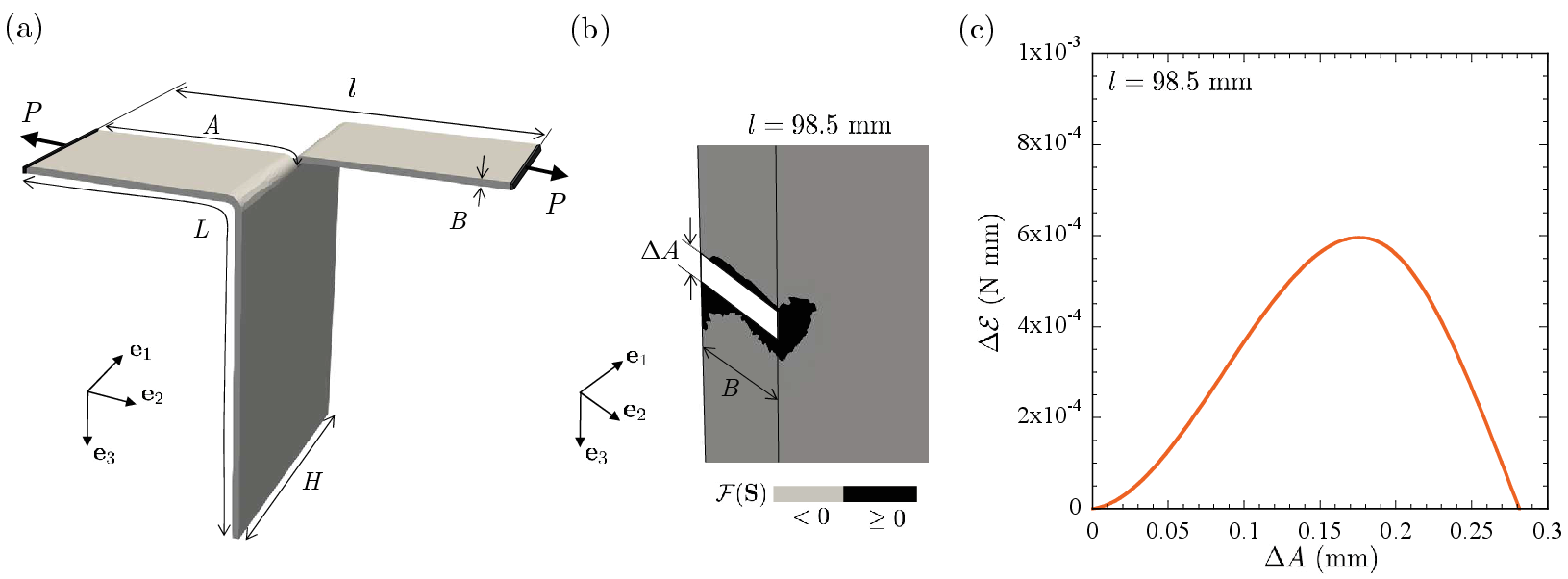}
\caption{{\small (a) Schematic of the trousers test on synthetic rubber. (b) Contour plot of a magnified region where the strength surface is violated ($\mathcal{F}(\mathbf{S})\geq0$), illustrating the considered test add-cracks, at the separation between the grips  $l=l_{cr}=98.5$ mm. (c) The corresponding energy difference $\Delta\mathcal{E}=\mathcal{E}-\mathcal{E}\vert{}_{\Delta A=0}$ plotted as a function of the add-crack length $\Delta A$.}}\label{Fig23}
\end{figure}
%
%
\begin{figure}[t!]
\centering
\includegraphics[width=0.9\linewidth]{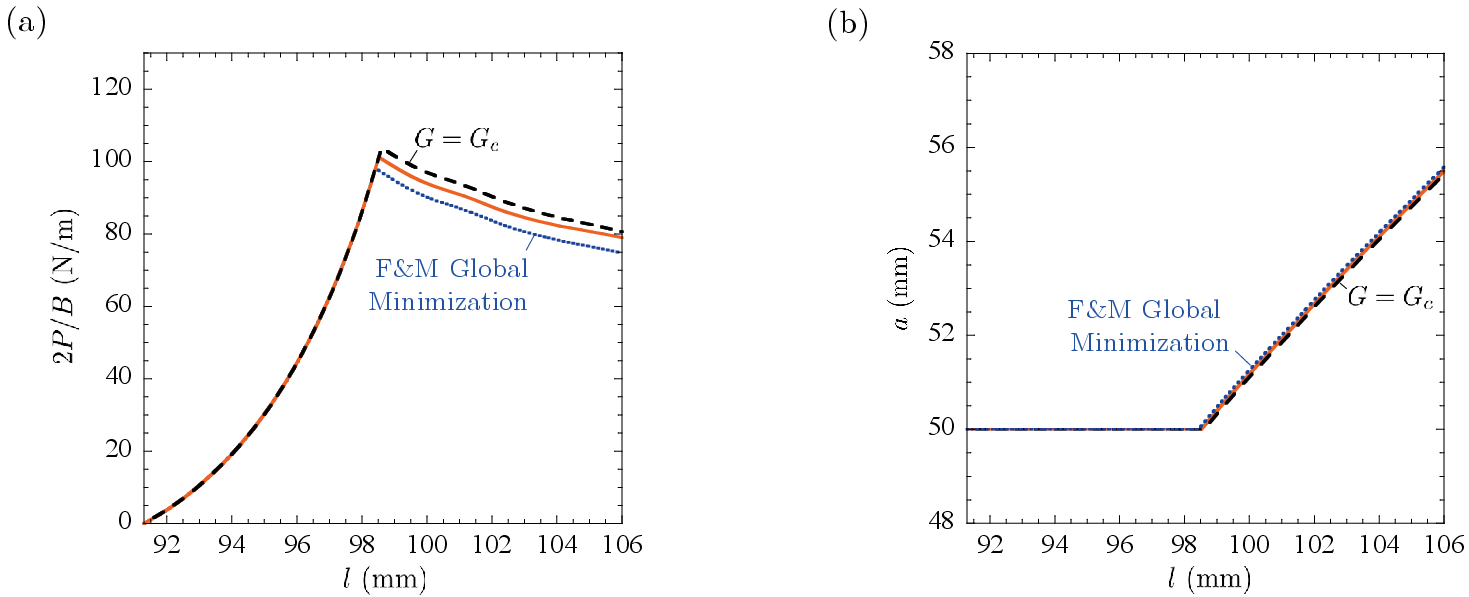}
\caption{{\small Theoretical predictions (solid lines) for the trousers test. (a) The normalized force $2P/B$ as a function of the applied separation $l$ between the grips. (b) The associated evolution of crack length $a$. For direct comparison, both plots include the predictions (dashed and dotted lines) associated with the classical Griffith criterion, $G=G_c$, and the Francfort-Marigo global energy minimization.}}\label{Fig24}
\end{figure}
%

The application of the proposed theory (\ref{Potential-Energy})-(\ref{Gamma-Admi}) follows the procedure used for the preceding circle. We incrementally apply the grip-to-grip separation $l$ and computationally identify the regions of strength violation, $\mathcal{F}(\mathbf{S}) \geq 0$. These regions localize tightly near the crack front. For each separation increment, we calculate the length $\Delta A$ of a straight through-thickness add-crack (with surface area $\mathcal{H}^2(\Gamma)=\Delta A B$) that minimizes the energy (\ref{Fracture-Energy-Min}). Again, restricting our scope to these specific topologies is justified since kinked paths are not energy minimizers and, owing to the thinness of the specimen ($B=1$ mm), through-thickness curved fronts barely alter $\mathcal{E}$. Figure \ref{Fig23} presents a magnified contour plot of the strength-violation region at $l=98.5$ mm, overlaying the test add-cracks evaluated within. As shown in Fig.~\ref{Fig23}(c), this specific separation corresponds to the critical value $l_{cr}$ at which an add-crack first minimizes (\ref{Fracture-Energy-Min}). Importantly, the minimizing crack is small but distinctly finite, measuring $\Delta A=0.28$ mm.

Subsequent calculations on the updated configuration --- now with a crack length of $a=A+\Delta A=50.28$ mm --- show a second discrete advancement of $\Delta A=0.28$ mm once the grip-to-grip separation reaches $l=98.86$ mm. This stable step-wise crack propagation  --- and, once again, \emph{not}  via a time-continuous smooth growth as commonly assumed ---  persists under continued loading. These findings, plotted in Fig.~\ref{Fig24}, are in agreement with the reference results.

As in the previous case, we have included in Fig.~\ref{Fig24} the theoretical predictions from both the classical Griffith criterion and the global energy minimization of \cite{Francfort98} for direct comparison. We reiterate the fundamental distinction here: the classical Griffith criterion restricts add-cracks to be infinitesimally small, whereas the global energy minimization approach of \cite{Francfort98} imposes no add-crack size constraints. The conclusions and implications mirror those noted for the double cantilever beam test.

\section{Final comments}\label{Sec: Final Comments}

The ability to predict when and where fracture nucleates and propagates under arbitrary loading conditions has long stood as a major hurdle in solid mechanics. The broad range of results presented in the preceding section suggests that the proposed theory provides a satisfactory answer. Further validation should confirm the generality of the theory. Specifically, validation for problems where fracture is driven by body forces --- such as gravity --- should prove valuable, as very few benchmark studies of this type currently exist.

Of the three intrinsic material properties entering the theory --- the elastic energy density $W(\mathbf{F})$, the strength surface $\mathcal{F}(\mathbf{S})=0$, and the fracture toughness $G_c$ --- the strength surface is arguably the most challenging to measure in its entirety through direct experiments, although, in all fairness, characterizing the complete $W(\mathbf{F})$ for a nonlinear elastic material presents a comparable challenge. This stems from the practical difficulty of subjecting specimens to spatially uniform stress states that span the entire stress space. Nevertheless, because the strength surface strictly constrains where fracture can occur, experiments featuring non-uniform stress fields can now be leveraged. Specifically, the entire strength surface $\mathcal{F}(\mathbf{S})=0$ can be identified via inverse calibration, tuning its parameters so that theoretical predictions match macroscopic experimental observations. 

Deriving the time-continuous limit of the proposed sharp formulation is key. So is the development of phase-field and other diffuse regularizations. In particular, it would be highly valuable to establish whether the phase-field theory introduced in \citep*{KFLP18} --- which directly motivated the theory proposed here --- provides a mathematically rigorous regularization of the sharp formulation, one where the driving force $c_{\texttt{e}}$ introduced in that formulation plays the role of a penalty term that enforces the strength surface constraint.

Developing an efficient and robust numerical implementation of the proposed theory represents another crucial priority. The physical constraint imposed by the strength surface should render the computations highly tractable, even when explicitly tracking sharp cracks. This holds particularly true for initial-boundary-value problems that can be treated in 2D.


\section*{Acknowledgements}

The work of the first two authors was supported by the National Science Foundation through the Grants DMS--2308169 and CMMI--2132528. This support is gratefully acknowledged. This work used Expanse  at SDSC through allocation MCH260079 from the Advanced Cyberinfrastructure Coordination Ecosystem: Services and Support (ACCESS) program, which is supported by U.S. National Science Foundation grants 2138259, 2138286, 2138307, 2137603, and 2138296.


\begin{thebibliography}{82}
\expandafter\ifx\csname natexlab\endcsname\relax\def\natexlab#1{#1}\fi
\providecommand{\url}[1]{\texttt{#1}}
\providecommand{\href}[2]{#2}
\providecommand{\path}[1]{#1}
\providecommand{\DOIprefix}{doi:}
\providecommand{\ArXivprefix}{arXiv:}
\providecommand{\URLprefix}{URL: }
\providecommand{\Pubmedprefix}{pmid:}
\providecommand{\doi}[1]{\href{http://dx.doi.org/#1}{\path{#1}}}
\providecommand{\Pubmed}[1]{\href{pmid:#1}{\path{#1}}}
\providecommand{\bibinfo}[2]{#2}
\ifx\xfnm\relax \def\xfnm[#1]{\unskip,\space#1}\fi
\bibitem[{Ambrosio et~al.(2000)Ambrosio, Fusco and Pallara}]{AFP}
\bibinfo{author}{Ambrosio, L.}, \bibinfo{author}{Fusco, N.},
  \bibinfo{author}{Pallara, D.}, \bibinfo{year}{2000}.
\newblock \bibinfo{title}{Functions of Bounded Variation and Free Discontinuity
  Problems}.
\newblock \bibinfo{publisher}{Oxford University Press},
  \bibinfo{address}{Oxford}.
\bibitem[{Ambrosio and Tilli(2004)}]{ambrosio.tilli}
\bibinfo{author}{Ambrosio, L.}, \bibinfo{author}{Tilli, P.},
  \bibinfo{year}{2004}.
\newblock \bibinfo{title}{Topics on Analysis in Metric Spaces}.
\newblock \bibinfo{publisher}{Oxford University Press},
  \bibinfo{address}{Oxford}.
\bibitem[{Beer et~al.(2019)Beer, Johnston, Dewolf and
  Mazurek}]{BeerJohnston2019}
\bibinfo{author}{Beer, F.}, \bibinfo{author}{Johnston, E.R.},
  \bibinfo{author}{Dewolf, J.}, \bibinfo{author}{Mazurek, D.},
  \bibinfo{year}{2019}.
\newblock \bibinfo{title}{Mechanics of Materials}.
\newblock \bibinfo{publisher}{McGraw-Hill Education}.
\bibitem[{Bourdin et~al.(2008)Bourdin, Francfort and Marigo}]{Bourdin08}
\bibinfo{author}{Bourdin, B.}, \bibinfo{author}{Francfort, G.A.},
  \bibinfo{author}{Marigo, J.J.}, \bibinfo{year}{2008}.
\newblock \bibinfo{title}{The variational approach to fracture}.
\newblock \bibinfo{journal}{Journal of Elasticity} \bibinfo{volume}{91},
  \bibinfo{pages}{5--148}.
\bibitem[{Breedlove et~al.(2024)Breedlove, Chen, Lindeman and
  Lopez-Pamies}]{BCLLP24}
\bibinfo{author}{Breedlove, E.}, \bibinfo{author}{Chen, C.},
  \bibinfo{author}{Lindeman, D.}, \bibinfo{author}{Lopez-Pamies, O.},
  \bibinfo{year}{2024}.
\newblock \bibinfo{title}{Cavitation in elastomers: {A} review of the evidence
  against elasticity}.
\newblock \bibinfo{journal}{Journal of the Mechanics and Physics of Solids}
  \bibinfo{volume}{188}, \bibinfo{pages}{105678}.
\bibitem[{Budiansky and Rice(1973)}]{BudyRice1973}
\bibinfo{author}{Budiansky, B.}, \bibinfo{author}{Rice, J.R.},
  \bibinfo{year}{1973}.
\newblock \bibinfo{title}{Conservation laws and energy-release rates}.
\newblock \bibinfo{journal}{Journal of Applied Mechanics} \bibinfo{volume}{40},
  \bibinfo{pages}{201--203}.
\bibitem[{Cauchy(1823)}]{Cauchy1823}
\bibinfo{author}{Cauchy, A.L.}, \bibinfo{year}{1823}.
\newblock \bibinfo{title}{Recherches sur l’\'equilibre et le mouvement
  int\'erieur des corps solides ou fluides, \'elastiques ou non \'elastiques}.
\newblock \bibinfo{journal}{Bulletin ile la Soci\'et\'e Philomatique} ,
  \bibinfo{pages}{9--13}.
\bibitem[{Cauchy(1828)}]{Cauchy1828}
\bibinfo{author}{Cauchy, A.L.}, \bibinfo{year}{1828}.
\newblock \bibinfo{title}{Sur les \'equations qui expriment l'\'equilibre ou
  les lois du mouvement int\'erieur d'un corps solide \'elastique ou non
  \'elastique}.
\newblock \bibinfo{journal}{Exercices de Math\'ematiques} ,
  \bibinfo{pages}{195--226}.
\bibitem[{Chambolle et~al.(2009)Chambolle, Francfort and
  Marigo}]{Chambolle2009}
\bibinfo{author}{Chambolle, A.}, \bibinfo{author}{Francfort, G.A.},
  \bibinfo{author}{Marigo, J.J.}, \bibinfo{year}{2009}.
\newblock \bibinfo{title}{When and how do cracks propagate?}
\newblock \bibinfo{journal}{Journal of the Mechanics and Physics of Solids}
  \bibinfo{volume}{57}, \bibinfo{pages}{1614--1622}.
\bibitem[{Chambolle et~al.(2010)Chambolle, Francfort and
  Marigo}]{Chambolle2010}
\bibinfo{author}{Chambolle, A.}, \bibinfo{author}{Francfort, G.A.},
  \bibinfo{author}{Marigo, J.J.}, \bibinfo{year}{2010}.
\newblock \bibinfo{title}{Revisiting energy release rates in brittle fracture}.
\newblock \bibinfo{journal}{Journal of Nonlinear Science} \bibinfo{volume}{20},
  \bibinfo{pages}{395--424}.
\bibitem[{Chambolle et~al.(2008)Chambolle, Giacomini and
  Ponsiglione}]{Chambolle2008}
\bibinfo{author}{Chambolle, A.}, \bibinfo{author}{Giacomini, A.},
  \bibinfo{author}{Ponsiglione, M.}, \bibinfo{year}{2008}.
\newblock \bibinfo{title}{Crack initiation in brittle materials}.
\newblock \bibinfo{journal}{Archive for Rational Mechanics and Analysis}
  \bibinfo{volume}{188}, \bibinfo{pages}{309--349}.
\bibitem[{Cornetti et~al.(2006)Cornetti, Pugno, Carpinteri and
  Taylor}]{Cornetti2006}
\bibinfo{author}{Cornetti, P.}, \bibinfo{author}{Pugno, N.},
  \bibinfo{author}{Carpinteri, A.}, \bibinfo{author}{Taylor, D.},
  \bibinfo{year}{2006}.
\newblock \bibinfo{title}{Finite fracture mechanics: {A} coupled stress and
  energy failure criterion}.
\newblock \bibinfo{journal}{Engineering Fracture Mechanics}
  \bibinfo{volume}{73}, \bibinfo{pages}{2021--2033}.
\bibitem[{Crouzeix and Raviart(1973)}]{CR1973}
\bibinfo{author}{Crouzeix, M.}, \bibinfo{author}{Raviart, P.A.},
  \bibinfo{year}{1973}.
\newblock \bibinfo{title}{Conforming and nonconforming finite element methods
  for solving the stationary {S}tokes equations}.
\newblock \bibinfo{journal}{R.A.I.R.O.} \bibinfo{volume}{R3},
  \bibinfo{pages}{33--76}.
\bibitem[{{Dal Maso} et~al.(2005){Dal Maso}, Francfort and Toader}]{DMFT}
\bibinfo{author}{{Dal Maso}, G.}, \bibinfo{author}{Francfort, G.A.},
  \bibinfo{author}{Toader, R.}, \bibinfo{year}{2005}.
\newblock \bibinfo{title}{Quasistatic crack growth in nonlinear elasticity}.
\newblock \bibinfo{journal}{Archive for Rational Mechanics and Analysis}
  \bibinfo{volume}{176}, \bibinfo{pages}{165--225}.
\bibitem[{Davenport and Smith(1993)}]{Davenport93}
\bibinfo{author}{Davenport, J.C.W.}, \bibinfo{author}{Smith, D.J.},
  \bibinfo{year}{1993}.
\newblock \bibinfo{title}{A study of superimposed fracture modes {I, II, and
  III} on {PMMA}}.
\newblock \bibinfo{journal}{Fatigue and Fracture of Engineering Materials and
  Structures} \bibinfo{volume}{16}, \bibinfo{pages}{1125--1133}.
\bibitem[{Doitrand et~al.(2024)Doitrand, Duminy, Girard and
  Chen}]{Doitrand2024}
\bibinfo{author}{Doitrand, A.}, \bibinfo{author}{Duminy, T.},
  \bibinfo{author}{Girard, H.}, \bibinfo{author}{Chen, X.},
  \bibinfo{year}{2024}.
\newblock \bibinfo{title}{A review of the coupled criterion}.
\newblock \bibinfo{journal}{Journal of Theoretical, Computational and Applied
  Mechanics} , \bibinfo{pages}{11072}.
\bibitem[{Erdogan and Sih(1963)}]{Erdogan63}
\bibinfo{author}{Erdogan, G.}, \bibinfo{author}{Sih, G.C.},
  \bibinfo{year}{1963}.
\newblock \bibinfo{title}{On the crack extension in plates under plane loading
  and transverse shear}.
\newblock \bibinfo{journal}{Journal of Basic Engineering} \bibinfo{volume}{85},
  \bibinfo{pages}{519--527}.
\bibitem[{Evans and Gariepy(1992)}]{evans.gariepy}
\bibinfo{author}{Evans, L.C.}, \bibinfo{author}{Gariepy, R.F.},
  \bibinfo{year}{1992}.
\newblock \bibinfo{title}{Measure Theory and Fine Properties of Functions}.
\newblock \bibinfo{publisher}{CRC Press}, \bibinfo{address}{Boca Raton}.
\bibitem[{Francfort et~al.(2019)Francfort, Giacomini and Lopez-Pamies}]{FGLP19}
\bibinfo{author}{Francfort, G.A.}, \bibinfo{author}{Giacomini, A.},
  \bibinfo{author}{Lopez-Pamies, O.}, \bibinfo{year}{2019}.
\newblock \bibinfo{title}{Fracture with healing: {A} first step towards a new
  view of cavitation}.
\newblock \bibinfo{journal}{Analysis and PDE} \bibinfo{volume}{12},
  \bibinfo{pages}{417--447}.
\bibitem[{Francfort and Larsen(2003)}]{francfort.larsen}
\bibinfo{author}{Francfort, G.A.}, \bibinfo{author}{Larsen, C.J.},
  \bibinfo{year}{2003}.
\newblock \bibinfo{title}{Existence and convergence for quasi-static evolution
  in brittle fracture}.
\newblock \bibinfo{journal}{Communications on Pure and Applied Mathematics}
  \bibinfo{volume}{56}, \bibinfo{pages}{1465--1500}.
\bibitem[{Francfort and Marigo(1998)}]{Francfort98}
\bibinfo{author}{Francfort, G.A.}, \bibinfo{author}{Marigo, J.J.},
  \bibinfo{year}{1998}.
\newblock \bibinfo{title}{Revisiting brittle fracture as an energy minimization
  problem}.
\newblock \bibinfo{journal}{Journal of the Mechanics and Physics of Solids}
  \bibinfo{volume}{46}, \bibinfo{pages}{1319--1342}.
\bibitem[{Goldstein and Salganik(1974)}]{Goldstein74}
\bibinfo{author}{Goldstein, R.V.}, \bibinfo{author}{Salganik, R.L.},
  \bibinfo{year}{1974}.
\newblock \bibinfo{title}{Brittle fracture of solids with arbitrary cracks}.
\newblock \bibinfo{journal}{International Journal of Fracture}
  \bibinfo{volume}{10}, \bibinfo{pages}{507--523}.
\bibitem[{Green(1838)}]{Green1838}
\bibinfo{author}{Green, G.}, \bibinfo{year}{1838}.
\newblock \bibinfo{title}{On the laws of reflection and refraction of light at
  the common surface of two non-crystallized media}.
\newblock \bibinfo{journal}{Transactions of the Cambridge Philosophical
  Society} \bibinfo{volume}{7}, \bibinfo{pages}{245--269}.
\bibitem[{Griffith(1921)}]{Griffith21}
\bibinfo{author}{Griffith, A.A.}, \bibinfo{year}{1921}.
\newblock \bibinfo{title}{The phenomena of rupture and flow in solids}.
\newblock \bibinfo{journal}{Philos. Trans. R. Soc. Lond. Ser. A}
  \bibinfo{volume}{221}, \bibinfo{pages}{163--198}.
\bibitem[{Griffith(1924)}]{Griffith1924}
\bibinfo{author}{Griffith, A.A.}, \bibinfo{year}{1924}.
\newblock \bibinfo{title}{The theory of rupture}, in:
  \bibinfo{booktitle}{Proceedings of the First International Congress for
  Applied Mechanics}, \bibinfo{address}{Delft}.
\bibitem[{Hua et~al.(2023)Hua, Li, Zhu, A.Li, Huang, Gan and Dong}]{Hua23}
\bibinfo{author}{Hua, W.}, \bibinfo{author}{Li, J.}, \bibinfo{author}{Zhu, Z.},
  \bibinfo{author}{A.Li}, \bibinfo{author}{Huang, J.}, \bibinfo{author}{Gan,
  Z.}, \bibinfo{author}{Dong, S.}, \bibinfo{year}{2023}.
\newblock \bibinfo{title}{A review of mixed mode {I-II} fracture criteria and
  their applications in brittle or quasi-brittle fracture analysis}.
\newblock \bibinfo{journal}{Theoretical and Applied Fracture Mechanics}
  \bibinfo{volume}{124}, \bibinfo{pages}{103741}.
\bibitem[{Irwin(1948)}]{Irwin48}
\bibinfo{author}{Irwin, G.R.}, \bibinfo{year}{1948}.
\newblock \bibinfo{title}{Fracture dynamics}, in:
  \bibinfo{booktitle}{Fracturing of Metals. American Society For Metals},
  \bibinfo{address}{Cleveland}. pp. \bibinfo{pages}{147--166}.
\bibitem[{Irwin(1957)}]{Irwin57}
\bibinfo{author}{Irwin, G.R.}, \bibinfo{year}{1957}.
\newblock \bibinfo{title}{Analysis of stresses and strains near the end of a
  crack traversing a plate}.
\newblock \bibinfo{journal}{Journal of Applied Mechanics} \bibinfo{volume}{24},
  \bibinfo{pages}{361--364}.
\bibitem[{Kamarei et~al.(2025)Kamarei, Breedlove and Lopez-Pamies}]{KLP25}
\bibinfo{author}{Kamarei, F.}, \bibinfo{author}{Breedlove, E.},
  \bibinfo{author}{Lopez-Pamies, O.}, \bibinfo{year}{2025}.
\newblock \bibinfo{title}{Validating {G}riffith fracture propagation in the
  phase-field approach to fracture: {T}he case of {M}ode {III} by means of the
  trousers test}.
\newblock \bibinfo{journal}{International Journal of Fracture}
  \bibinfo{volume}{249}, \bibinfo{pages}{51}.
\bibitem[{Kamarei et~al.(2026a)Kamarei, {Bueno De Castro}, Roesler and
  Lopez-Pamies}]{KBRLP26}
\bibinfo{author}{Kamarei, F.}, \bibinfo{author}{{Bueno De Castro}, S.},
  \bibinfo{author}{Roesler, J.R.}, \bibinfo{author}{Lopez-Pamies, O.},
  \bibinfo{year}{2026}a.
\newblock \bibinfo{title}{Fracture under body forces: {A}n accessible test and
  analysis}.
\newblock \bibinfo{journal}{Extreme Mechanics Letters} \bibinfo{volume}{In
  press}.
\bibitem[{Kamarei et~al.(2024)Kamarei, Kumar and Lopez-Pamies}]{KKLP24}
\bibinfo{author}{Kamarei, F.}, \bibinfo{author}{Kumar, A.},
  \bibinfo{author}{Lopez-Pamies, O.}, \bibinfo{year}{2024}.
\newblock \bibinfo{title}{The poker-chip experiments of synthetic elastomers
  explained}.
\newblock \bibinfo{journal}{Journal of the Mechanics and Physics of Solids}
  \bibinfo{volume}{188}, \bibinfo{pages}{105683}.
\bibitem[{Kamarei et~al.(2026b)Kamarei, Zeng, Dolbow and
  Lopez-Pamies}]{KZDLP26}
\bibinfo{author}{Kamarei, F.}, \bibinfo{author}{Zeng, B.},
  \bibinfo{author}{Dolbow, J.E.}, \bibinfo{author}{Lopez-Pamies, O.},
  \bibinfo{year}{2026}b.
\newblock \bibinfo{title}{Nine circles of elastic brittle fracture: {A} series
  of challenge problems to assess fracture models}.
\newblock \bibinfo{journal}{Computer Methods in Applied Mechanics and
  Engineering} \bibinfo{volume}{448}, \bibinfo{pages}{118449}.
\bibitem[{von K{\'a}rm{\'a}n(1911)}]{vonKarman1911}
\bibinfo{author}{von K{\'a}rm{\'a}n, T.}, \bibinfo{year}{1911}.
\newblock \bibinfo{title}{Festigkeitsversuche unter allseitigem druck
  [{S}trength tests under hydrostatic pressure]}.
\newblock \bibinfo{journal}{Zeitschrift des Vereines Deutscher Ingenieure}
  \bibinfo{volume}{55}, \bibinfo{pages}{1749--1757}.
\bibitem[{Knowles and Sternberg(1972)}]{KS1972}
\bibinfo{author}{Knowles, J.K.}, \bibinfo{author}{Sternberg, E.},
  \bibinfo{year}{1972}.
\newblock \bibinfo{title}{On a class of conservation laws in linearized and
  finite elastostatics}.
\newblock \bibinfo{journal}{Archive for Rational Mechanics and Analysis}
  \bibinfo{volume}{44}, \bibinfo{pages}{187--211}.
\bibitem[{Knowles and Sternberg(1973)}]{KS1973}
\bibinfo{author}{Knowles, J.K.}, \bibinfo{author}{Sternberg, E.},
  \bibinfo{year}{1973}.
\newblock \bibinfo{title}{An asymptotic finite-deformation analysis of the
  elastostatic field near the tip of a crack}.
\newblock \bibinfo{journal}{Journal of Elasticity} \bibinfo{volume}{3},
  \bibinfo{pages}{67--107}.
\bibitem[{Kumar et~al.(2020)Kumar, Bourdin, Francfort and
  Lopez-Pamies}]{KBFLP20}
\bibinfo{author}{Kumar, A.}, \bibinfo{author}{Bourdin, B.},
  \bibinfo{author}{Francfort, G.A.}, \bibinfo{author}{Lopez-Pamies, O.},
  \bibinfo{year}{2020}.
\newblock \bibinfo{title}{Revisiting nucleation in the phase-field approach to
  brittle fracture}.
\newblock \bibinfo{journal}{Journal of the Mechanics and Physics of Solids}
  \bibinfo{volume}{142}, \bibinfo{pages}{104027}.
\bibitem[{Kumar et~al.(2018a)Kumar, Francfort and Lopez-Pamies}]{KFLP18}
\bibinfo{author}{Kumar, A.}, \bibinfo{author}{Francfort, G.A.},
  \bibinfo{author}{Lopez-Pamies, O.}, \bibinfo{year}{2018}a.
\newblock \bibinfo{title}{Fracture and healing of elastomers: {A}
  phase-transition theory and numerical implementation}.
\newblock \bibinfo{journal}{Journal of the Mechanics and Physics of Solids}
  \bibinfo{volume}{112}, \bibinfo{pages}{523--551}.
\bibitem[{Kumar et~al.(2024)Kumar, Liu, Dolbow and Lopez-Pamies}]{KLDLP24}
\bibinfo{author}{Kumar, A.}, \bibinfo{author}{Liu, Y.},
  \bibinfo{author}{Dolbow, J.E.}, \bibinfo{author}{Lopez-Pamies, O.},
  \bibinfo{year}{2024}.
\newblock \bibinfo{title}{The strength of the {B}razilian fracture test}.
\newblock \bibinfo{journal}{Journal of the Mechanics and Physics of Solids}
  \bibinfo{volume}{182}, \bibinfo{pages}{105473}.
\bibitem[{Kumar and Lopez-Pamies(2020)}]{KLP20}
\bibinfo{author}{Kumar, A.}, \bibinfo{author}{Lopez-Pamies, O.},
  \bibinfo{year}{2020}.
\newblock \bibinfo{title}{The phase-field approach to self-healable fracture of
  elastomers: {A} model accounting for fracture nucleation at large, with
  application to a class of conspicuous experiments}.
\newblock \bibinfo{journal}{Theoretical and Applied Fracture Mechanics}
  \bibinfo{volume}{107}, \bibinfo{pages}{102550}.
\bibitem[{Kumar and Lopez-Pamies(2021)}]{KLP21}
\bibinfo{author}{Kumar, A.}, \bibinfo{author}{Lopez-Pamies, O.},
  \bibinfo{year}{2021}.
\newblock \bibinfo{title}{The poker-chip experiments of {G}ent and {L}indley
  (1959) explained}.
\newblock \bibinfo{journal}{Journal of the Mechanics and Physics of Solids}
  \bibinfo{volume}{150}, \bibinfo{pages}{104359}.
\bibitem[{Kumar et~al.(2018b)Kumar, Ravi-Chandar and Lopez-Pamies}]{KRLP18}
\bibinfo{author}{Kumar, A.}, \bibinfo{author}{Ravi-Chandar, K.},
  \bibinfo{author}{Lopez-Pamies, O.}, \bibinfo{year}{2018}b.
\newblock \bibinfo{title}{The configurational-forces view of fracture and
  healing in elastomers as a phase transition}.
\newblock \bibinfo{journal}{International Journal of Fracture}
  \bibinfo{volume}{213}, \bibinfo{pages}{1--16}.
\bibitem[{Kumar et~al.(2022)Kumar, Ravi-Chandar and Lopez-Pamies}]{KRLP22}
\bibinfo{author}{Kumar, A.}, \bibinfo{author}{Ravi-Chandar, K.},
  \bibinfo{author}{Lopez-Pamies, O.}, \bibinfo{year}{2022}.
\newblock \bibinfo{title}{The revisited phase-field approach to brittle
  fracture: {A}pplication to indentation and notch problems}.
\newblock \bibinfo{journal}{International Journal of Fracture}
  \bibinfo{volume}{237}, \bibinfo{pages}{83--100}.
\bibitem[{Lam\'e(1852)}]{Lame1852}
\bibinfo{author}{Lam\'e, G.}, \bibinfo{year}{1852}.
\newblock \bibinfo{title}{Leçons Sur la Th\'eorie Math\'ematique de
  l'\'elasticit\'e des Corps Solides}.
\newblock \bibinfo{publisher}{Mallet-Bachelier}.
\bibitem[{Lam\'e and Clapeyron(1831)}]{Lame1831}
\bibinfo{author}{Lam\'e, G.}, \bibinfo{author}{Clapeyron, E.},
  \bibinfo{year}{1831}.
\newblock \bibinfo{title}{M\'emoire sur l'\'equilibre int\'erieur des corps
  solides homog\`enes}.
\newblock \bibinfo{journal}{Journal f\"ur die reine und angewandte Mathematik}
  \bibinfo{volume}{7}, \bibinfo{pages}{381--413}.
\bibitem[{Larsen(2021)}]{Larsen21}
\bibinfo{author}{Larsen, C.J.}, \bibinfo{year}{2021}.
\newblock \bibinfo{title}{Variational fracture with boundary loads}.
\newblock \bibinfo{journal}{Applied Mathematics Letters} \bibinfo{volume}{121},
  \bibinfo{pages}{107437}.
\bibitem[{Larsen(2024)}]{Larsen24}
\bibinfo{author}{Larsen, C.J.}, \bibinfo{year}{2024}.
\newblock \bibinfo{title}{A local variational principle for fracture}.
\newblock \bibinfo{journal}{Journal of the Mechanics and Physics of Solids}
  \bibinfo{volume}{187}, \bibinfo{pages}{105625}.
\bibitem[{Larsen et~al.(2024)Larsen, Dolbow and Lopez-Pamies}]{LDLP24}
\bibinfo{author}{Larsen, C.J.}, \bibinfo{author}{Dolbow, J.E.},
  \bibinfo{author}{Lopez-Pamies, O.}, \bibinfo{year}{2024}.
\newblock \bibinfo{title}{A variational formulation of {G}riffith phase-field
  fracture with material strength}.
\newblock \bibinfo{journal}{International Journal of Fracture}
  \bibinfo{volume}{247}, \bibinfo{pages}{319--327}.
\bibitem[{Lef\`evre and Lopez-Pamies(2017)}]{LLP2017}
\bibinfo{author}{Lef\`evre, V.}, \bibinfo{author}{Lopez-Pamies, O.},
  \bibinfo{year}{2017}.
\newblock \bibinfo{title}{Nonlinear electroelastic deformations of dielectric
  elastomer composites: {II} --- {N}on-{G}aussian elastic dielectrics}.
\newblock \bibinfo{journal}{Journal of the Mechanics and Physics of Solids}
  \bibinfo{volume}{99}, \bibinfo{pages}{438--470}.
\bibitem[{Lef\`evre et~al.(2015)Lef\`evre, Ravi-Chandar and
  Lopez-Pamies}]{LRLP15}
\bibinfo{author}{Lef\`evre, V.}, \bibinfo{author}{Ravi-Chandar, K.},
  \bibinfo{author}{Lopez-Pamies, O.}, \bibinfo{year}{2015}.
\newblock \bibinfo{title}{Cavitation in rubber: {A}n elastic instability or a
  fracture phenomenon?}
\newblock \bibinfo{journal}{International Journal of Fracture}
  \bibinfo{volume}{192}, \bibinfo{pages}{1--23}.
\bibitem[{Leguillon(2002)}]{Leguillon2002}
\bibinfo{author}{Leguillon, D.}, \bibinfo{year}{2002}.
\newblock \bibinfo{title}{Strength or toughness? {A} criterion for crack onset
  at a notch}.
\newblock \bibinfo{journal}{European Journal of Mechanics-A/Solids}
  \bibinfo{volume}{21}, \bibinfo{pages}{61--72}.
\bibitem[{Lopez-Pamies and Kamarei(2025)}]{LPK25}
\bibinfo{author}{Lopez-Pamies, O.}, \bibinfo{author}{Kamarei, F.},
  \bibinfo{year}{2025}.
\newblock \bibinfo{title}{When and where do large cracks grow? {G}riffith
  energy competition constrained by material strength}.
\newblock \bibinfo{journal}{Extreme Mechanics Letters} \bibinfo{volume}{81},
  \bibinfo{pages}{102417}.
\bibitem[{Love(1906)}]{Love1906}
\bibinfo{author}{Love, A.E.H.}, \bibinfo{year}{1906}.
\newblock \bibinfo{title}{A Treatise on the Mathematical Theory of Elasticity}.
\newblock \bibinfo{publisher}{Cambridge University Press}.
\bibitem[{Mahajan and Ravi-Chandar(1989)}]{Ravi89}
\bibinfo{author}{Mahajan, R.V.}, \bibinfo{author}{Ravi-Chandar, K.},
  \bibinfo{year}{1989}.
\newblock \bibinfo{title}{An experimental investigation of mixed-mode
  fracture}.
\newblock \bibinfo{journal}{International Journal of Fracture}
  \bibinfo{volume}{41}, \bibinfo{pages}{235--252}.
\bibitem[{Mantič(2014)}]{Mantic2014}
\bibinfo{author}{Mantič, V.}, \bibinfo{year}{2014}.
\newblock \bibinfo{title}{Prediction of initiation and growth of cracks in
  composites. {Co}upled stress and energy criterion of the finite fracture
  mechanics}, in: \bibinfo{booktitle}{16th European Conference on Compopiste
  Materials}, pp. \bibinfo{pages}{1--16}.
\bibitem[{Mohr(1900)}]{Mohr1900}
\bibinfo{author}{Mohr, O.}, \bibinfo{year}{1900}.
\newblock \bibinfo{title}{Welche umst{\"a}nde bedingen die
  elastizit{\"a}tsgrenze und den bruch eines materiales? [{W}hat circumstances
  determine the elastic limit and fracture of a material?]}.
\newblock \bibinfo{journal}{Zeitschrift des Vereines deutscher Ingenieure}
  \bibinfo{volume}{44}, \bibinfo{pages}{1524--1530,1572--1574}.
\bibitem[{Navier(1827)}]{Navier1827}
\bibinfo{author}{Navier, C.L.}, \bibinfo{year}{1827}.
\newblock \bibinfo{title}{M\'emoire sur les lois de l'\'equilibre et du
  mouvement des corps solides \'elastiques}.
\newblock \bibinfo{journal}{M\'emoires de l'Acad\'emie Royale des Sciences de
  l'Institut de France} \bibinfo{volume}{7}, \bibinfo{pages}{375--393}.
\bibitem[{Navier(1864)}]{Navier1864}
\bibinfo{author}{Navier, C.L.}, \bibinfo{year}{1864}.
\newblock \bibinfo{title}{R{\'e}sum{\'e} des le{\c{c}}ons donn{\'e}es {\`a}
  l'{\'E}cole des ponts et chauss{\'e}es sur l'application de la m{\'e}canique
  {\`a} l'{\'e}tablissement des constructions et des machines}.
\newblock \bibinfo{edition}{3} ed., \bibinfo{publisher}{Dunod},
  \bibinfo{address}{Paris}.
\bibitem[{Nuismer(1975)}]{Nuismer75}
\bibinfo{author}{Nuismer, R.J.}, \bibinfo{year}{1975}.
\newblock \bibinfo{title}{An energy release rate criterion for mixed mode
  fracture}.
\newblock \bibinfo{journal}{International Journal of Fracture}
  \bibinfo{volume}{11}, \bibinfo{pages}{245--250}.
\bibitem[{Orowan(1948)}]{Orowan48}
\bibinfo{author}{Orowan, E.}, \bibinfo{year}{1948}.
\newblock \bibinfo{title}{Fracture and strength of solids}.
\newblock \bibinfo{journal}{Reports on Progress in Physics}
  \bibinfo{volume}{12}, \bibinfo{pages}{185--232}.
\bibitem[{Palaniswamy and Knauss(1978)}]{Knauss78}
\bibinfo{author}{Palaniswamy, K.}, \bibinfo{author}{Knauss, W.G.},
  \bibinfo{year}{1978}.
\newblock \bibinfo{title}{On the problem of crack extension in brittle solids
  under general loading}, in: \bibinfo{booktitle}{Mechanics Today}, pp.
  \bibinfo{pages}{87--148}.
\bibitem[{Panasyuk et~al.(1965)Panasyuk, Berezhnitskiy and
  Kovchik}]{Panasyuk65-RUS}
\bibinfo{author}{Panasyuk, V.V.}, \bibinfo{author}{Berezhnitskiy, L.T.},
  \bibinfo{author}{Kovchik, S.Y.}, \bibinfo{year}{1965}.
\newblock \bibinfo{title}{Propagation of an arbitrarily oriented rectilinear
  crack during extension of a plate}.
\newblock \bibinfo{journal}{Prikladnaya Mekhanika} \bibinfo{volume}{1},
  \bibinfo{pages}{48--55}.
\bibitem[{Qian and Fatemi(1996)}]{Fatemi96}
\bibinfo{author}{Qian, J.}, \bibinfo{author}{Fatemi, A.}, \bibinfo{year}{1996}.
\newblock \bibinfo{title}{Mixed mode fatigue crack growth: {A} literature
  survey}.
\newblock \bibinfo{journal}{Engineering Fracture Mechanics}
  \bibinfo{volume}{55}, \bibinfo{pages}{969--990}.
\bibitem[{Rankine(1858)}]{Rankine1858}
\bibinfo{author}{Rankine, W.J.M.}, \bibinfo{year}{1858}.
\newblock \bibinfo{title}{A Manual of Applied Mechanics}.
\newblock \bibinfo{publisher}{Richard Griffin and Company}.
\bibitem[{Rice(1966)}]{Rice1966}
\bibinfo{author}{Rice, J.R.}, \bibinfo{year}{1966}.
\newblock \bibinfo{title}{An examination of the fracture mechanics energy
  balance from the point of view of continuum mechanics}, in:
  \bibinfo{booktitle}{Proceedings of the First International Conference on
  Fracture}, \bibinfo{address}{Sendai}. pp. \bibinfo{pages}{309--340}.
\bibitem[{Rice(1968)}]{Rice1968}
\bibinfo{author}{Rice, J.R.}, \bibinfo{year}{1968}.
\newblock \bibinfo{title}{A path independent integral and approximate analysis
  of strain concentration by notches and cracks}.
\newblock \bibinfo{journal}{Journal of Applied Mechanics} \bibinfo{volume}{35},
  \bibinfo{pages}{379--386}.
\bibitem[{Richard(1985)}]{Richard85}
\bibinfo{author}{Richard, H.A.}, \bibinfo{year}{1985}.
\newblock \bibinfo{title}{Fracture predictions for cracks exposed to
  superimposed normal and shear stresses [{B}ruchvorhersagen bei \"uberlagerter
  {N}ormal-und {S}chubbeanspruchung von {R}issen]}.
\newblock \bibinfo{journal}{VDI-Forschungsheft} \bibinfo{volume}{631},
  \bibinfo{pages}{60}.
\bibitem[{Richard et~al.(2014)Richard, Schramm and Schirmeisen}]{Richard2014}
\bibinfo{author}{Richard, H.A.}, \bibinfo{author}{Schramm, B.},
  \bibinfo{author}{Schirmeisen, N.H.}, \bibinfo{year}{2014}.
\newblock \bibinfo{title}{Cracks on mixed mode loading --- {T}heories,
  experiments, simulations}.
\newblock \bibinfo{journal}{International Journal of Fatigue}
  \bibinfo{volume}{62}, \bibinfo{pages}{93--103}.
\bibitem[{Rivlin(1948a)}]{Rivlin48I}
\bibinfo{author}{Rivlin, R.S.}, \bibinfo{year}{1948}a.
\newblock \bibinfo{title}{Large elastic deformations of isotropic materials.
  {I}. {F}undamental concepts}.
\newblock \bibinfo{journal}{Philosophical Transactions of the Royal Society of
  London. Series A} \bibinfo{volume}{240}, \bibinfo{pages}{459--490}.
\bibitem[{Rivlin(1948b)}]{Rivlin48IV}
\bibinfo{author}{Rivlin, R.S.}, \bibinfo{year}{1948}b.
\newblock \bibinfo{title}{Large elastic deformations of isotropic materials.
  {IV}. {F}urther developments of the general theory}.
\newblock \bibinfo{journal}{Philosophical Transactions of the Royal Society of
  London. Series A} \bibinfo{volume}{241}, \bibinfo{pages}{379--397}.
\bibitem[{Rivlin and Thomas(1953)}]{RT53}
\bibinfo{author}{Rivlin, R.S.}, \bibinfo{author}{Thomas, A.G.},
  \bibinfo{year}{1953}.
\newblock \bibinfo{title}{Rupture of rubber. {I}. {C}haracteristic energy for
  tearing}.
\newblock \bibinfo{journal}{Journal of Polymer Science} \bibinfo{volume}{10},
  \bibinfo{pages}{291--318}.
\bibitem[{Royer(1986)}]{Royer86}
\bibinfo{author}{Royer, J.}, \bibinfo{year}{1986}.
\newblock \bibinfo{title}{A specimen geometry for plane mixed mode}.
\newblock \bibinfo{journal}{Engineering Fracture Mechanics}
  \bibinfo{volume}{23}, \bibinfo{pages}{763--775}.
\bibitem[{Saha et~al.(2024)Saha, Dolbow and Lopez-Pamies}]{SDLP24}
\bibinfo{author}{Saha, S.}, \bibinfo{author}{Dolbow, J.E.},
  \bibinfo{author}{Lopez-Pamies, O.}, \bibinfo{year}{2024}.
\newblock \bibinfo{title}{A {G}riffith description of fracture for
  non-monotonic loading with application to fatigue}.
\newblock \bibinfo{journal}{Journal of the Mechanics and Physics of Solids}
  \bibinfo{volume}{191}, \bibinfo{pages}{105754}.
\bibitem[{Saha et~al.(2026a)Saha, Manaughand, Moore, Roesler and
  Lopez-Pamies}]{SMMRLP26}
\bibinfo{author}{Saha, S.}, \bibinfo{author}{Manaughand, B.},
  \bibinfo{author}{Moore, B.J.}, \bibinfo{author}{Roesler, J.R.},
  \bibinfo{author}{Lopez-Pamies, O.}, \bibinfo{year}{2026}a.
\newblock \bibinfo{title}{A guide to fully characterize the fracture properties
  of cementitious materials from simple experiments}.
\newblock \bibinfo{journal}{Submitted} .
\bibitem[{Saha et~al.(2026b)Saha, Roesler and Lopez-Pamies}]{SRLP26}
\bibinfo{author}{Saha, S.}, \bibinfo{author}{Roesler, J.R.},
  \bibinfo{author}{Lopez-Pamies, O.}, \bibinfo{year}{2026}b.
\newblock \bibinfo{title}{Breaking four-point and three-point bending tests}.
\newblock \bibinfo{journal}{International Journal of Solids and Structures}
  \bibinfo{volume}{339}, \bibinfo{pages}{114169}.
\bibitem[{Sih(1974)}]{Sih74}
\bibinfo{author}{Sih, G.C.}, \bibinfo{year}{1974}.
\newblock \bibinfo{title}{Strain-energy-density factor applied to mixed mode
  crack problems}.
\newblock \bibinfo{journal}{International Journal of Fracture}
  \bibinfo{volume}{10}, \bibinfo{pages}{305--321}.
\bibitem[{Tada et~al.(1973)Tada, Paris and Irwin}]{Tada73}
\bibinfo{author}{Tada, H.}, \bibinfo{author}{Paris, P.},
  \bibinfo{author}{Irwin, G.}, \bibinfo{year}{1973}.
\newblock \bibinfo{title}{The stress analysis of cracks handbook 3rd edition}.
\newblock \bibinfo{publisher}{The American Society of Mechanical Engineers},
  \bibinfo{address}{New York}.
\bibitem[{Theocaris and Andrianopoulos(1982)}]{Theocaris82}
\bibinfo{author}{Theocaris, P.S.}, \bibinfo{author}{Andrianopoulos, N.},
  \bibinfo{year}{1982}.
\newblock \bibinfo{title}{A modified strain-energy density criterion applied to
  crack propagation}.
\newblock \bibinfo{journal}{Journal of Applied Mechanics} \bibinfo{volume}{49},
  \bibinfo{pages}{81--86}.
\bibitem[{Tian et~al.(1982)Tian, Lu and Zhu}]{Tian82}
\bibinfo{author}{Tian, D.C.}, \bibinfo{author}{Lu, D.Q.}, \bibinfo{author}{Zhu,
  J.J.}, \bibinfo{year}{1982}.
\newblock \bibinfo{title}{Crack propagation under combined stresses in three
  dimensional medium}.
\newblock \bibinfo{journal}{Engineering Fracture Mechanics}
  \bibinfo{volume}{16}, \bibinfo{pages}{5--16}.
\bibitem[{Truesdell and Noll(1965)}]{TN1965}
\bibinfo{author}{Truesdell, C.}, \bibinfo{author}{Noll, W.},
  \bibinfo{year}{1965}.
\newblock \bibinfo{title}{The Non-Linear Field Theories of Mechanics}.
\newblock \bibinfo{publisher}{Springer}.
\bibitem[{Ueda et~al.(1983)Ueda, Ikeda, Yao and Aoki}]{Ueda83}
\bibinfo{author}{Ueda, Y.}, \bibinfo{author}{Ikeda, K.}, \bibinfo{author}{Yao,
  T.}, \bibinfo{author}{Aoki, M.}, \bibinfo{year}{1983}.
\newblock \bibinfo{title}{Characteristics of brittle fracture under general
  combined modes including those under bi-axial tensile loads}.
\newblock \bibinfo{journal}{Engineering Fracture Mechanics}
  \bibinfo{volume}{18}, \bibinfo{pages}{1131--1158}.
\bibitem[{Williams and Ewing(1972)}]{Williams72}
\bibinfo{author}{Williams, J.G.}, \bibinfo{author}{Ewing, P.D.},
  \bibinfo{year}{1972}.
\newblock \bibinfo{title}{Fracture under complex stress --- {T}he angled crack
  problem}.
\newblock \bibinfo{journal}{International Journal of Fracture Mechanics,}
  \bibinfo{volume}{8}, \bibinfo{pages}{441--446}.
\bibitem[{Wu(1978)}]{Wu78}
\bibinfo{author}{Wu, C.H.}, \bibinfo{year}{1978}.
\newblock \bibinfo{title}{Fracture under combined loads by
  maximum-energy-release-rate criterion}.
\newblock \bibinfo{journal}{Journal of Applied Mechanics} \bibinfo{volume}{45},
  \bibinfo{pages}{553--558}.

\end{thebibliography}
\end{document}